\documentclass[a4paper,fleqn]{cas-sc}

\usepackage[numbers,sort&compress]{natbib}

\usepackage[justification=centering]{caption}
\usepackage{subcaption}
\usepackage{placeins}
\usepackage{float}
\usepackage{amsmath,amsfonts}
\usepackage{bm}
\usepackage{xcolor}
\usepackage{booktabs}

\usepackage{nomencl}
\makenomenclature

\graphicspath{{figs/}}

\usepackage{framed} 
\usepackage{multicol} 

\usepackage{nomencl} 
\makenomenclature
\renewcommand\nomgroup[1]{%
  \ifthenelse{\equal{#1}{A}}{%
    \item[\textbf{Acronyms}]}{
  \ifthenelse{\equal{#1}{R}}{%
    \item[\textbf{Roman Symbols}]}{
  \ifthenelse{\equal{#1}{G}}{%
    \item[\textbf{Greek Symbols}]}{
 {
  \ifthenelse{\equal{#1}{S}}{%
    \item[\textbf{Subscripts/superscripts}]}{
  \ifthenelse{\equal{#1}{X}}{%
    \item[\textbf{Other Symbols}]}{
  {}}}}}}}}
  
\renewcommand*\nompreamble{\begin{multicols}{2}}
\renewcommand*\nompostamble{\end{multicols}}

\begin{document}

\let\WriteBookmarks\relax
\def\floatpagepagefraction{1}
\def\textpagefraction{.001}
\shorttitle{Dimensional lattice Boltzmann method for transport phenomena simulation}
\shortauthors{Martins, Alvariño and Cabezas-Gómez}

\title [mode = title]{Dimensional lattice Boltzmann method for transport phenomena simulation without conversion to lattice units} 




\author[1]{Ivan Talão Martins}[orcid=0000-0003-3961-5638]
\ead{ivanmartins@usp.br}
\cormark[1]

\author[2]{Pablo Fariñas Alvariño}[orcid=0000-0002-9598-5249]
\ead{pablo.farinas@udc.es}

\author[1]{Luben Cabezas-Gómez}[orcid=0000-0002-9550-9453]
\ead{lubencg@sc.usp.br}

\address[1]{Department of Mechanical Engineering, Heat Transfer Research Group, São Carlos School of Engineering (EESC), University of São Paulo (USP), São Carlos 13566-590, Brazil}

\address[2]{Universidade da Coruña, Campus Industrial de Ferrol, Ferrol, 15403, A Coruña, Spain}

\cortext[cor1]{Corresponding author}

\begin{abstract}
In this paper it is proposed a dimensional Lattice Boltzmann Method (LBM) of wide application for simulating fluid flow and heat transfer problems. The proposed LBM consists in the numerical solution of the discrete lattice Boltzmann equation (LBE) using directly the variables in physical units, without the necessity of employing any particular unit conversion system. This includes the integration of LBE in the discrete physical domain using the spatial and time intervals and all the involved quantities in physical units. The dimensional LBM is proposed for both the single and multiple relaxation time schemes, considering the BGK (Bhatnagar-Gross-Krook) and MRT (Multiple-relaxation-time) collision operators, respectively. Several simple tests problems treating different physical phenomena such as one dimensional heat diffusion with source term, two dimensional forced convection with developed and developing flow in a channel under an oscillating and constant heat flux, two-phase stationary bubble in a liquid phase and two-phase dynamic layered Poiseuille flow, both under very high density and viscosity ratios, are simulated. The solutions of three additional problems considering one-dimensional advection-diffusion, isothermal channel flow and natural convection, are briefly explored in Appendices \ref{sec:appendixB}, \ref{sec:appendixC} and \ref{sec:appendixD}. All the numerical solutions were compared with analytical solutions, when available, or with finite difference solutions, otherwise, showing a very good agreement. The proposed method was also compared with the traditional LBM for the treated problems, showing the same accuracy. Besides the simulation of the applied problems employing physical units directly, which is more logical for applied problems, the proposed LBM allowed the solution of transport phenomena for more severe operational conditions. This includes the simulation of the two multiphase problems with liquid/gas density and gas/liquid kinematic viscosity ratios of about $43300$ and $470$ respectively, employing the Allen-Canh phase field model, commonly found in open literature. With base on the obtained results it is estimated that the proposed method could enhance the LBM use as simulation tool for the wide transport phenomena were it founds application.

\end{abstract}


\maketitle

\nomenclature[r]{$f_i$}{discrete density distribution function $[kg\ m^{-3}]$}%
\nomenclature[r]{$f_i^{eq}$}{discrete equilibrium density distribution function $[kg\ m^{-3}]$}%
\nomenclature[r]{$\mathbf{x}$}{position vector $[m]$}%
\nomenclature[r]{$\mathbf{c}$}{discrete velocity vector $[m\ s^{-1}]$}%
\nomenclature[r]{$c$}{lattice speed $[m\ s^{-1}]$}%
\nomenclature[r]{$t$}{time $[s]$}%
\nomenclature[r]{$[\mathbf{M}]$}{transformation matrix}%
\nomenclature[r]{$\mathbf{M}$}{moments which compound the transformation matrix rows}%
\nomenclature[r]{$h_i$}{discrete distribution function for interface-tracking LBE}%
\nomenclature[r]{$\mathbf{f}$}{vector with the distribution functions}%
\nomenclature[r]{$\mathbf{f^{eq}}$}{vector with the equilibrium distribution functions}%
\nomenclature[r]{$\mathbf{u}$}{velocity $[m\ s^{-1}]$}%
\nomenclature[r]{$w$}{weight functions}%
\nomenclature[r]{$c_s$}{sound speed $[m\ s^{-1}]$}%
\nomenclature[r]{$\mathbf{m}$}{vector with the moments of the distribution functions}%
\nomenclature[r]{$\mathbf{F}$}{external force field $[N\ m^{-3}]$}%
\nomenclature[r]{$\mathbf{F_m}$}{vector with moments of the forcing}%
\nomenclature[r]{$\hat{F_i}$}{forcing term for the momentum LBE $[kg\ s^{-1}m^{-3}]$}%
\nomenclature[r]{$\mathbf{F_b}$}{external body force $[N\ m^{-3}]$}%
\nomenclature[r]{$\mathbf{F_s}$}{surface tension force $[N\ m^{-3}]$}%
\nomenclature[r]{$\mathbf{L}$}{vector with the sizes of the domain $[m]$}%
\nomenclature[r]{$c_p$}{specific heat at constant pressure $[J\ kg^{-1}K^{-1}]$}%
\nomenclature[r]{$\dot{q}$}{heat generation term $[K\ s^{-1}]$}%
\nomenclature[r]{$q'''$}{volumetric heat generation $[W\ m^{-3}]$}%
\nomenclature[r]{$q''$}{heat flux $[W\ m^{-2}]$}%
\nomenclature[r]{$z_i$}{discrete pressure distribution function $[Pa]$}%
\nomenclature[r]{$M$}{mobility $[m^2\ s^{-1}]$}%
\nomenclature[r]{$p$}{pressure $[Pa]$}%
\nomenclature[r]{$W$}{interface width $[m]$}%
\nomenclature[r]{$\mathbf{n}$}{vector normal to the phase-interface}%
\nomenclature[r]{$s_i$}{term for $z^{eq}$ calculation}%
\nomenclature[r]{$C$}{conversion factors}%
\nomenclature[r]{$T$}{temperature $[K\ \mbox{or}\  ^{\circ}C]$}%
\nomenclature[r]{$E_2$}{global error, given by L-2 norm $[\%]$}%
\nomenclature[r]{$V$}{volume $[m^3]$}%
\nomenclature[r]{$R$}{radius $[m]$ }%
\nomenclature[r]{$g_i$}{discrete temperature distribution function $[K\ \mbox{or}\  ^{\circ}C]$}%
\nomenclature[r]{$L$}{domain length $[m]$}%
\nomenclature[r]{$H$}{domain height $[m]$}%
\nomenclature[r]{$d$}{diameter $[m]$}%
\nomenclature[r]{$k$}{thermal conductivity $[W\ m^{-1}K^{-1}]$}%
\nomenclature[r]{$\overline{h}$}{average heat transfer coefficient $[W\ m^{-1}K^{-1}]$}%
\nomenclature[r]{$I_e$}{electrical current $[A]$}%
\nomenclature[r]{$res$}{electrical resistivity $[\Omega m]$}%
\nomenclature[r]{$\dot{Q}'''$}{heat power per unit of volume $[W\ m^{-3}]$}%
\nomenclature[r]{$Per$}{perimeter $[m]$}%
\nomenclature[r]{$A_c$}{cross section area $[m^2]$}%
\nomenclature[r]{$u$}{speed in $x$ direction $[m\ s^{-1}]$}%
\nomenclature[r]{$Nu$}{Nusselt number}%
\nomenclature[r]{$Re$}{Reynolds number}%
\nomenclature[r]{$Ra$}{Rayleigh number}%
\nomenclature[r]{$Br$}{Brinkman number}%
\nomenclature[r]{$L_e$}{Hydrodynamic entrance length}%
\nomenclature[r]{$L_h$}{spaces between the oscillating heat flux}%
\nomenclature[r]{$q_s''$}{mean heat flux at channel wall $[W\ m^{-2}]$}%
\nomenclature[r]{$S$}{source term $[kg\ m^{-3}s^{-1}]$}%
\nomenclature[r]{$T_{\infty}$}{air mean temperature away of the fuse $[^{\circ}C]$}%
\nomenclature[r]{$P_{in}$}{Pressure inside the bubble $[Pa]$}%
\nomenclature[r]{$P_{out}$}{Pressure outside the bubble $[Pa]$}%
\nomenclature[r]{$x_c,y_c$}{center bubble coordinates $[m]$}%
\nomenclature[r]{$x,y$}{cartesian coordinates $[m]$}%

\nomenclature[g]{$[\mathbf{\Lambda}]$}{collision matrix}%
\nomenclature[g]{$\rho$}{density $[kg\ m^{-3}]$}%
\nomenclature[g]{$\nu$}{kinematic viscosity $[m^2\ s^{-1}]$}%
\nomenclature[g]{$\zeta$}{dynamic viscosity $[Pa\ s]$}%
\nomenclature[g]{$\eta$}{bulk viscosity $[Pa\ s]$}%
\nomenclature[g]{$\Omega$}{collision operator $[kg\ m^{-3}s^{-1}]$}%
\nomenclature[g]{$\tau$}{relaxation time for the momentum LBE $[s]$}%
\nomenclature[g]{$\omega$}{relaxation frequency $[s^{-1}]$}%
\nomenclature[g]{$\alpha$}{thermal diffusivity $[m^2\ s^{-1}]$}%
\nomenclature[g]{$\lambda$}{constant for heat flux thermal BCs }%
\nomenclature[g]{$\phi$}{order parameter}%
\nomenclature[g]{$\mu$}{chemical potential $[N\ m^{-2}]$}%
\nomenclature[g]{$\Psi$}{total free energy $[J]$}%
\nomenclature[g]{$\psi$}{volumetric free energy (or potential) $[J\ m^{-3}]$}%
\nomenclature[g]{$\beta$}{constant for $\mu$ calculation $[N\ m^{-2}]$}%
\nomenclature[g]{$\kappa$}{constant for $\mu$ calculation $[N]$}%
\nomenclature[g]{$\sigma$}{surface tension $[N\ m^{-1}]$}%
\nomenclature[g]{$\gamma$}{coordinate perpendicular to the phase-interface $[ m]$}%
\nomenclature[g]{$\chi$}{solution values (both numerical or analytical)}%
\nomenclature[g]{$\Delta x$}{discrete space interval $[m]$}%
\nomenclature[g]{$\Delta t$}{discrete time increment $[s]$}%
\nomenclature[g]{$\Delta P$}{pressure difference between inside and outside the bubble $[Pa]$}%
\nomenclature[g]{$\Delta T$}{temperature variation $[K]$}%
\nomenclature[g]{$\beta_{exp}$}{thermal expansion coefficient $[K^{-1}]$}%
\nomenclature[g]{$\Delta x_{next}$}{next discretization level $[m]$}%
\nomenclature[g]{$\Delta x_{next2}$}{next of the next discretization level $[m]$}%

\nomenclature[a]{BC}{Boundary condition}%
\nomenclature[a]{BGK}{Bhatnagar-Gross-Krook}%
\nomenclature[a]{EoS}{Equation of State}%
\nomenclature[a]{FD}{Finite Difference}%
\nomenclature[a]{FDM}{Finite Difference Method}%
\nomenclature[a]{NSE}{Navier-Stokes equations}%
\nomenclature[a]{LBM}{Lattice Boltzmann Method}%
\nomenclature[a]{LBE}{Lattice Boltzmann Equation}%
\nomenclature[a]{MRT}{Multiple-relaxation-time}%
\nomenclature[a]{Dim.}{Dimensional}%
\nomenclature[a]{Conv.}{Conventional}%
\nomenclature[a]{Sat.}{Saturation}%

\nomenclature[s]{MRT}{referent to the MRT collision operator}%
\nomenclature[s]{BGK}{referent to the BGK collision operator}%
\nomenclature[s]{$*$}{post-collision variables}%
\nomenclature[s]{$dim$}{referent to the dimensional LBM}%
\nomenclature[s]{$conv$}{referent to the conventional LBM}%
\nomenclature[s]{$i,j$}{discrete velocity directions}%
\nomenclature[s]{$f_i$}{referent to the density distr. function}%
\nomenclature[s]{$w$}{variables at the boundary wall}%
\nomenclature[s]{$b$}{referent to the boundary node}%
\nomenclature[s]{$b-1$}{next node after the boundary node}%
\nomenclature[s]{$tan$}{tangential part of the vector}%
\nomenclature[s]{$t$}{refers to time dimension}%
\nomenclature[s]{$T$}{refers to the thermal LBM}%
\nomenclature[s]{$g_i$}{refers to the temperature distr. function}%
\nomenclature[s]{$h_i$}{refers to $h_i$ distr. function}%
\nomenclature[s]{$\overline{i}$}{velocity direction opposite to $i$}%
\nomenclature[s]{$z,z_i$}{refers to $z_i$ distr. function}%
\nomenclature[s]{$\phi$}{refers to interface-tracking LBE}%
\nomenclature[s]{$l$}{refers to the liquid phase}%
\nomenclature[s]{$g$}{refers to the gas phase}%
\nomenclature[s]{$ref$}{quantity of reference}%
\nomenclature[s]{$num$}{refers to numerical solution}%
\nomenclature[s]{$ini$}{initial state variables}%
\nomenclature[s]{$m$}{average quantity}%
\nomenclature[s]{$FDM$}{value for/from FDM}%
\nomenclature[s]{$LBM$}{value for/from LBM}%
\nomenclature[s]{$var$}{refers to oscillating heat flux channel}%
\nomenclature[s]{$const$}{refers to constant heat flux channel}%
\nomenclature[s]{$\rho$}{refers to the zeroth moment of the transformation matrix}%
\nomenclature[s]{$e$}{refers to the energy moment of the transformation matrix}%
\nomenclature[s]{$\epsilon$}{refers to the energy square moment of the transformation matrix}%
\nomenclature[s]{$q_x$}{refers to the $x$ energy flux moment of the transformation matrix}%
\nomenclature[s]{$q_y$}{refers to the $y$ energy flux moment of the transformation matrix}%
\nomenclature[s]{$J_x$}{refers to the $x$ mass flux moment of the transformation matrix}%
\nomenclature[s]{$J_y$}{refers to the $y$ mass flux moment of the transformation matrix}%
\nomenclature[s]{$p_{xx},p_{xy}$}{refers to the stress tensor moment of the transformation matrix}%

\printnomenclature

\section{Introduction}


Nowadays, the lattice Boltzmann method (LBM) has been extensively used to simulate a wide range of transport phenomena such as fluid flow and heat transfer \citep{Chen_1998}, flows in porous media \citep{Kang_2002,Liu_2014,WANG_2016_2}, heat transfer with nanofluids \citep{Xuan_2005,Sheikholeslami_2014,WU_2017,KHOSHTARASH_2023}, multiphase flows with liquid-liquid or liquid-gas systems \citep{Rotman_Keller_1988,Shan_Chen_1993,Swift_1996,YAN_2011}, phase-change phenomena for thermal multiphase systems, both liquid-gas or solid-liquid \citep{Shan_Chen_1994,Fabritiis_1998,Miller_2001,Safari_2013,JARAMILLO_20221,LI_2021}, and many others. These transport phenomena are of great importance for several engineering applications considering petroleum, energy, nuclear, electronic and refrigeration industries, to mention a few.

The LBM is considered a mesoscopic method that consists in the numerical solution of the discrete Boltzmann transport equation in phase space and time, called the lattice Boltzmann equation (LBE). The LBE, firstly proposed by \cite{McNamara_Zanetti_1988}, is employed for finding the values of the particle distribution function in the discrete domain, allowing the calculation of the desired macroscopic quantities, such as density, velocity, concentration, temperature and others from its statistical moments \cite{McNamara_Zanetti_1988, Kruger2017, Succi}. Thus, the LBM allows to recover the macroscopic conservative laws through the consideration of the mesoscopic physical phenomena by the numerical solution of LBE, as proven in \cite{Wolf-Gladrow} by using the Chapman-Enskog analysis \citep{Chapman_Enskog}. 


Traditionally the LBM is solved in the so called lattice units or lattice scales, commonly considering unitary spatial and temporal increments, for convenience. These lattice units are related with the corresponding physical quantities that describe the physical problem to be solved in the macroscopic scales. Two main methods employed for establishing the relations between the physical and the lattice units for a particular problem are the dimensional analysis based on the use of Buckingham $\Pi$ theorem (firstly proposed by \cite{buckingham1914physically}) and the scaling method (or the principle of corresponding states for employing the thermodynamic equations of state (EoS)) \citep{Huang_2019,Mohamad_paper_2021}. The $\Pi$ theorem is commonly applied to infer dimensionless groups $\Pi$ from units of input variables and parameters in the absence of known governing equations \citep{bakarji2022dimensionally}. When the governing equations are known, the scale analysis by using the non-dimensionalization process allows to find the representative dimensionless numbers, as the equation coefficients in terms of specific references for the problem variables \citep{bakarji2022dimensionally}. These references must be known and constant \citep{Mohamad_paper_2021}.

Several works addressed and proposed different approaches for unit conversion in the LBM, see for example the following works to cite some of them \cite{Sukop,Su_2016,Kruger2017,Huang_2019,Mohamad_paper_2021,Wang_2022}. In general some works use the Buckingham $\Pi$ theorem to relate the physical scale with the lattice scale \citep{Sukop} for solving various problems related with the simulation of multiphase flows. Other authors \citep{Su_2016} made use of the scale analysis alone  for finding the leading dimensionless numbers and simulating melting and solidification processes. Other works, as pointed out by \cite{Mohamad_paper_2021}, can made use of the Buckingham $\Pi$ theorem and scale analysis together for considering in the conversion unit process properties such as specific heat capacity, viscosity, thermal diffusivity, etc, that have not reference state. In a recent study \cite{Huang_2019} proposed the use of Planck units as a reference for performing the conversion between physical and lattice units, respectively. The method was successfully applied for simulating the forced convection in tube banks considering heat sources, but requires the realization of more steps in the conversion unit process. 

\cite{Mohamad_paper_2021} proposed a general procedure for simulating diverse fluid flow and heat transfer tests problems, including a two-dimensional stationary droplet with the pseudopotential model. The procedure is based on the use of the same basic reference parameters of the physical scale and lattice scale to perform the conversion process. The methodology allows certain flexibility while ensures the the stability of the solution. Very recently \cite{Wang_2022} proposed a conversion strategy for simulating the liquid-vapor phase change with the pseudopotential method. The authors proposed the determination of conversion relations of fundamental units from the surface tension and EoS parameters related to fluid properties, in order to deduce the conversion relations of other quantities. The authors simulated a single bubble nucleation process recovering the latent heat of the fluid and the correct superheating temperature in physical units. 

However, the use of any of the proposed procedures implies in various previous computational steps before starting to simulate the problem with the LBM, and also in more calculations for post-processing the output simulated data. The procedures also require a carefully analysis and will depend on the particular solved problem, having a certain degree of complexity. In fact, \cite{Cates_2005}, simulated binary fluid mixtures in the presence of colloid particles and stated that the LBM cannot fully solve the hierarchy of length, energy and time-scales that arise in typical flows of complex fluids. Thus, it should be decided what physics to solve and what to leave unsolved, above all when colloidal particles were present in one or both of two fluid phases. Then, it is very important to chose the most relevant dimensionless numbers for a proper simulation of the macroscopic problem.

In the present paper it is proposed a procedure to solve the LBM using directly the variables in physical units, without the necessity of employing any particular unit conversion approach. The procedure is applied for the solution of four applied problems involving one-dimensional heat conduction with heat source; convective heat transfer considering a developed and a developing two-dimensional channel flow, both with constant and oscillating heat flux; the static two-phase flow problem of a bubble surrounded by liquid (both air-water and saturated liquid-vapor water systems), and the solution of the layered Poiseuille dynamic flow for the same two-phase systems than for the static problem. For these multiphase problems, it is considered a phase-field LBM based on the conservative Allen-Cahn equation for the interface-tracking. 


Furthermore, the solutions of three other classical problems are also presented in Appendices \ref{sec:appendixB}, \ref{sec:appendixC} and \ref{sec:appendixD}. All the LBM solutions are compared with analytical solutions when available, with the finite difference method (FDM) when necessary, or with benchmark solutions from literature, for both the proposed LBM and the traditional version of it, solved in lattice units. Considering the accuracy of the obtained results, it can be affirmed that the main paper novelty is the proposition of dimensional LBM that works in physical units, which was not found in the open literature and can allow a simpler and more applied implementation of the LBM.

The paper is divided in the following sections. In section \ref{sec:mathematical-modeling} are presented all the LBM models employed for the simulations of the analyzed problems. The proposed dimensional LBM model is presented in section \ref{sec:dimensional}, while the paper results are provided in section \ref{sec:problem-definition}. Finally the paper conclusions are presented in section \ref{sec:conclusions}.


\section{Mathematical modelling}
\label{sec:mathematical-modeling}

In this section it is explained the LBM employed for the numerical simulations presented in Sec. ~\ref{sec:problem-definition}.

\subsection{Lattice Boltzmann Method for fluid flow}
\label{standard_MLB}

The LBM is based on the discretization of the Boltzmann transport equation in the phase space formed by the velocity space, physical space and time \citep{Kruger2017}. Considering a second order discretization in time, the general LBE can be given by Eq. \ref{eq_lattice_boltzmann_geral}, which provides the evolution of the discrete distribution functions $f_i$ in space and time for each discrete velocity direction $i$. 

\begin{equation}
    f_i(\mathbf{x}+\mathbf{c_{i}} \Delta t, t + \Delta t) - f_i(\mathbf{x},t)  
    = \Delta t \left[ \Omega_i(\mathbf{x},t) +  S_{f_i}(\mathbf{x},t) \right]
    \label{eq_lattice_boltzmann_geral} \;.
\end{equation}

In Eq. \ref{eq_lattice_boltzmann_geral}, $S_{f_i}$ represents the source term, related with the presence of external forces in the case of fluid motion simulations. $\Delta t$ and $\Delta x$ are the discrete time and space intervals, while $\mathbf{c_i}$ are the discrete particle velocities in each $i$ direction and whose values depend on the selected velocity scheme. Usually the velocity schemes are defined as D$d$Q$q$, following \cite{Qian_1992}, where $d$ represents the spatial dimension of the simulation (one, two or three-dimensional) and $q$ is the considered number of discrete velocities.

The variable $\Omega_i$ of Eq. \ref{eq_lattice_boltzmann_geral} stands for the collision operator, which takes into account the effects of particle collisions and can be modeled in several forms. The simplest collision operator that allows the simulation of Navier-Stokes equations (NSE) is the Bhatnagar-Gross-Krook (BGK) operator \citep{BGK}, defined as: $\Omega^{BGK}_i = -(f_i - f_i^{eq})/\tau$, where $\tau$ consists in the relaxation time and $f_i^{eq}$ is the equilibrium distribution function, which represents when the system is in the equilibrium state. Considering now the BGK operator, the LBE can be re-written as shown in Eq. \ref{eq_lattice_boltzmann_BGK}.

\begin{equation}
    f_i(\mathbf{x}+\mathbf{c_{i}} \Delta t, t + \Delta t) - f_i(\mathbf{x},t)  
    = - \frac{\Delta t}{\tau} \left[f_i(\mathbf{x},t) - f_i^{eq}(\mathbf{x},t)  \right ] + S_{f_i}(\mathbf{x},t)\Delta t 
    \label{eq_lattice_boltzmann_BGK} \;.
\end{equation}

The equilibrium distribution function, $f_i^{eq}$, is given by Eq. \ref{feq_geral} for a general case \citep{Guo_book_2013,Kruger2017}. In this equation, $\rho$ represents the fluid density, $c_s$ is the lattice sound speed and $w_i$ are the weights for each velocity direction $i$. The values of these last two variables also depend of the chosen velocity scheme.

\begin{equation}
    f^{eq}_i(\mathbf{x},t) = w_i  \rho\left[1 + \frac{\textbf{c}_i\cdot \textbf{u}}{c_s^2} + \frac{(\textbf{c}_i \cdot \textbf{u})^2}{2c_s^4} - \frac{\textbf{u}\cdot \textbf{u}}{2c_s^2} \right ]
    \label{feq_geral}\;.
\end{equation}

In the case of the two-dimensional $D2Q9$ velocity scheme, used in this work, $c_s = c/\sqrt{3}$ and the respective velocity set ($\mathbf{c_i},{w_i}$) \citep{Qian_1992} is defined as shown in Eqs. \ref{velocity_D2Q9} and \ref{weights_D2Q9}. The variable $c$ used in all the presented relations is the lattice speed, defined as $c = \Delta x/\Delta t$.


\begin{equation} 
\label{velocity_D2Q9}
\mathbf{c_i} = c
\begin{cases} 
      (0,0), ~~~~~~~~~~~~~~~~~~~~~~~~~~~~~~~~~~~~~~~~~ i = 0,  \\
      (1,0), (0,1), (-1,0), (0,-1), ~~~~~~ i = 1,...,4, \\
      (1,1), (-1,1), (-1,-1), (1,-1), ~ i = 5,...,8. \\
   \end{cases}
\end{equation}

\begin{equation} 
\label{weights_D2Q9}
{w}_i =
\begin{cases} 
      4/9, ~~~~~~~~~~~~~~~~~~~~~~~~~~ i = 0,  \\
      1/9, ~~~~~~~~~~~~~~~~~~~~~~~~~~ i = 1,...,4, \\
      1/36 ~~~~~~~~~~~~~~~~~~~~~~~~~~ i = 5,...,8. \\
   \end{cases}
\end{equation}		

It is important to mention that, by the Chapman-Enskog analysis \citep{Chapman_Enskog}, it is possible to recover the NSE with sufficient degree of precision, establishing a link between the LBE and the NSE. This link is expressed by the relation of the relaxation time $\tau$ with the kinematic viscosity of the fluid $\nu$, which is represented by the following expression, 

\begin{equation}
    \nu = (\tau - 0.5\Delta t)c_s^2
    \label{eq_kinematic_viscosity} \;.
\end{equation}

When there are external forces acting in the domain, their effects can be considered by the inclusion of the source term $S_{f_i}$ in the LBE, see Eq. \ref{eq_lattice_boltzmann_geral}. There are several schemes for modeling the source term in the literature, but in this paper it was selected the proposed by \cite{guo-forces}, which is the most used according to \cite{Mohamad_forces}. This scheme avoids the presence of undesired derivatives in the continuity and momentum equations due to time discretization artifacts, which can occur with other schemes \citep{Kruger2017}. Thus, the source term for a given external field of forces $\mathbf{F}$ can be given by Eq. \ref{forcing_term}, where $\hat{F_i}$ is the forcing term. It is important to note that the dimensions of $\mathbf{F}$ are force per unit of volume ($N\ m^{-3}$). 

\begin{equation}
    S_{f_i} =\left(1 - \frac{\Delta t}{2\tau}\right)\hat{F_i} = \left(1 - \frac{\Delta t}{2\tau}\right)w_i\left[ \frac{\mathbf{c_i}-\mathbf{u}}{c_s^2} + \frac{(\mathbf{c_i} \cdot \mathbf{u})\mathbf{c_i}}{c_s^4}\right]\cdot \mathbf{F}
    \label{forcing_term}
\end{equation}

Nevertheless, in several more complex flows problems the BGK collision operator may not be sufficient to guarantee good accuracy or stability. Having this in mind, the multiple-relaxation-time (MRT) collision operator, introduced into LBE by \cite{Higuera_1989}, can be used to increase the stability of the method, mainly when low values of $\tau$ are involved \citep{Kruger2017}. The principles of this operator is to perform the collision step in the space of the distribution functions moments. This collision operator can be generally defined as $\Omega^{MRT}_i=-\left[\mathbf{M}^{-1}\mathbf{\Lambda}\mathbf{M}\right]_{ij}(f_j-f_j^{eq})$, being $[\mathbf{M}]$ the transformation matrix and $[\mathbf{\Lambda}]$ the collision matrix. 

The transformation matrix is responsible to calculate the moments $\mathbf{m}$ of the distribution functions, such as $\mathbf{m} = [\mathbf{M}] \mathbf{f}$. Similarly, the equilibrium moments can be obtained as $\mathbf{m^{eq}} = [\mathbf{M}] \mathbf{f^{eq}}$. Considering the $D2Q9$ velocity scheme, the non-dimensional transformation matrix can be given by Eq. \ref{M_matrix} \citep{Lallemand_2000}. In this matrix, each row is related with one moment of the distribution function, being $e$ the energy (2nd order moment), $\epsilon$ the energy squared (4th order moment), $J_x$ and $J_y$ the mass fluxes (1st order moments), $q_x$ and $q_y$ the energy fluxes (3rd order momnets) and $p_{xx}$ and $p_{xy}$ the components of the stress tensor (2nd order moments).
 
\begin{equation}
    [\mathbf{M}]= \left(\begin{matrix} \mathbf{M}_{\rho}\\
    \mathbf{M}_e\\
    \mathbf{M}_{\epsilon}\\
    \mathbf{M}_{J_x}\\
    \mathbf{M}_{q_x}\\
    \mathbf{M}_{J_y}\\
    \mathbf{M}_{q_y}\\
    \mathbf{M}_{p_{xx}}\\
    \mathbf{M}_{p_{xy}}\\\end{matrix}\right) = \left (\begin{matrix}
    1 & 1 & 1 & 1 & 1 & 1 & 1 & 1 & 1 \\
    -4 & -1 & -1 & -1 & -1 & 2 & 2 & 2 & 2\\
    4 & -2 & -2 & -2 & -2 & 1 & 1 & 1 & 1 \\
    0 & 1 & 0 & -1 & 0 & 1 & -1 & -1 & 1 \\
    0 & -2 & 0 & 2 & 0 & 1 & -1 & -1 & 1 \\
    0 & 0 & 1 & 0 & -1 & 1 & 1 & -1 & -1 \\
    0 & 0 & -2 & 0 & 2 & 1 & 1 & -1 & -1 \\
    0 & 1 & -1 & 1 & -1 & 0 & 0 & 0 & 0 \\
    0 & 0 & 0 & 0 & 0 & 1 & -1 & 1 & -1 
    \end{matrix}\right)
    \label{M_matrix}
\end{equation}

Despite of being represented by $\Omega^{MRT}_i$, it is common for the MRT collision step to be completely performed in the moment space. Then, the post-collision functions $f_i^*$ can be recuperated by $\mathbf{f^*} = [\mathbf{M}]^{-1}\mathbf{m^*}$, which are later used to perform the streaming process: $f_i(\mathbf{x}+\mathbf{c}_i \Delta t,t + \Delta t) = f_i^*(\mathbf{x},t)$.

The \cite{guo-forces} force scheme can be also adapted to the MRT operator \citep{Kuzmin_PhD,Kuzmin_Mohamad}. In this case, the collision process in the moment space can be represented by Eq. \ref{colisao_MRT}, being $\mathbf{F_m} = [\mathbf{M}] \mathbf{\hat{F}} $ the moments of the forcing term.

\begin{equation}
    \mathbf{m^*} = \mathbf{m} - \Delta t [\mathbf{\Lambda}]
\left(\mathbf{m} - \mathbf{m^{eq}} \right ) + \Delta t\left(1 - \frac{\Delta t}{2}[\mathbf{\Lambda}] \right) \mathbf{F_m}
    \label{colisao_MRT}
\end{equation}

The collision matrix can be defined as a diagonal matrix $[\mathbf{\Lambda}] = \mbox{diag}(\omega_0,...,\omega_{q-1})$, in which the main diagonal are composed by the relaxation frequencies $\omega_i$, related with each moment of the distribution functions. In this matrix, the frequencies associated with conserved moments are zero, because they are not affected by the collision process \citep{Lallemand_2000}. 

Thus, for the $D2Q9$ velocity scheme the collision matrix can be defined as $[\mathbf{\Lambda}] = \mbox{diag}(0,\omega_e,\omega_{\epsilon},0,\omega_q,0,\omega_q,\omega_{\nu},\omega_{\nu})$ \citep{Kruger2017}. The two last frequencies are related with the kinematic viscosity of the fluid, being defined as $\omega_{\nu} = 1/\tau$. Also, the relaxation frequency related with the energy can be associated with the bulk viscosity of the fluid as $\eta = (\omega_e^{-1} - 0.5\Delta t)c_s^2$ \citep{Guo_book_2013}. The other frequencies can be chosen without significant effects in the transport coefficients, considering a second order approach of the transport equations, but values between $1$ and $2$ are recommended \citep{Lallemand_2000}. However, these authors also recommended that the 3rd order relaxation parameter can be related with the 2nd order ones as $\omega_q = 3(2-\omega_{\nu})/(3 - \omega_{\nu})$. 

The macroscopic quantities of interest can be obtained taking the zero and first moments of the distribution function $f_i$ \citep{Succi}. Considering the force scheme, these moments can be calculated by Eqs. \ref{zero_moment_f} and \ref{first_moment_f}, respectively, for both BGK and MRT collision operators.
 
\begin{equation}
    \rho(\mathbf{x},t) = \sum_{i=0}^{q-1} f_i (\mathbf{x},t)
    \label{zero_moment_f}
\end{equation}

\begin{equation}
    \mathbf{u}(\mathbf{x},t)\rho (\mathbf{x},t) = \sum_{i=0}^{q-1} \mathbf{c_i} f_i (\mathbf{x},t) + \frac{\Delta t}{2}  \mathbf{F}(\mathbf{x},t)
    \label{first_moment_f}
\end{equation}

In the simulations performed in this paper, the scheme adopted for formulating the boundary conditions is the \textit{link-wise}. In this scheme the boundaries lie on lattice links, being positioned at a distance of $0.5\Delta x$ in relation to the position of the domain physical boundaries \cite{Kruger2017}.

Five main boundary conditions (BC) were used for the simulations of fluid flow, namely: inlet with prescribed velocity, outlet at atmosphere pressure, periodic boundaries, fixed non-slip walls and symmetric boundaries. For both inlet and fixed walls, it was considered the bounce-back scheme \citep{Ladd_1994,Ladd_2001}, which can be represented by Eq. \ref{BB_vel}. In this equation, $\overline{i}$ represents the opposite directions of $i$, $\mathbf{u_w}$ and $\rho_w$ are the boundary velocity and density, respectively, and $\mathbf{x_b}$ is the position of the boundary node. In the case of fixed walls, the BC is reduced to $f_{\overline{i}}(\mathbf{x_b},t + \Delta t) = f_i^*(\mathbf{x_b},t)$.

\begin{equation}
    f_{\overline{i}}(\mathbf{x_b},t + \Delta t) = f_i^*(\mathbf{x_b},t) - 2w_i\rho_w \frac{\mathbf{c_i}\cdot \mathbf{u_w}}{c_s^2}
    \label{BB_vel}
\end{equation}

For the fluid outlet, it was used the anti-bounce-back scheme \citep{anti_BB}, that can be expressed by Eq. \ref{anti_BB_vel}. To fix the atmosphere pressure with this BC, the $\rho_w$ value is assumed equal to the fluid density at atmospheric pressure (usually it is used the average density of the fluid, considered in an equilibrium state), and the velocity is calculated by extrapolation: $\mathbf{u_w} \approx \mathbf{u}(\mathbf{x_b}) + 0.5[\mathbf{u}(\mathbf{x_b}) - \mathbf{u}(\mathbf{x_{b-1}})]$. In this case $\mathbf{x_{b-1}}$ represents the next node into the domain in the normal direction of the boundary \citep{Kruger2017}.

\begin{equation}
    f_{\overline{i}}(\mathbf{x_b},t + \Delta t) = -f_i^*(\mathbf{x_b},t) + 2w_i\rho_w\left[1 + \frac{(\mathbf{c_i}\cdot \mathbf{u_w})^2}{2c_s^4} - \frac{\mathbf{u_w}\cdot \mathbf{u_w}}{2c_s^2} \right]
    \label{anti_BB_vel}
\end{equation}

The symmetric boundaries are applied using Eq. \ref{simetria} \citep{Kruger2017}. In this relation, $j$ are the indices of the population with same tangential velocity than $i$, but with opposite normal velocity as $(c_{j,tan};c_{j,n}) = (c_{i,tan};-c_{i,n})$. For example, considering a top boundary as symmetric for the $D2Q9$ velocity scheme, there are three unknown distribution functions: $f_4$, $f_7$ and $f_8$. So, $f_6$ has the same tangential velocity than $f_7$ (being $\mathbf{c_6} = c(-1,1)$ and $\mathbf{c_7} = c(-1,-1)$), but opposite normal velocity, thus $f_7(\mathbf{x_b}+\mathbf{c_{7,tan}}\Delta t,t + \Delta t) = f_6^*(\mathbf{x_b},t)$. Being the $f_7$ tangential velocity $c_{7,tan} = -c$, the final relation is $f_7(\mathbf{x_b}-(\Delta x,0),t + \Delta t) = f_6^*(\mathbf{x_b},t)$. The same procedure can be applied for the other functions.

\begin{equation}
    f_j(\mathbf{x_b}+\mathbf{c_{j,tan}}\Delta t,t + \Delta t) = f_i^*(\mathbf{x_b},t)
    \label{simetria}
\end{equation}

Lastly, for the periodic boundaries, the leaving distribution functions at one side are the unknown functions which arrive at the opposite boundary. Then, this BC can be represented by $f_i(\mathbf{x_b},t + \Delta t) = f_i^*(\mathbf{x_b} + \mathbf{L}-\mathbf{c_i}\Delta t,t)$, being $\mathbf{L}$ the size of the domain at the normal direction of the boundary.

\subsection{Thermal Lattice Boltzmann Method}
\label{thermal_LBM}

Because of its generality, the LBM can be also used to simulate heat transfer problems. There are several methods proposed in the literature to deal with heat transfer \citep{Guo_book_2013}. The chosen to be used in ours simulations is the double-distribution-function model. 

In this model, it is defined another distribution function $g_i$ for the temperature field $T$ \citep{Guo_book_2013,Inamuro_Yoshiko_Suzuki}. Thus, while the \textit{momentum} evolution is simulated by Eq. \ref{eq_lattice_boltzmann_BGK}, the evolution of the temperature field is calculated by Eq. \ref{eq_lattice_boltzmann_BGK_T} for the BGK operator \citep{Chen_Zhangs}.

\begin{equation}
    g_i(\mathbf{x}+\mathbf{c_{i}} \Delta t, t + \Delta t) - g_i(\mathbf{x},t)  
    = - \frac{\Delta t}{\tau_T} \left[g_i(\mathbf{x},t) - g_i^{eq}(\mathbf{x},t)  \right ] + S_{g_i}(\mathbf{x},t)\Delta t 
    \label{eq_lattice_boltzmann_BGK_T}
\end{equation}

By the Chapman-Enskog analysis it is possible to recover the energy conservation equation, given a relation between the relaxation time of the thermal LBE, $\tau_T$, and the thermal diffusivity of the fluid $\alpha$: $\alpha = (\tau_T - 0.5\Delta t)c_s^2$. This relation represents the link between the LBE and the energy conservation equation.

In this case, the source term is related with the volumetric heat generation instead of external forces, and can be similarly formulated by Eq. \ref{forcing_term_g} \citep{He_temp_source,Kruger2017}. It is important to note that $\dot{q}$ has units of $[K\ s^{-1}]$, because it is defined  as $\dot{q} = q'''/(\rho c_p) $, were $q'''$ is the volumetric heat generation in ($W\ m^{-3}$) and $c_p$ stands for the specific heat at constant pressure of the substance or material in ($J\ kg^{-1}K^{-1}$).



\begin{equation}
    S_{g_i} =  \left(1 - \frac{\Delta t}{2\tau}\right) w_i\dot{q}
    \label{forcing_term_g}
\end{equation}

The equilibrium distribution function, $g_i^{eq}$, considered in Eq. \ref{eq_lattice_boltzmann_BGK_T}, is related with the temperature and can be defined by Eq. \ref{geq_geral}.

\begin{equation}
    g^{eq}_i = w_i  T\left(1 + \frac{\textbf{c}_i\cdot \textbf{u}}{c_s^2}\right )
    \label{geq_geral}
\end{equation}

It should be mention that in the previous relation it was used a linear velocity-dependent form of the equilibrium distribution function \citep{Kruger2017}. But if necessary, it is also possible to consider a second order form, which is given by: $g^{eq}_i = w_i T \left[1 + \frac{\textbf{c}_i\cdot \textbf{u}}{c_s^2} + \frac{(\textbf{c}_i \cdot \textbf{u})^2}{2c_s^4} - \frac{\textbf{u}\cdot \textbf{u}}{2c_s^2} \right ]$. For the simulations performed in this work, it was used only the first order equilibrium distribution function (Eq. \ref{geq_geral}), being enough for the desired results.

The macroscopic temperature can be found from the zero moment of the distribution function $g_i$. However, differently of the flow field, in the presence of volumetric heat generation it is necessary to add an extra term for the temperature calculation, as shown in Eq. \ref{zero_moment_g} \citep{Seta_2013}.

\begin{equation}
    T(\mathbf{x},t) = \sum_{i=0}^{q-1} g_i (\mathbf{x},t) + \frac{\Delta t}{2}\dot{q}
    \label{zero_moment_g}
\end{equation}

The MRT collision operator was also used for the thermal LBM. In this case, considering the $D2Q9$ velocity set, the collision matrix can assume also a diagonal form. In this paper it will be used the same relaxation parameter as employed in \cite{Martins_2022}: $[\mathbf{\Lambda}_T] = \mbox{diag}(0,1,1,\omega_T,1,\omega_T,1,1,1)$, being $\omega_T = 1/\tau_T$. The transformation matrices remain equal to those used for the simulation of fluid flow.


In the case of the heat transfer simulations with the thermal LBM, it were considered five kinds of boundary conditions: inlet with prescribed temperature, fixed walls with prescribed heat flux, fixed walls with prescribed temperature, outlet and symmetric boundaries.

For inlets with fixed temperature it was used the anti-Bounce-Back scheme \citep{Zhang_BC_T,Li_mei_klausner_2013,Fei_Luo}. This BC is given by Eq. \ref{anti_BB_T}, where $g_i^*$ is the post-collision distribution function, $\overline{i}$ is the opposite direction to $i$, $\mathbf{x_b}$ represents the coordinates of the boundary node and the subscript $w$ indicates the variables values at the boundary wall.  

\begin{equation}
    g_{\overline{i}}(\mathbf{x_b},t + \Delta t) = -g_i^*(\mathbf{x_b},t) + 2w_iT_w\left[1 + \frac{(\mathbf{c_i}\cdot \mathbf{u_w})^2}{2c_s^4} - \frac{\mathbf{u_w}\cdot \mathbf{u_w}}{2c_s^2} \right]
    \label{anti_BB_T}
\end{equation}

The same BC can be also used to treat fixed walls with prescribed temperatures \citep{Kruger2017}. In this case, the boundary velocity $\mathbf{u_w}$ is set to zero and the BC reduces itself to $ g_{\overline{i}}(\mathbf{x_b},t + \Delta t) = -g_i^*(\mathbf{x_b},t) + 2w_iT_w$.

The fixed wall Neumann BC (prescribed heat flux) was modeled with the scheme given by \cite{Li_mei_klausner_2017} for the $D2Q9$ velocity set with the halfway boundary, based on the modification of the scheme proposed by \cite{Yoshida_Nagaoka_2010} for the $D2Q5$. This BC can be given by Eq. \ref{BB_T}, where $\lambda_i$ \ is a constant and $q''$ is the heat flux normal to the boundary (units of $[W\ m^{-2}]$). There are many possible values of $\lambda_i$ according to the authors, so in this paper it will be selected $\lambda_i = 2w_i/c_s^2$, giving $\lambda_{1,2,3,4} = 4/6$ and $\lambda_{5,6,7,8} = 1/6$.

\begin{equation}
    g_{\overline{i}}(\mathbf{x_b},t + \Delta t) = g_i^*(\mathbf{x_b},t) + \lambda_i \left(\frac{\Delta t}{\Delta x}\right)\left( \frac{q''}{\rho c_p}\right)
    \label{BB_T}
\end{equation}

The symmetry boundary condition is the same used for the fluid moment distribution functions, as previously explained in section \ref{standard_MLB}.

For the outlets, it was considered a first-order extrapolation scheme: $ g_i(\mathbf{x_b},t + \Delta t) = g_i(\mathbf{x_{b-1}},t+\Delta t)$. It is important to mention that the first order scheme was used instead of the second order because it is more stable than the other \citep{Mohamad_book}. 

\subsection{Lattice Boltzmann Method for two-phase systems}
\label{multiphase_LBM}

There are several models in the literature proposed to simulate multi-phase and multi-component systems with the Lattice Boltzmann method \citep{Rotman_Keller_1988,Shan_Chen_1993,Shan_Chen_1994,Swift_1996,Guo_book_2013,Sukop,LI_2016_rev}. In this paper, the \cite{Liang_2018} model is briefly described and then applied for simulating two-phase systems with the new proposed dimensional approach. This multi-phase model is based on the evolution of two distribution functions: $h_i$ and $z_i$, to capture the interface movement and the pressure field, respectively. 

The authors developed a model similar to the proposed by \cite{Ren_2016} and \cite{WANG_2016_2}, based on the use of the conservative Allen-Cahn equation for the interface tracking, given by Eq. \ref{Allen_cahn} \citep{Chiu_Lin}. In this equation,  $\phi$ is the order parameter, being responsible to identify the region occupied by each phase, assuming $\phi = 1$ for the liquid region and $\phi = 0$ for the gas one. $W$ is the interface width (the model considers a diffuse interface between the different phases), $M$ stands for the mobility and $\mathbf{n}$ is the normal direction to the interface, that can be calculated as $\mathbf{n} = \nabla \phi / |\nabla \phi|$.

\begin{equation}
    \partial_t \phi + \nabla \cdot (\phi \mathbf{u}) = \nabla \cdot \left[ M \left(\nabla \phi  - \frac{4\phi (1 - \phi)}{W}\mathbf{n}\right)\right]
	\label{Allen_cahn}
\end{equation} 

The LBE responsible for the evolution of $h_i$ in the simulations (interface movement) is provided by Eq. \ref{eq_lattice_boltzmann_interface}, considering the BGK collision operator. The source term $S_{h_i}$ is defined in such a way that the Eq. \ref{Allen_cahn} can be recovered by the Chapman-Enskog analysis. Thus, this term is given by Eq. \ref{source_term_interface}. 

\begin{equation}
    h_i(\mathbf{x}+\mathbf{c_{i}} \Delta t, t + \Delta t) - h_i(\mathbf{x},t)  
    = - \frac{\Delta t}{\tau_{\phi}} \left[h_i(\mathbf{x},t) - h_i^{eq}(\mathbf{x},t)  \right ] + S_{h_i}(\mathbf{x},t)\Delta t 
    \label{eq_lattice_boltzmann_interface}
\end{equation}

\begin{equation}
    S_{h_i} = \left (1 - \frac{\Delta t}{2\tau_{\phi}} \right)w_i \frac{ \mathbf{c_i} \cdot \left [ \partial_t (\phi \mathbf{u}) + c_s^2 \frac{4\phi(1-\phi)}{W}\mathbf{n} \right ]}{c_s^2}
    \label{source_term_interface}
\end{equation}

The temporal derivative in Eq. \ref{source_term_interface} can be calculated by the explicit Euler's scheme: $\partial_t (\phi \mathbf{u})\approx [\phi(t) \mathbf{u}(t) - \phi(t-\Delta t) \mathbf{u}(t-\Delta t) ]/\Delta t$ \citep{Liang_2019}. In addition, $\tau_{\phi}$ is the relaxation time for the interface tracking LBE, which can be related with the mobility as $M = (\tau_{\phi} - 0.5\Delta t)c_s^2$. The equilibrium distribution function is calculated considering a first order expansion by Eq. \ref{heq_liang}.

\begin{equation}
	h^{eq}_i = \phi w_i \left(1 + \frac{\textbf{c}_i\cdot \textbf{u}}{c_s^2}\right)
	\label{heq_liang}
\end{equation}

Similarly to the traditional LBM, the macroscopic quantities are obtained by the moments of the distribution functions. Thus, the order parameter is calculated by the zero moment of $h_i$, using Eq. \ref{zero_moment_h}, and both the macroscopic density and kinematic viscosity can be calculated as a linear function of $\phi$, such as $\rho = \rho_g + \phi(\rho_l - \rho_g)$ and $\nu = \nu_g + \phi(\nu_l - \nu_g)$. In these relations the subscript $g$ refers to gas properties and $l$, to the liquid ones.

\begin{equation}
    \phi(\mathbf{x},t) = \sum_{i=0}^{q-1} h_i (\mathbf{x},t)
    \label{zero_moment_h}
\end{equation}

In the case of the pressure evolution, for developing the LBE, the authors performed an incompressible transformation similar to that proposed by \cite{HCZ}, obtaining the Eq. \ref{eq_lattice_boltzmann_pressure} for the BGK collision operator. However, it is important to mention that the MRT operator can also be used to perform the collision process for $z_i$ functions, applying Eq. \ref{colisao_MRT} for the moments of these functions. In the case of $D2Q9$, a possible value to the collision matrix can be $[\mathbf{\Lambda}] = \mbox{diag}(1,1,1,1,3(2-1/\tau)/(3 - 1/\tau),1,1,1/\tau,1/\tau)$.

\begin{equation}
    z_i(\mathbf{x}+\mathbf{c_{i}} \Delta t, t + \Delta t) - z_i(\mathbf{x},t)  
    = - \frac{\Delta t}{\tau} \left[z_i(\mathbf{x},t) - z_i^{eq}(\mathbf{x},t)  \right ] + S_{z_i}(\mathbf{x},t)\Delta t 
    \label{eq_lattice_boltzmann_pressure}
\end{equation}

For calculating the equilibrium distribution functions, the authors proposed the Eq. \ref{zeq_liang} in order to satisfy the divergence free condition. In this equation, $s_i$ represents the term calculated by Eq. \ref{s_i_Liang} \citep{Liang_2014,Liang_2016}. Also, in Eq. \ref{zeq_liang} $p$ is defined as the total pressure.

\begin{equation}
    z_i^{eq} = 
    \begin{cases}
    	\frac{p}{c_s^2} (w_i - 1) + \rho s_i(\mathbf{u}),\mbox{ if } i=0\\
    	\frac{p}{c_s^2} w_i + \rho s_i(\mathbf{u}),\mbox{ if } i\neq 0
    	\end{cases}
	\label{zeq_liang}
\end{equation}

\begin{equation}
	s_i (\mathbf{u}) = w_i \left(\frac{\textbf{c}_i\cdot \textbf{u}}{c_s^2} + \frac{(\textbf{c}_i \cdot \textbf{u})^2}{2c_s^4} - \frac{\textbf{u}\cdot \textbf{u}}{2c_s^2} \right)
	\label{s_i_Liang}
\end{equation}

In order to recover the NSE for a two-phase system, represented by Eq. \ref{NS} ($\zeta$ is the dynamic viscosity), the relaxation time is related again with the kinematic viscosity as $\nu = (\tau - 0.5\Delta t)c_s^2$. In this case, two body forces are employed. $\mathbf{F_b}$ is defined as the body force acting over the domain and $\mathbf{F_s}$ is the force related with the surface tension. There are several forms in the literature for calculating this last force. However, in this paper it will be used the potential form related to the chemical potential, $\mu$, and expressed as: $\mathbf{F_s} = \mu \nabla \phi$ \citep{Jacqmin_1999,Zou_He_2013}. Therefore, the forcing therm for the pressure evolution LBE is defined by Eq. \ref{forcing_term_pressure_evolution} \citep{Liang_2019}. 

\begin{equation}
\frac{\partial (\rho \mathbf{u})}{\partial t} + \nabla \cdot (\rho \mathbf{u} \mathbf{u}) = -\nabla p + \nabla \cdot [\zeta (\nabla \mathbf{u} + \nabla \mathbf{u}^T)] + \mathbf{F_s} + \mathbf{F_b} 
    \label{NS}
\end{equation}

\begin{equation}
    S_{z_i} =\left(1 - \frac{\Delta t}{2\tau}\right)\hat{F_i^z} = \left(1 - \frac{\Delta t}{2\tau}\right)w_i\left[ \frac{\mathbf{c_i} \cdot (\mathbf{F_s} + \mathbf{F_b})}{c_s^2} + \frac{(\mathbf{u} \nabla \rho):(\mathbf{c_i}\mathbf{c_i})}{c_s^2}\right]
    \label{forcing_term_pressure_evolution}
\end{equation}

Now, the macroscopic quantities related with the pressure and momentum of the fluid are given by Eqs. \ref{p_Liang} and \ref{u_Liang}, respectively. It is important to mention that in those equations the pressure is dependent on the velocity. Thus, it must be determined after calculating the velocity by Eq. \ref{u_Liang}.

\begin{equation}
	p(\mathbf{x},t) = \frac{c_s^2}{(1 - w_0)}\left [\sum_{i=1}^{q-1} z_i(\mathbf{x},t) + \frac{\Delta t}{2} \mathbf{u}(\mathbf{x},t)  \cdot \nabla \rho + \rho s_0(\mathbf{u}) \right ]
	\label{p_Liang}
\end{equation}

\begin{equation}
	\mathbf{u}(\mathbf{x},t) \rho(\mathbf{x},t)  = \sum_{i=0}^{q-1} \mathbf{c_i} z_i(\mathbf{x},t) + \frac{\Delta t}{2}(\mathbf{F_s} + \mathbf{F_b})
	\label{u_Liang}
\end{equation}

The knowledge of the chemical potential, $\mu$, is very important for performing the simulations with the presented LBM. This thermodynamic property can be calculated from the free energy, $\Psi$, of the two-phase system \citep{Reichl}. The free energy is a function of density, and knowing that the relation between $\rho$ and $\phi$ is linear, the chemical potential can be found as: $\mu = \partial_{\phi} \Psi(\phi)$. The total free energy of the system can be calculated by Eq. \ref{free_energy}, where $\psi(\phi)$ stands for the volumetric free energy (or the potential) and $\kappa$ is a constant related with the strength of the surface tension \citep{Jamet}.

\begin{equation}
    \Psi = \int_V \psi(\phi) + \frac{\kappa}{2}|\nabla \phi|^2 dV
    \label{free_energy}
\end{equation}

In regions close to the critical point, some simplifications of the fluid equation of state (EoS) can be made \citep{Rowlinson_widom}, resulting in the following expression: $\psi(\phi) \approx \beta \phi^2 (1 - \phi)^2$, where $\beta$ is a constant. Even though this relationship was developed for regions nearest the critical point, this simplification has been largely used by the multi-phase LBM models based on the mean-field theory \citep{Zheng_2006,Lee_Liu_2010,Fakhari_2010,Fakhari_2016}. 

Then, considering that $\mu = \partial_{\phi} \Psi(\phi) = \partial_{\phi}\psi - \kappa \nabla^2 \phi$ \citep{Bray_1994}, the chemical potential can be calculated according to Eq. \ref{chemical_potential}. The constants $\kappa$ and $\beta$ are related with both the interface thickness ($W$) and the surface tension of the fluid ($\sigma$) such as $\kappa = \frac{3}{2}\sigma W$ and $\beta = \frac{12\sigma}{W}$. 

\begin{equation}
    \mu = \partial_{\phi}\psi - \kappa \nabla^2 \phi = 4\beta \phi(\phi - 1)(\phi - 0.5) - \kappa \nabla^2 \phi
    \label{chemical_potential}
\end{equation}

In addition, the equilibrium of a planar interface between two phases can be modeled by Eq. \ref{diffuse_interface}, being $\gamma(\mathbf{x})$ the coordinate perpendicular to the interface. In order to avoid instabilities, the macroscopic density and $\phi$ are both initialized with this equilibrium profile at the beginning of the simulations. 

\begin{equation}
    \phi(\mathbf{x}) = 0.5 + 0.5\tanh{\left(\frac{2\gamma(\mathbf{x})}{W} \right)}
    \label{diffuse_interface}
\end{equation}

In previous equations, like Eq. \ref{chemical_potential}, it is evident the necessity of computing the second order derivative of $\phi$. From the linear relation between $\rho$ and $\phi$, the density gradients can be re-written as $\nabla \rho = (\rho_l - \rho_g) \nabla \phi$. Thus, the density gradient can be calculated through a first derivative of $\phi$. Following \cite{Liang_2018}, the spatial gradients and the Laplacian of $\phi$ can be calculated using a second-order isotropic central scheme given by Eqs. \ref{gradient} and \ref{laplacian}, respectively.

\begin{equation}
   \nabla \phi (\mathbf{x}) = \sum_{i \neq 0}\frac{w_i \mathbf{c_i}\phi(\mathbf{x} + \mathbf{c_i}\Delta t)}{c_s^2\Delta t} \label{gradient}
\end{equation}

\begin{equation}
   \nabla^2 \phi (\mathbf{x}) = \sum_{i \neq 0}\frac{2w_i \left [\phi(\mathbf{x} + \mathbf{c_i}\Delta t)-\phi(\mathbf{x})\right]}{c_s^2\Delta t^2} \label{laplacian}
\end{equation}

In relation to the interface normal vector, it is important to mention that its calculation in the numerical code must be performed carefully. This vector is defined as $\mathbf{n}=\nabla \phi/|\nabla \phi|$, thereby there are nodes were $|\nabla \phi| = 0$ and, consequently, the division by this term can assume extremely high values. Therefore, the referred division was performed only where $|\nabla \phi| \neq 0$, otherwise $\mathbf{n}=\mathbf{0}$. 

\section{The dimensional LBM}
\label{sec:dimensional}

It is common in the LBM the consideration of the variables in dimensionless lattice units instead of in physical ones \citep{Succi,Mohamad_book,Kruger2017}. This requires the use of dimensionless numbers for representing all physical parameters. The non-dimensionalization process generally involves the choice of reference unit scales, which are defined by independent conversion factors and also by the employment of similarity laws, in order to obtain the dimensionless values for all the physical quantities involved  \citep{Kruger2017}. There are several methods proposed in the literature for performing this non-dimensionalization process, as well for mapping the physical properties of a particular system to the lattice scales and vice-versa. Some of these methods are presented by \cite{Su_2016,Huang_2019,Mohamad_paper_2021,Wang_2022}. These works give an idea of the complexity involved in this conversion unit process, which is greater for the simulation of multiphase and multicomponent problems.

To differentiate between the LBM proposed in this paper and the traditional LBM, which is based on the non-dimensionalization process, in this work the former is called "dimensional LBM", while the latter, the "conventional LBM". The numerical simulations accomplished by the conventional LBM were developed with the models presented in section \ref{sec:mathematical-modeling}. The procedure followed to define the independent conversion factors was similar to the proposed by \cite{Kruger2017}. As length, time and mass are fundamental quantities, first the respective conversion factors related to them ($C_t$, $C_x$ and $C_m$) are chosen for performing the non-dimensionalization process. Then, using these conversion factors some non-dimensional variables (represented by $\tilde{}$) can be given as $\tilde{t} = t/C_t$, $\tilde{x} = x/C_x$, $\tilde{\mathbf{u}} = \mathbf{u}/C_u$ and $\tilde{\rho} = \rho/C_{\rho}$, where $C_{\rho} = C_m/C_x^3$ and $C_u = C_x/C_t$. The same procedure is applied to determine the conversion factors for other quantities, such as pressure, surface tension, etc, which are defined according to the physical units of each variable. However, for the non-dimensionalization of the temperature it is usually set a reference temperature $T_{ref}$, in such a way that $\tilde{T} = (T - T_{ref})/C_T$. 

In addition, it is important to highlight that the values of $C_t$, $C_x$ and $C_m$ are generally set in such a way that the non-dimensional density, the discrete space and discrete time intervals become $\Delta \tilde{t}=\Delta \tilde{x}=\tilde{\rho}=1.0$, being $\Delta \tilde{t} = \Delta t/C_t$ and $\Delta \tilde{x} = \Delta x/C_x$. To obtain these unitary values, the three independent conversion factors must be defined as $C_t = \Delta t$, $C_x = \Delta x$ and $C_m = \rho C_x^3$. An important observation to be made is that, using the conventional LBM, the forces and the heat generation terms must be converted to dimensionless quantities before being applied into the source terms calculation (Eqs. \ref{forcing_term} and \ref{first_moment_f} for the momentum LBE, Eqs. \ref{forcing_term_g} and \ref{zero_moment_g} for the thermal LBM and Eqs. \ref{source_term_interface}, \ref{forcing_term_pressure_evolution} and \ref{u_Liang} for the multiphase LBM). In this work it is employed the common uniform square lattice which considers the same spatial discrete interval in all coordinates (e.g., for a two-dimensional case, $\Delta y = \Delta x$). 

Therefore, in the present work it is introduced the "dimensional LBM", consisting in the application of the LBM in its dimensional form. In other words, the non-dimensionalization process is ignored and the values of $\Delta x$, $\Delta t$, $c = \Delta x/ \Delta t$ and of the macroscopic variables (density, velocity, temperature, viscosity, and others) are all kept in physical units, preferentially in the SI unit system. This proposed approach completely avoids the use of the unit conversion procedure, usually employed for: mapping the physical units of the data input into lattice ones, performing the numerical simulations in lattice units, and mapping the data output from the lattice space back to the physical one. Therefore, it is not necessary to select neither to employ any conversion parameters and the entire simulation, including data input and output, is performed in physical units.

For the simulations performed in the paper, it were used both the BGK and MRT collision operators. In the case of the BGK operator, the proposed dimensional LBM uses all the relations as were presented in section \ref{sec:mathematical-modeling}, keeping all the values in physical units ($\Delta t$, $\Delta x$, $c$, and all others). However, for the MRT collision operator, the transformation and the collision matrices must be mapped to the physical scale using the correct physical units, because they are originally defined in their dimensionless forms in lattice units. 

To dimensionalize the transformation matrix $[\mathbf{M}]$, it should be modified multiplying each row by the correct physical dimension of the respective moment. This is done multiplying the rows by $c^n$, being $n$ given by the moment order related with each row. Therefore, the dimensional form of $[\mathbf{M}]_{dim}$ for $D2Q9$ velocity set is given by Eq. \ref{M_matrix_dim}. The matrix $[\mathbf{M}]_{dim}^{-1}$ can be obtained naturally, just performing the inversion of the dimensional transformation matrix, and is showed in appendix \ref{sec:appendixA}.

\begin{equation}
    [\mathbf{M}]_{dim}= \left(\begin{matrix} c^0 \cdot \mathbf{M}_{\rho}\\
    c^2 \cdot\mathbf{M}_e\\
    c^4 \cdot\mathbf{M}_{\epsilon}\\
    c^1 \cdot\mathbf{M}_{J_x}\\
    c^3 \cdot\mathbf{M}_{q_x}\\
    c^1 \cdot\mathbf{M}_{J_y}\\
    c^3 \cdot\mathbf{M}_{q_y}\\
    c^2 \cdot\mathbf{M}_{p_{xx}}\\
    c^2 \cdot\mathbf{M}_{p_{xy}}\\\end{matrix}\right) = \left (\begin{matrix}
    1 & 1 & 1 & 1 & 1 & 1 & 1 & 1 & 1 \\
    -4c^2 & -c^2 & -c^2 & -c^2 & -c^2 & 2c^2 & 2c^2 & 2c^2 & 2c^2\\
    4c^4 & -2c^4 & -2c^4 & -2c^4 & -2c^4 & c^4 & c^4 & c^4 & c^4 \\
    0 & c & 0 & -c & 0 & c & -c & -c & c \\
    0 & -2c^3 & 0 & 2c^3 & 0 &c^3 & -c^3 & -c^3 & c^3 \\
    0 & 0 & c & 0 & -c & c & c & -c & -c \\
    0 & 0 & -2c^3 & 0 & 2c^3 & c^3 & c^3 & -c^3 & -c^3 \\
    0 & c^2 & -c^2 & c^2 & -c^2 & 0 & 0 & 0 & 0 \\
    0 & 0 & 0 & 0 & 0 & c^2 & -c^2 & c^2 & -c^2 
    \end{matrix}\right)
    \label{M_matrix_dim}
\end{equation}

Besides the previous changes, the relaxation rates that are not directly related with the relaxation time must also be dimensionalized for the collision matrix $[\mathbf{\Lambda}]_{dim}$. Thus, considering again the $D2Q9$ velocity scheme, the dimensional collision matrix for the momentum LBE (dimensional) can be given as explained in section \ref{standard_MLB}, but the relation for $\omega_q$ must be redefined by $\omega_q = (3/\Delta t)(2/\Delta t -\omega_{\nu})/(3/\Delta t - \omega_{\nu})$ and the arbitrary values ($\omega_{\epsilon}$) should vary between $1/\Delta t$ and $2/\Delta t$, instead of $1$ and $2$.

Now, for the thermal dimensional LBM, the dimensional collision matrix can be defined as $[\mathbf{\Lambda}_T]_{dim} = \mbox{diag}(0,1/\Delta t,1/\Delta t,\omega_T, 1/\Delta t,\omega_T,1/\Delta t,1/\Delta t,1/\Delta t)$, being still $\omega_T = 1/\tau_T$. Similarly, for the momentum equation of the dimensional multiphase LBM the following relations should be applied: $[\mathbf{\Lambda}]_{dim} = \mbox{diag}(1/\Delta t,1/\Delta t,1/\Delta t,1/\Delta t,\omega_{q2}, \break 1/\Delta t, 1/\Delta t, 1/\tau,1/\tau)$, with $\omega_{q2} = (3/\Delta t)(2/\Delta t -\omega_{\nu})/(3/\Delta t - \omega_{\nu})$.

In the main, the proposed dimensional LBM consists in a rather simple modification of the conventional LBM, but the implications are very considerable. The methodology allows the solution of applied transport phenomena problems in their own physical units without the necessity of a difficult, and some times cumbersome, intermediate step of unit conversion between physical and lattice scales. This enables the setting of the numerical solution using the input variables in their natural physical units, and also the control of the numerical solution convergence, stability and accuracy by changing only the $\Delta x$ and $\Delta t$ values. 

It should be noted that the dimensional LBM may present the same stability, accuracy and numerical order of the standard LBM, but its use is simpler and direct. These issues are not addressed in the present paper, where it is shown the application and validity of the proposed procedure through the obtainment of very accurate solution of various physical problems.

\section{Results}
\label{sec:problem-definition}

In this section are presented the simulation results obtained with the proposed dimensional LBM for four main problems treated in the subsections to come. The results are compared with those obtained with the conventional LBM to access the correctness of the proposed LBM. Also, in order to evaluate the method performance and accuracy, the obtained LBM solutions were compared with available analytical or numerical finite difference (FD) solutions.

The comparisons between LBM and reference solutions were performed using the global error defined by $L_2$ relative error norm \citep{CFD_1,CFD_2} defined by Eq. \ref{error_norm_2}. For the case of FD solutions, it was performed a convergence study to guarantee the good quality of the reference solutions. To attain this aim it was established that the global errors between two simulations, one with a grid size of $\Delta x$ and the other with $\Delta x_{next} = \Delta x /2$ (reference solution), must be less or equal to 0.01\%, meaning $E_2 \leq 0.01\%$.

\begin{equation}
E_2(\%) = 100\sqrt{\frac{\sum_{\mathbf{x}} (\chi_{ref} - \chi_{num})^2}{\sum_{\mathbf{x}} \chi_{ref}^2}}
    \label{error_norm_2}
\end{equation}

The LBM is naturally a transient numerical method. Then, in order to check if the numerical solutions reached the steady-state it was applied the condition given by Eq. \ref{steady_state_cond}. In this relation, $\chi$ represents the main variable of the problem (velocity for the Poiseuille flow, or temperature for the forced convection problems, for example) to be checked, and $\tilde{t} + 1000$ is the instant 1000 time steps after $\tilde{t}$.

\begin{equation}
 \mbox{max}\left[\chi(\mathbf{x},\tilde{t}+1000) - \chi(\mathbf{x},\tilde{t})\right] \leq 10^{-8}
    \label{steady_state_cond}
\end{equation}

For the resolution of the problems with the dimensional LBM, first it is set arbitrary values for $\Delta x$ and $\Delta t$. As the stability criteria are the same than for the conventional LBM, it is needed to be carefully about the $\tau$ values, which must not be too close to $\Delta t/2$. Then, if the simulation is unstable it is necessary to readjust the $\Delta x$ and $\Delta t$ in order to attain the required stability. For some more complex cases the stability of the simulations can be attained using the MRT operator instead of the BGK one. In the case of poor results, a grid refinement can be performed to obtain better precision. As the non-dimensionalization process is not performed for the dimensional LBM, the adjustment of $\Delta x$ and $\Delta t$ can be done directly and in a simpler way, similarly to the traditional numerical methods, such as finite difference or finite element method.

All the thermodynamic and transport properties employed for the fluids simulations were calculated using the free Coolprop python library \citep{CoolProp}. 

\subsection{One-dimensional heat diffusion}

The first problem is related with the determination of the axial temperature distribution of a fuse employed for preventing a break of a power electronic module due to a high current. This problem was taken from \cite{Nellis_Klein}, problem 3.8-1, and it is addressed here because this is an interesting engineering problem treating about one dimensional heat conduction. 

The fuse is a wire (with no insulation) with length $L = 0.08 m$ and diameter $d = 0.0015 m$. The surface of the fuse wire loses heat by convection to the air at $T_{\infty} = 20^oC$ with a heat transfer coefficient of $\overline{h} = 5W\ m^{-2}K^{-1}$. The fuse is made of an aluminum alloy with the following properties, which are assumed to be not dependent of the temperature: $\rho = 2700 kg\ m^{-3}$, $k = 150 W\ m^{-1}K^{-1}$, $c_p = 900 J\ kg^{-1}K^{-1}$ and electrical resistivity of $res = 1\cdot 10^{-7} \Omega\  m$. 

Initially, the fuse is at an uniform temperature of $T_{ini} = T_{\infty} = 20^oC$ when, at time $t = 0$, it is exposed to a current of $I_e = 100 A$, resulting in an uniform volumetric heat generation within the fuse material. Both ends of the fuse ($x=0$ and $x=L$) are kept at constant temperature of $T_w = 20^oC$. It is asked to found the axial temperature variation with time of the fuse, neglecting radial and angular temperature variations due to the small fuse diameter.  

The governing equation for the energy conservation in the fuse is given by Eq. \ref{conver_energy_fuse}. This equation is obtained considering that the heat source due to the current passage is modeled as $q''' = (16I_c^2res)/(\pi^2 d^4)$, and that the total heat lost by convection per unit of fuse volume is given by $\dot{Q}_{conv}''' = (\pi d L)\overline{h} (T - T_{\infty})/(0.25\pi d^2 L) = 4\overline{h} (T - T_{\infty})/d$.

\begin{equation}
    \frac{\partial T}{\partial t} = \alpha \frac{\partial^2 T}{\partial x^2} + \frac{q'''}{\rho c_p} - \frac{4\overline{h}(T - T_{\infty})}{\rho c_pd}
    \label{conver_energy_fuse}
\end{equation}


The problem was solved considering both transient and stead-state solutions. The numerical simulations with both LBM models were developed using the BGK collision operator and the $D1Q3$ velocity scheme, with lattice velocities of $\mathbf{c_0} = 0$, $\mathbf{c_1} = c$ and $\mathbf{c_2} = -c$, and the respective weights $w_0 = 4/6$, $w_1 = 1/6$ and $w_2 = 1/6$. The sound speed remains $c_s = c/\sqrt{3}$ \citep{Qian_1992}. The discrete time and space intervals employed in all simulations were $\Delta x = 4\cdot 10^{-4}m$ and $\Delta t = 2.5\cdot 10^{-4}s$. It was calculated a single source term for the LBM considering the sum of the volumetric heat generation and the convective losses by $\dot{q} = q'''/(\rho c_p) - 4\overline{h}(T - T_{\infty})/(\rho c_pd)$. In order to avoid implicitness with the LBM, the convection losses were calculated considering the temperature determined from the previous time. About the BCs, the both fixed temperatures at $x=0$ and $x=L$ were implemented with Eq. \ref{anti_BB_T}.

The transient LBM solutions were compared with numerical solutions obtained with the finite difference (FD) method, considering a forward time central space FD scheme for the solution of Eq. \ref{conver_energy_fuse}, with $\Delta x_{FDM} = 4.0 \cdot 10^{-4}m$ and $\Delta t_{FDM} = 2.5 \cdot 10^{-4}s$, satisfying $E_2 \leq 0.01\%$. The steady-state LBM results were compared with the analytical solution provided by Eq. \ref{sol_analitica_fusivel}, where $ m = \sqrt{(\overline{h}Per)/(k A_c)}$, and $Per$ and $A_c$ are the fuse perimeter and cross-sectional area, respectively.

\begin{equation}
    T(x) = \left(T_w - T_{\infty} - \frac{\dot{q}'''A_c}{\overline{h}Per} \right)\left[\left(1+\frac{e^{mL}-1}{e^{-mx}-e^{mx}}\right)e^{mx} + \left(\frac{1-e^{mL}}{e^{-mx}-e^{mx}}\right)e^{-mx} \right] + T_{\infty} + \frac{\dot{q}'''A_c}{\overline{h}Per}
    \label{sol_analitica_fusivel}
\end{equation}

\begin{figure}[h!]
    \centering
    \begin{subfigure}[b]{0.45\textwidth}
        \centering
        \includegraphics[width=\textwidth]{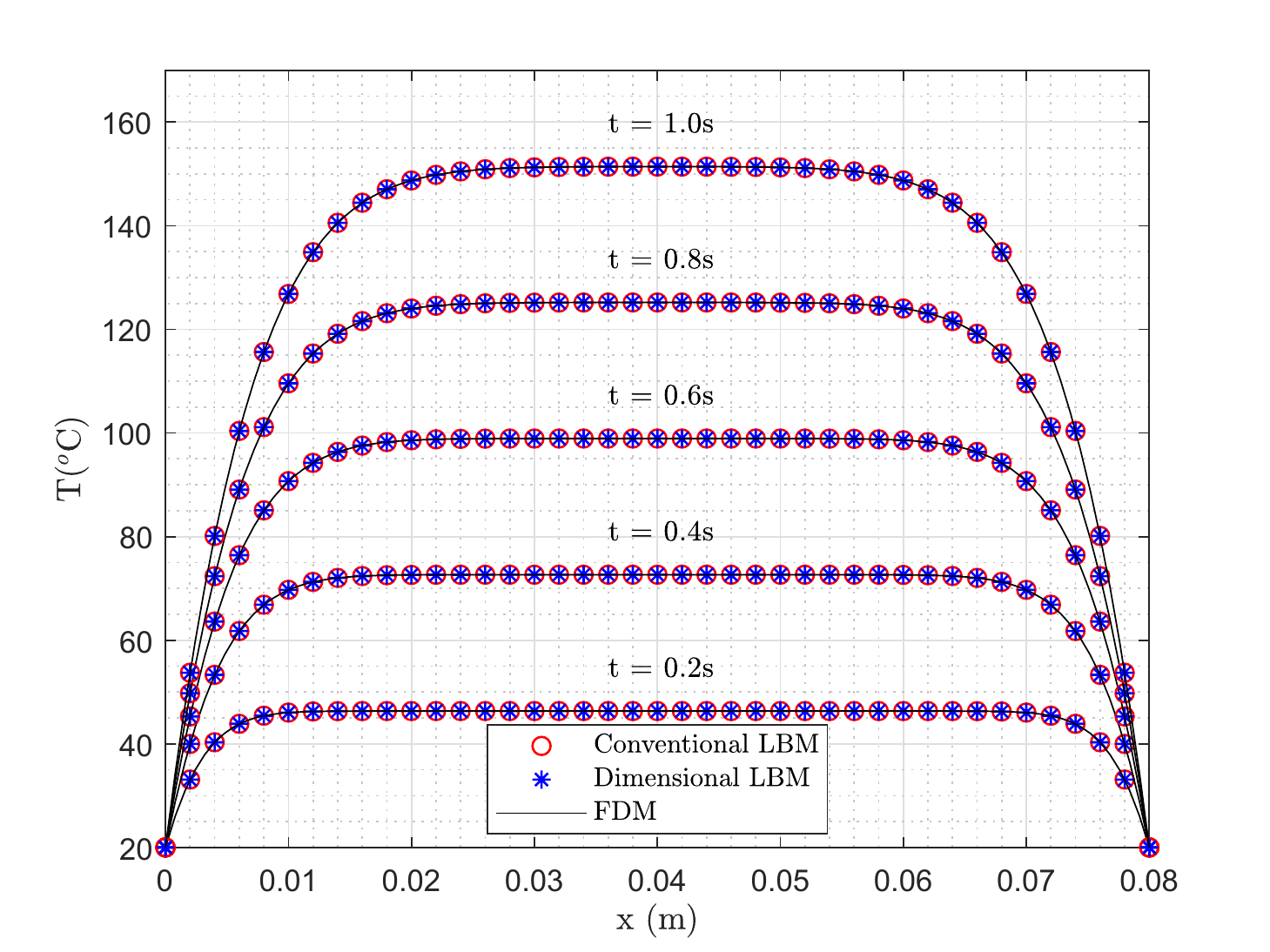}
        \caption{}\label{T_fusivel}
    \end{subfigure}
    \hfill
    \begin{subfigure}[b]{0.45\textwidth}
        \centering
        \includegraphics[width=\textwidth]{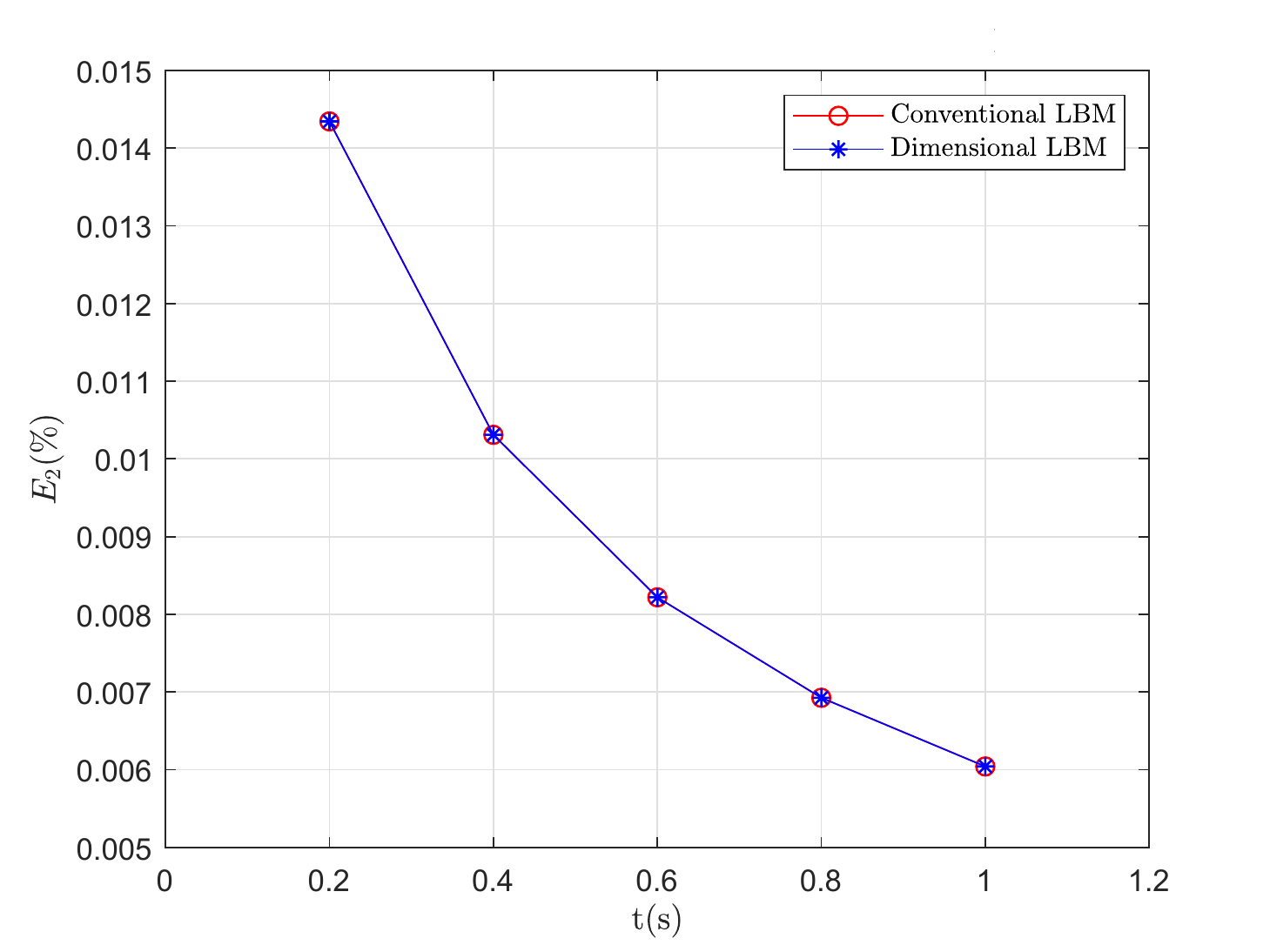}
        \caption{}\label{}
        \label{Erros_fusiveis}
    \end{subfigure}
       \caption{\centering (a) One-dimensional axial temperature variation of the fuse for some time steps, obtained by the FD scheme and the conventional and dimensional LBM. (b) Temporal variation of the global errors for the conventional and dimensional LBM, in comparison with the FD solutions.}
       \label{Resultados_fusiveis_transiente}
\end{figure}

The transient solutions for conventional and dimensional LBM and for the FD method at various times, as well as the global errors between each LBM and FD solutions are presented in Fig. \ref{T_fusivel} and Fig. \ref{Erros_fusiveis}, respectively. The transient solutions display the correct physical behavior, showing an increase of the fuse temperature with time, with the highest values at the fuse middle lengths, as expected. The relative errors decreased with time due to the attainment of the steady state regime, and are small for all cases. Both LBM models provided almost the same results.  

\begin{figure}[h!]
    \centering
        \centering
        \includegraphics[width=0.5\textwidth]{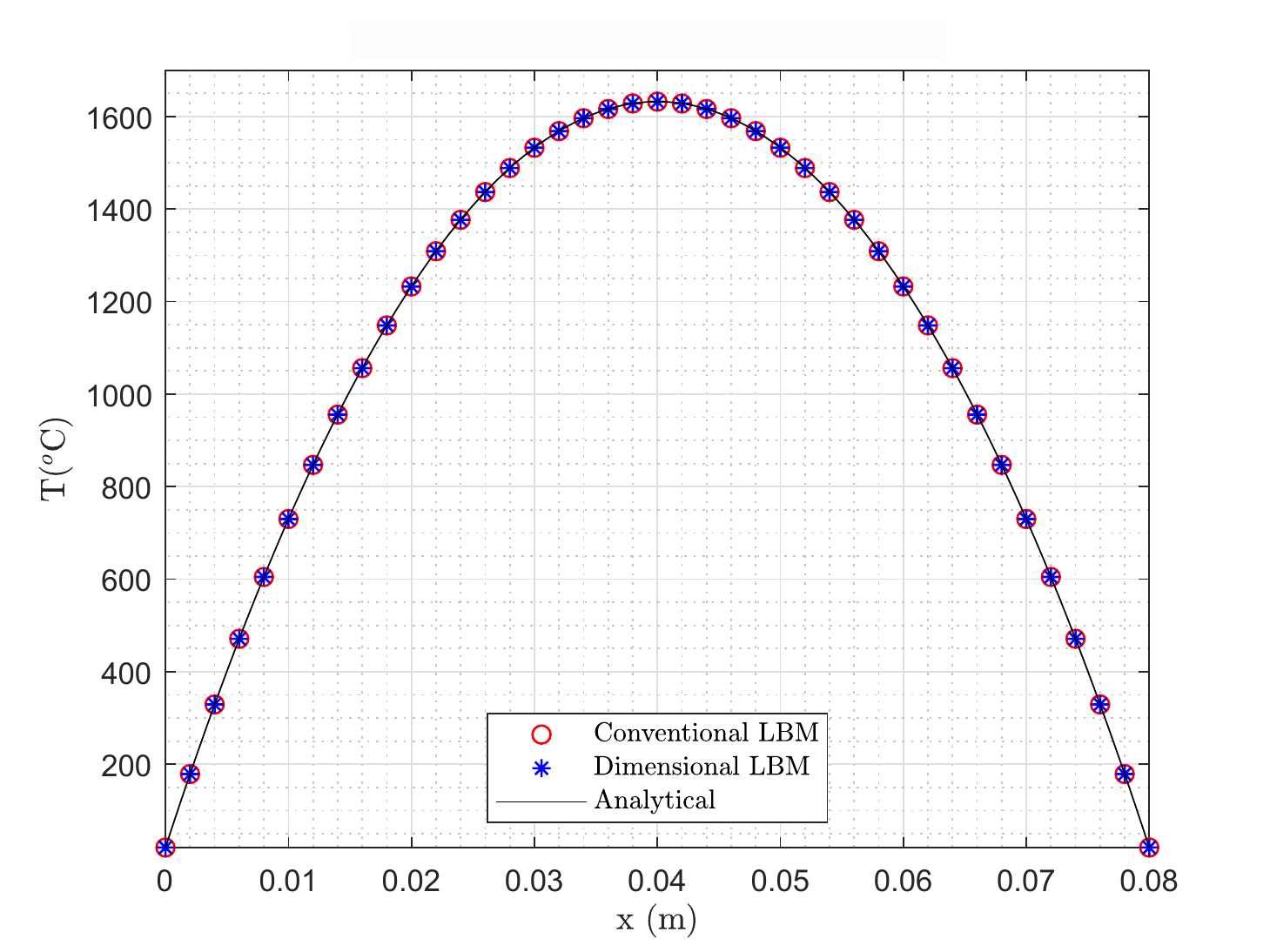}
       \caption{\centering Steady-state axial fuse temperature distribution for the numerical and analytical solutions.}
       \label{T_fusivel_permanente}
\end{figure}

The steady-state solutions obtained with LBM models and the analytical solution, given by Eq. \ref{sol_analitica_fusivel}, are presented in Fig. \ref{T_fusivel_permanente}. The stationary solutions show the correct temperature distribution with a maximum value at the middle of fuse length. This temperature distribution is shown for comparing the LBM and analytical solutions, because the fuse alloy will melts at approximately
$500^oC$. For the stationary solution the global relative errors were the same for the conventional and the dimensional LBM, being equal to $E_2^{dim} = E_2^{conv} = 0.0029 \%$.

In the present example it was found that the dimensional and the conventional LBM provided the same results and the same very good accuracy in relation to the reference solutions. This fact reveals that the proposed dimensional LBM can be used safely and that the non-dimensionalization process could be avoided without any changes over the results obtained with the method.  

\FloatBarrier

\subsection{Heated channel}

The second problem is related with the simulation of forced convection in a two-dimensional channel between two parallel plates, submitted to a heat flux at the plates. This case was inspired in a problem taken also from \cite{Nellis_Klein}, problem 5-11, and is addressed here because this is an interesting engineering problem with applications in heat exchangers, microchannels for electronic cooling and microfluidics. Also the problem concerns in the simulation of a forced convection where the LBM is employed for simulating the fluid flow and the energy conservation using the models presented in sections \ref{standard_MLB} and \ref{thermal_LBM}, respectively.

In this problem, three different cases of forced convection between two parallel plates will be considered. One involves the simulation of both thermally and hydrodynamically developed flow at the entrance of the channel, which is then submitted to an alternating heat flux at the top and bottom boundaries. The other involves a developing flow (thermally and hydro-dynamically) under a constant heat flux in both walls. The last problem consists in the consideration of the alternating heat flux at the walls applied to the developing flow. The results of simulations for the three cases will be analyzed for the steady-state solution.

The geometry of the channel is $H:L = 0.0005m:0.010m$, where $H$ and $L$ represent the channel height and length, respectively. The fluid is water, which is characterized by constant properties determined at the mean temperature of $301 K$ and summarized in Table \ref{water_properties_channel}. In this case it is considered the following BC: an imposed heat flux at bottom and top walls (Neumman BC), a constant temperature inflow BC at the left boundary and an outflow at atmospheric pressure BC at right wall.

\begin{table}[h]
\centering
\begin{tabular}{ll}
\hline
$\rho$   & 996.279 $kg\ m^{-3}$ \\
$\nu$    & 8.382e-7 $m^2\ s^{-1}$\\
$k$     & 0.611    $W\ m^{-1}K^{-1}$\\
$c_p$    & 4180.333 $J\ kg^{-1}K^{-1}\cdot m^{-3}$\\
$\alpha$ & 1.467e-7 $m^2\ s^{-1}$\\ \hline
\end{tabular}\caption{Thermodynamic and transport properties of water at $301K$ and $1$ atm, calculated from \cite{CoolProp}, used for the simulation of the forced convection in the heated channel.}\label{water_properties_channel}
\end{table}

The LBM boundary conditions were implemented considering the following issues. For the fluid flow it was used Eq. \ref{anti_BB_vel} for both inlet and stationary walls. The outlet BC was modeled using the Eq. \ref{anti_BB_vel} to consider the channel opened to atmosphere pressure. In the case of the temperature field simulation, the imposed heat flux in bottom wall was implemented by Eq. \ref{BB_T}, while the inlet wall BC was modeled using Eq. \ref{anti_BB_T}. For the outlet, the first order extrapolation scheme was considered, as explained in sec. \ref{thermal_LBM}. Because the channel geometry is symmetric in relation to its height (the heat flux is applied in both walls), it is simulated only half of the domain ($H/2$) and it is applied a symmetric BC for the domain top wall (resting at the $y$-center of the channel), as explained in sec. \ref{sec:mathematical-modeling}. For the results analysis, the solution is mirrored to consider the full channel.

In all the LBM simulations (with both models) it was used the $D2Q9$ velocity scheme, considering the BGK collision operator for the fluid flow and the MRT operator for the temperature field simulation. The MRT was employed for solving the energy conservation equation due to the stability and accuracy issues related with the simulation of the developing flow.

For the first case, it was considered that water enters both thermally and hydrodynamically developed. The velocity profile at the inlet is modeled by Eq. \ref{velocity_prof}, where $u_m = 0.2 m\ s^{-1}$ is the mean velocity of the fluid. The temperature profile in this inlet region is described by Eq. \ref{temp_prof}, being $T_{in} = 300 K$ the mean temperature at the inlet. The $y$ axes represents the direction along the channel height, varying from the bottom plate to the top plate as $0\le y \le H$. The $x$ axes points into the direction of the channel length, varying as $0\le x \le L$, from the channel inlet to outlet.

\begin{equation}
    u(y) = 6u_m\left(\frac{y}{H} -\frac{y^2}{H^2}\right)
    \label{velocity_prof}
\end{equation}

\begin{equation}
    T(y) = T_{in} + \frac{q_s''H}{k}\left[-\left(\frac{y}{H} \right)^4 + 2\left(\frac{y}{H}\right)^3 -\frac{y}{H} + 0.243\right]
    \label{temp_prof}
\end{equation}

It is assumed that both plates are submitted to a periodic heat flux given by Eq. \ref{var_flux}. The mean heat flux is kept equal to $q_s'' = 40000 W\ m^{-2}$, and the variation is about $\Delta q_s'' = 40000 W\ m^{-2}$, alternating between spaces of $L_h = 1mm$. The function signal in the heat flux definition ($\mbox{sign}$) returns $+1$ if the argument is positive, and $-1$, if it is negative. 

\begin{equation}
    q''(x) = q_s'' + \Delta q_s'' \mbox{sign}\left[\sin{\left (\frac{2\pi x}{L_h}\right)} \right]
    \label{var_flux}
\end{equation}

Given the conditions of the problem, the macroscopic equation which describes the energy conservation for the first case studied can be given by Eq. \ref{conver_energy_developed_channel}. Then, in order to evaluate the performance of the LBM models, this equation was again solved by a FD scheme to serve as a reference solution. This FD solution was obtained using the procedure presented in \cite{Nellis_Klein}, considering $\Delta x_{FDM} = \Delta y_{FDM} = 1.25 \cdot 10^{-6}m$ for the full channel, without the application of the symmetry BC. 

\begin{equation}
   u\frac{\partial T}{\partial x} = \alpha \left( \frac{\partial^2 T}{\partial x^2}+\frac{\partial^2 T}{\partial y^2}\right)
    \label{conver_energy_developed_channel}
\end{equation}

\begin{figure}[h]
    \centering
    \begin{subfigure}[b]{0.45\textwidth}
        \centering
        \includegraphics[width=\textwidth]{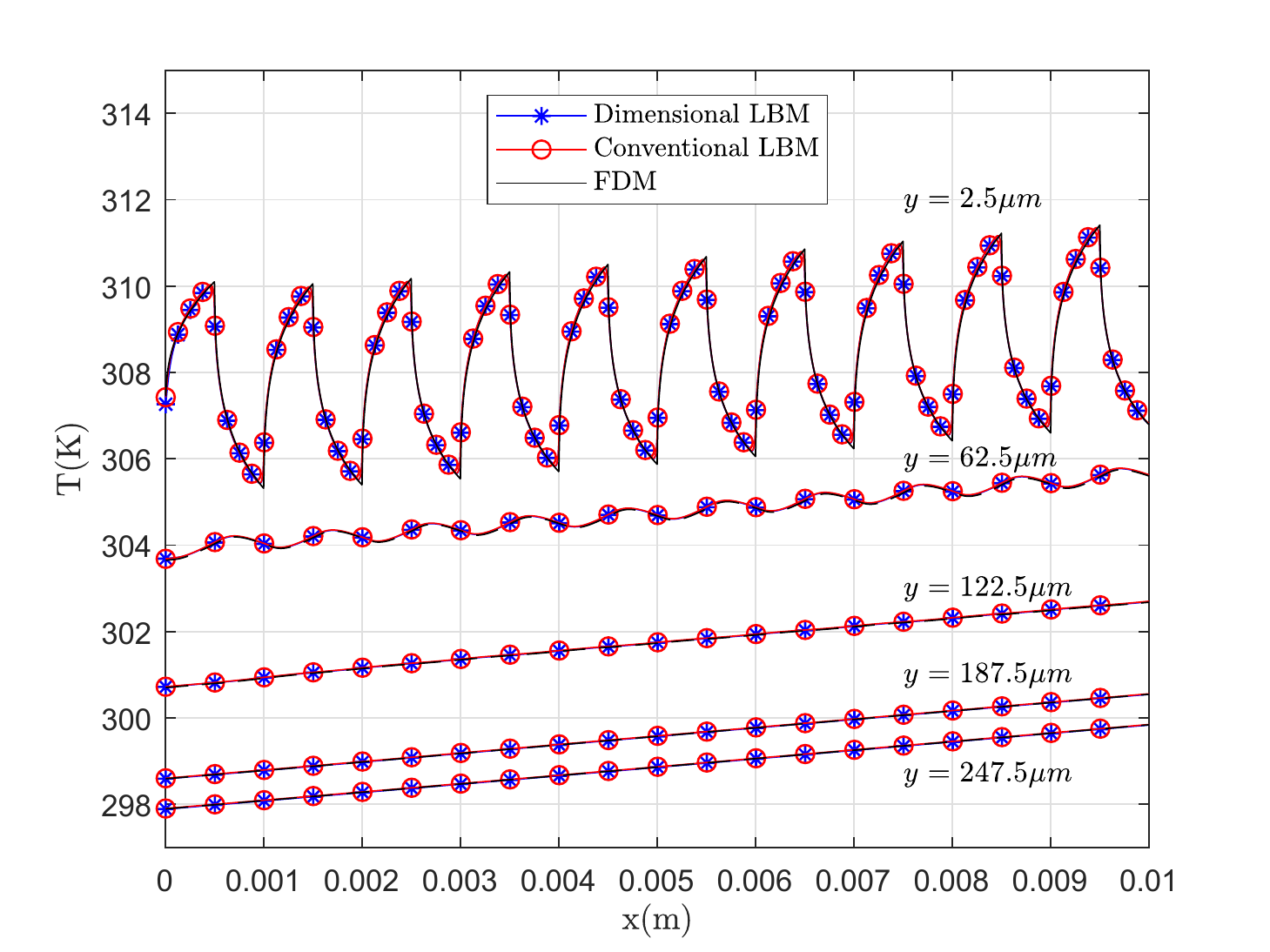}
        \caption{}\label{T_canal_senoidal_desenvolvido}
    \end{subfigure}
    \hfill
    \begin{subfigure}[b]{0.45\textwidth}
        \centering
        \includegraphics[width=\textwidth]{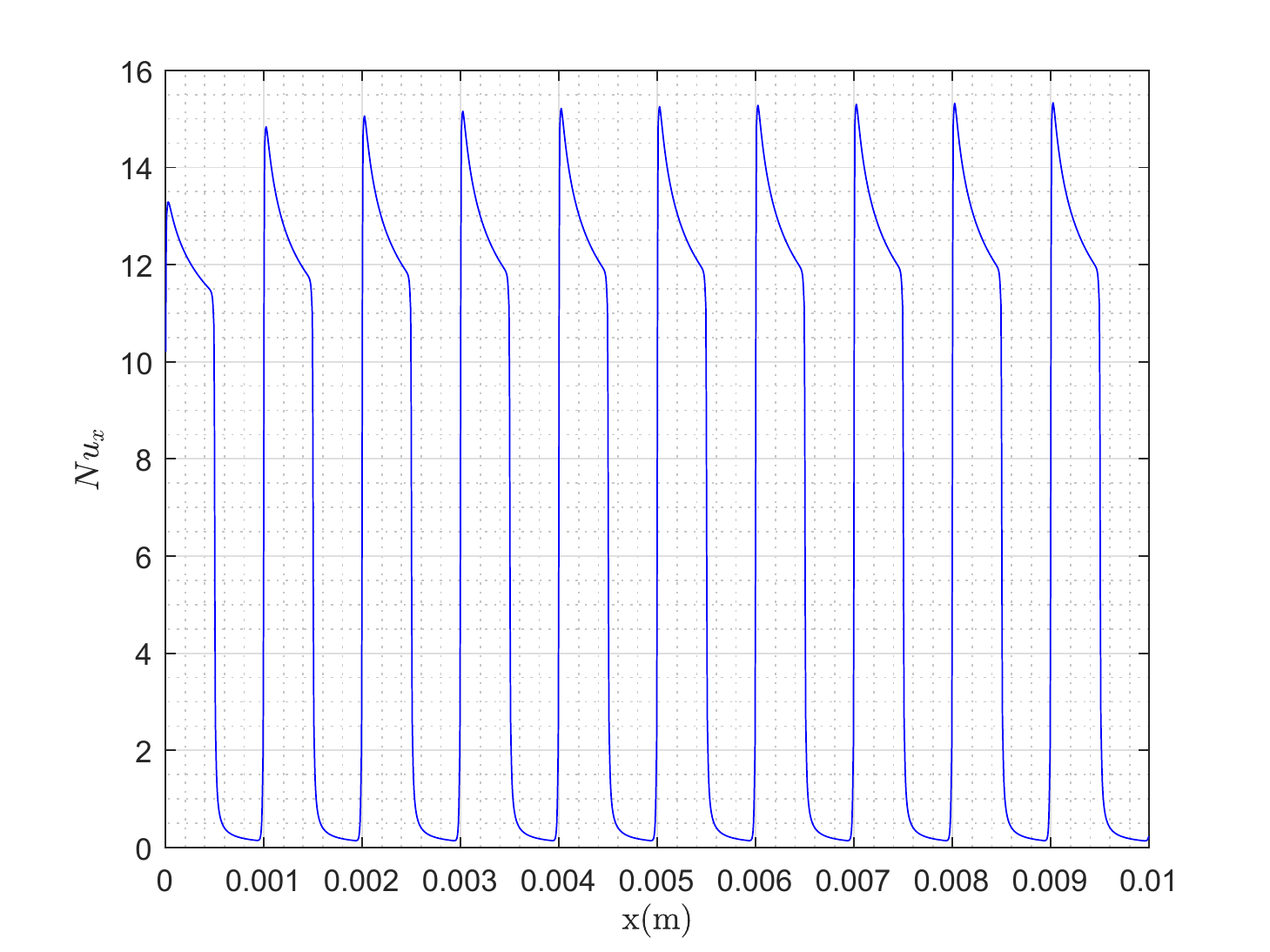}
        \caption{}\label{}
        \label{Nu_canal_senoidal_desenvolvido}
    \end{subfigure}
       \caption{\centering Temperature variation along the channel $x$-direction for several $y$-values, obtained with the FDM and the conventional and dimenisonal LBM. (b) Local Nusselt number at the channel walls for the dimensional LBM simulation.}
       \label{Temperaturas_canal_senoidal}
\end{figure}

The numerical solutions obtained are shown in Fig. \ref{Temperaturas_canal_senoidal}. For both LBM models it was used a spatial grid interval of $\Delta x = 5.0 \cdot 10^{-6}m$ and a time step of $\Delta t = 2.0 \cdot 10^{-6}s$. The LBM solutions correspond to the transient version of Eq. \ref{conver_energy_developed_channel}. The results displayed in Fig. \ref{T_canal_senoidal_desenvolvido} show a very good agreement between LBM results and also between LBM and FD results. The global errors measured for the LBM models, using the FD solution as reference, were $E_2^{dim} = 0.010\%$ and $E_2^{conv} = 0.012\%$, showing very good agreement with the expected solution, and a slightly small error for the dimensional LBM. Overall, the water temperature rises along its flow through the channel and becomes oscillating in the heights near to the heat flux sources. 

In Fig. \ref{Nu_canal_senoidal_desenvolvido} is shown the variation of local Nusselt number, $Nu$ with $x$ for the dimensional LBM. $Nu$ varies periodically with $x$, showing the same variation because the flow is completely developed. In this case the average Nusselt number is equal to $\overline{Nu} = 6.67$.

In the next second case it is considered a developing water flow in the channel under constant heat flux. Now the water enters with uniform velocity and temperature of $u_m = 0.02 m\ s^{-1}$ and $T_{in} = 300 K$, respectively, and are considered the same fluid properties. As the fluid flows through the channel, it is submitted in both plates to an uniform heat flux of $q'' = 40000 W\ m^{-2}$. As the flow is developing, the used discrete time and space interval decreased, $\Delta x = 2.50 \cdot 10^{-6}m$ and $\Delta t = 6.25 \cdot 10^{-7}s$, in order to obtain a convergent solution.

\begin{figure}[h]
    \centering
    \begin{subfigure}[b]{0.45\textwidth}
        \centering
        \includegraphics[width=\textwidth]{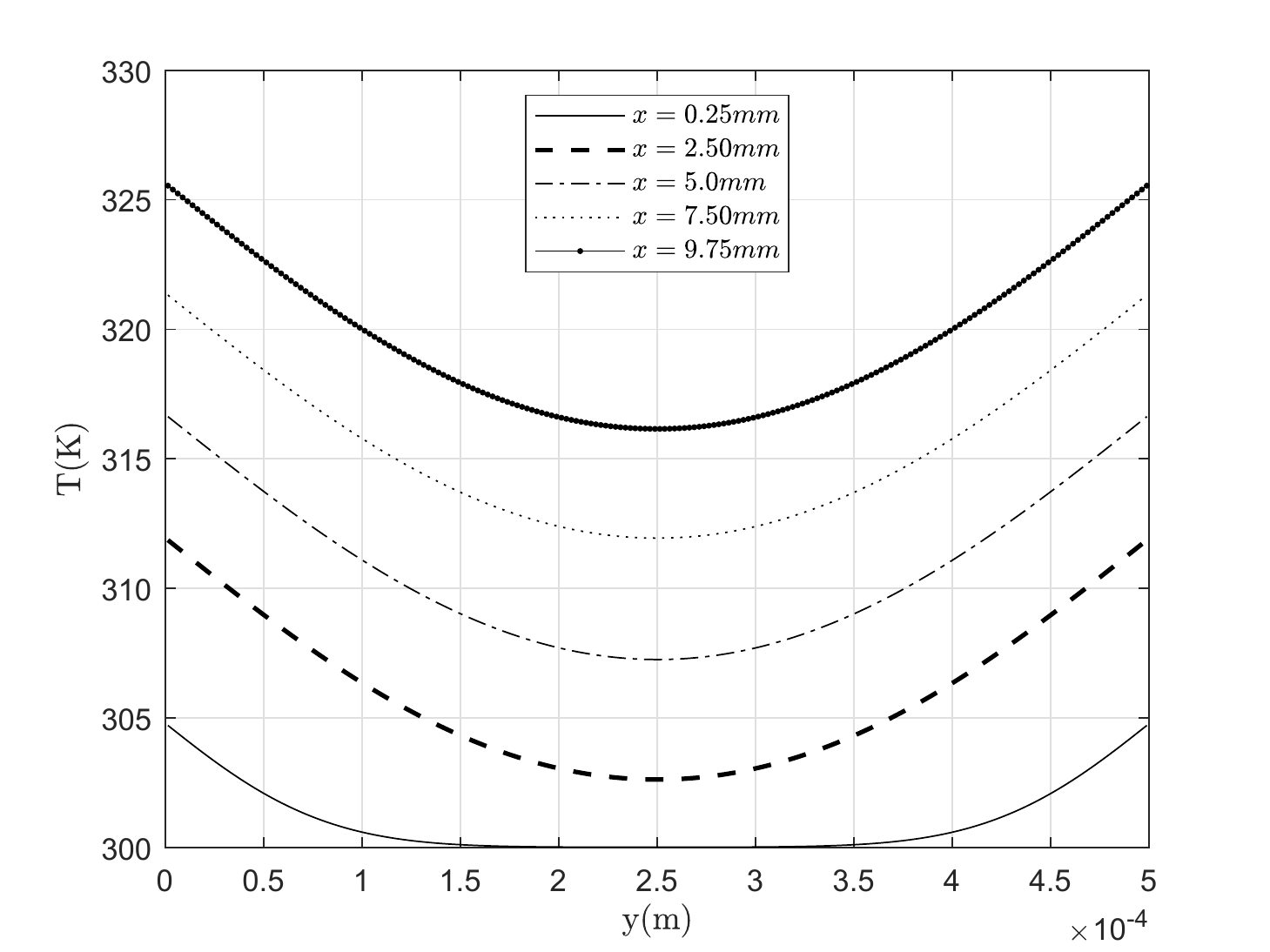}
        \caption{}\label{T_canal_desen}
    \end{subfigure}
    \hfill
    \begin{subfigure}[b]{0.45\textwidth}
        \centering
        \includegraphics[width=\textwidth]{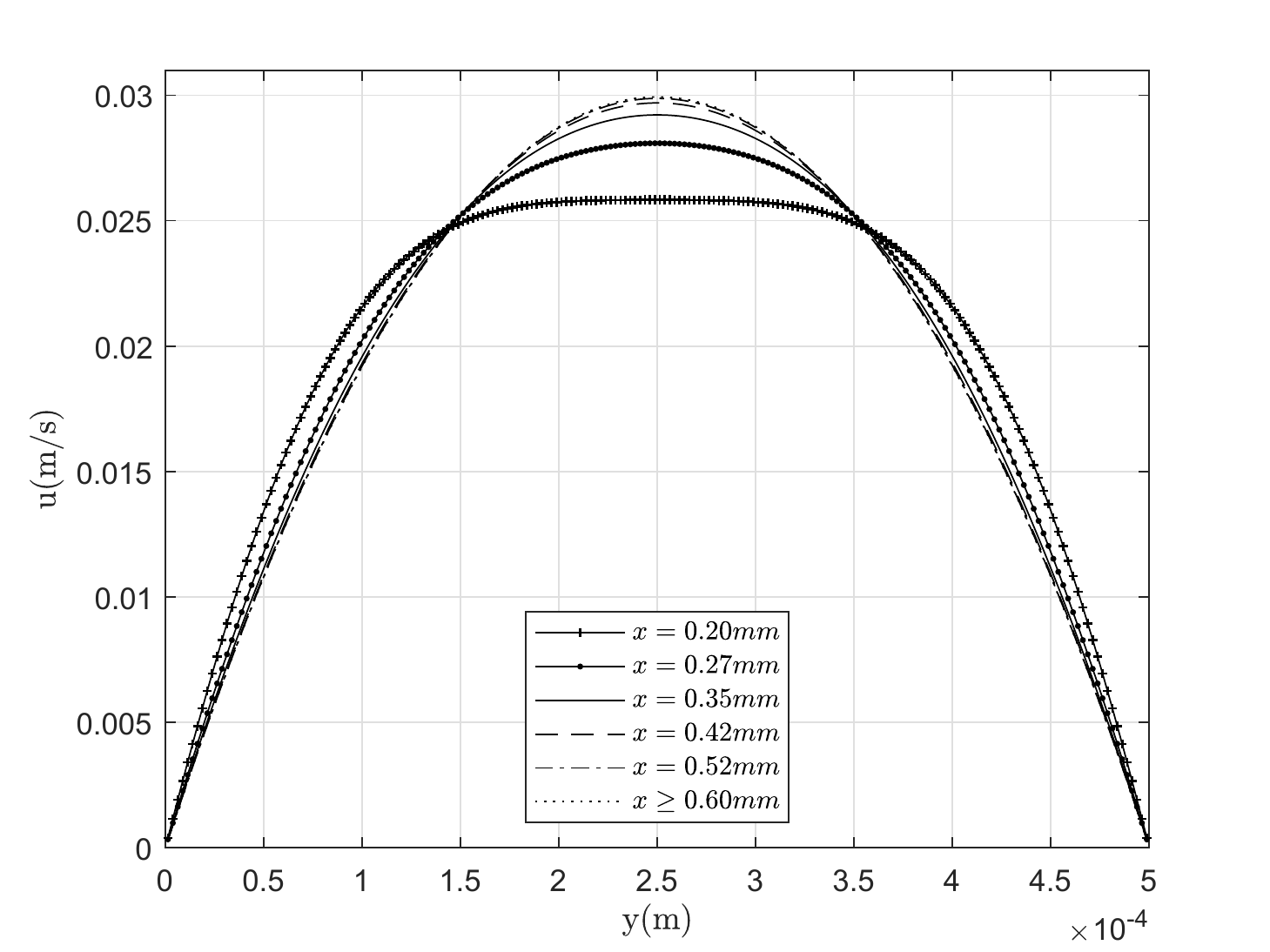}
        \caption{}\label{u_canal_desen}
    \end{subfigure}
       \caption{\centering Temperature (a) and velocity (b) $y$-profiles at several cross sections along the channel length for the developing flow.}
       \label{Temperaturas_canal_desen}
\end{figure}

The temperature and velocity profiles obtained by the dimensional LBM for several cross sections along the channel are shown in Fig. \ref{Temperaturas_canal_desen}. It is possible to perceive that after some channel length the velocity profile does not change, because the flow becomes hydro-dynamically developed. For a flow between parallel plates at low Reynolds numbers, giving an uniform profile at the inlet, the hydrodynamic entrance length can be calculated as $L_e= D_h(0.3125 + 0.011Re_D)$ \citep{Atkinson_1969}. In this case, for a channel Reynolds number of $Re_D = 23.86$, the predicted hydrodynamic entrance length is $L_e= 0.57mm$, and the measured from the LBM results is equal to $L_e \approx 0.60mm$. This simulated result is coherent with the expected ones from the analyzed relation. It should be noted that these relations are also theoretical estimations. Also, it is known that the developed velocity profile must follow the analytical relation, given by Eq. \ref{velocity_prof}. Comparing the profile for $x = 0.60mm$ and the analytical velocity profile, it was found a global error of about $E_2 = 0.1575\%$ for both conventional and dimensional LBM.

\begin{figure}[h]
    \centering
    \begin{subfigure}[b]{0.45\textwidth}
        \centering
        \includegraphics[width=\textwidth]{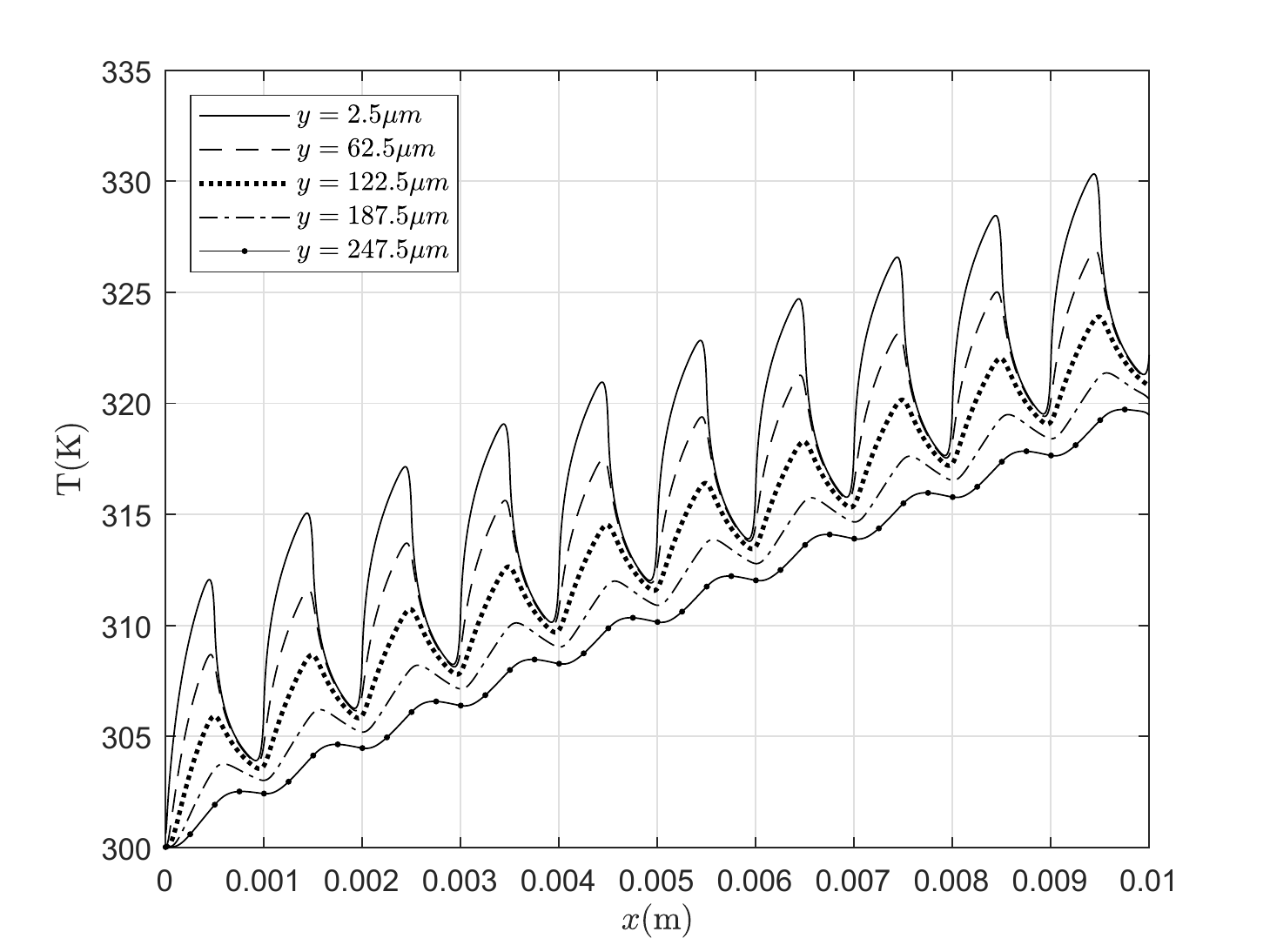}
        \caption{}\label{T_flux_var_under_develop}
    \end{subfigure}
    \hfill
    \begin{subfigure}[b]{0.45\textwidth}
        \centering
        \includegraphics[width=\textwidth]{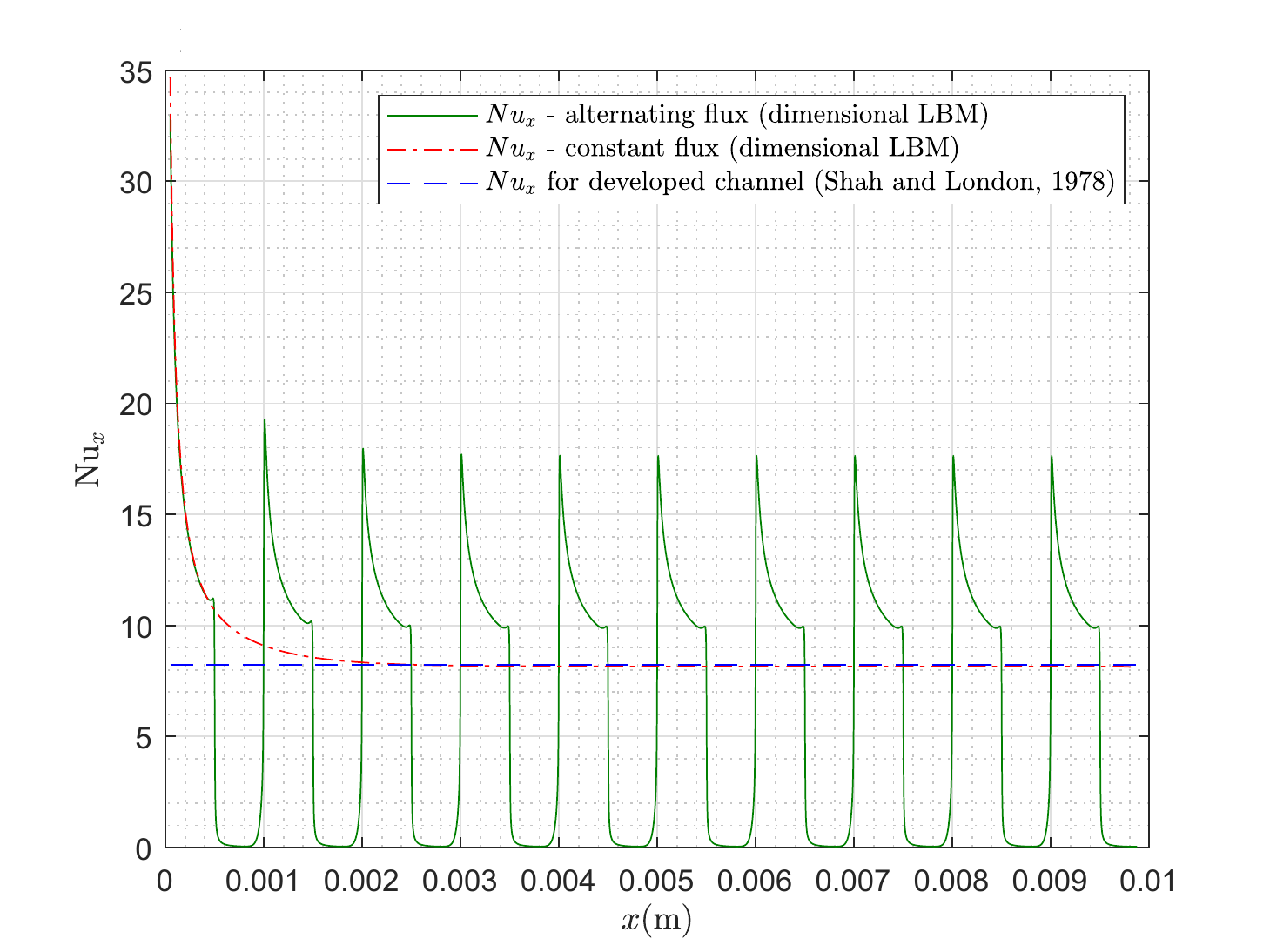}
        \caption{}\label{Nu_chan_under_develop}
    \end{subfigure}
    \caption{\centering (a) Temperature profiles along the $x$-direction at several $y$-values for the developing flow with oscillating heat flux. (b) Local Nusselt number for the developing flow, considering both constant and oscillating heat flux, and for the developed flow according to \citep{Shah_London}.}
       \label{T_and_Nu_under_develop}
\end{figure}

Similarly, at some $x$ value the temperature $y$-direction profiles stop to vary its shape, starting to just increase in module as the channel is heated, but maintaining the same $\Delta T$ in $y$. From this observation and analyzing the local Nusselt values shown in Fig. \ref{Nu_chan_under_develop}, which do not change after some $x$ value (approximately $2.5mm$), it is possible to conclude that the channel length is enough to both the velocity and the thermal profiles get fully developed. Therefore, the developed value of the local Nusselt number can be compared with the expected value from the literature. For a developed flow between two parallel plates with a constant and equal heat flux, the Nusselt number is $Nu = 8.24$ \citep{Shah_London}. In the results provided by the dimensional LBM it was obtained $Nu_{LBM} = 8.15$, resulting in a very good agreement with a relative error of $1,02\%$. The variation of the local Nusselt number in the channel walls is presented in Fig. \ref{Nu_chan_under_develop}.

In the last case, the developing channel flow is submitted to an alternating heat flux given by Eq. \ref{var_flux}, in order to compare the impact of this BC variation over the local Nusselt number. The spatial and time intervals used were the same as for the previous simulation ($\Delta x = 2.50 \cdot 10^{-6}m$ and $\Delta t = 6.25 \cdot 10^{-7}s$). The steady state temperature profiles obtained for this case are shown in Fig. \ref{T_flux_var_under_develop}. As the flow is developing it is noted a higher influence of the varying heat flux at the walls on the temperature profiles. As it is possible to see in Fig. \ref{Nu_chan_under_develop}, the Nusselt number of the developing channel flow submitted to an alternating heat flux reaches higher local values than that for the channel under a constant heat flux. However, the average Nusselt number calculated for each case were $\overline{Nu}_{const} = 8.92$ for the constant heat flux and $\overline{Nu}_{var} = 6.43$ for the varying one, being lower than the first due to intermittency of the wall heat flux.   

\begin{figure}[h]
    \centering
    \begin{subfigure}[b]{0.95\textwidth}
        \centering
        \includegraphics[width=\textwidth]{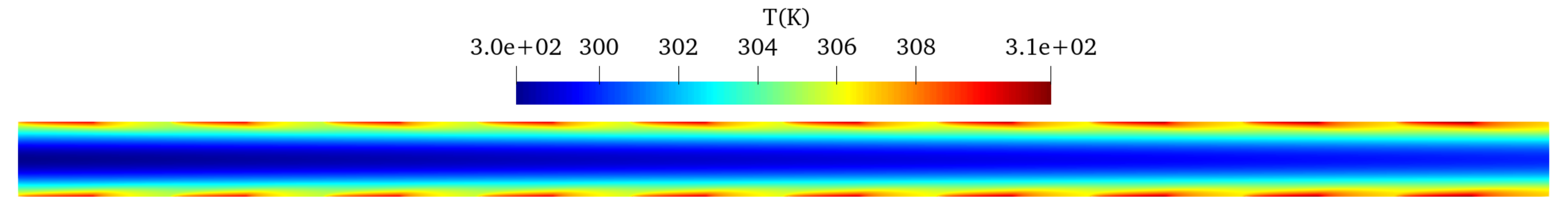}
        \caption{}\label{Temp_contours_channels_1}
    \end{subfigure}

    \begin{subfigure}[b]{0.95\textwidth}
        \centering
        \includegraphics[width=\textwidth]{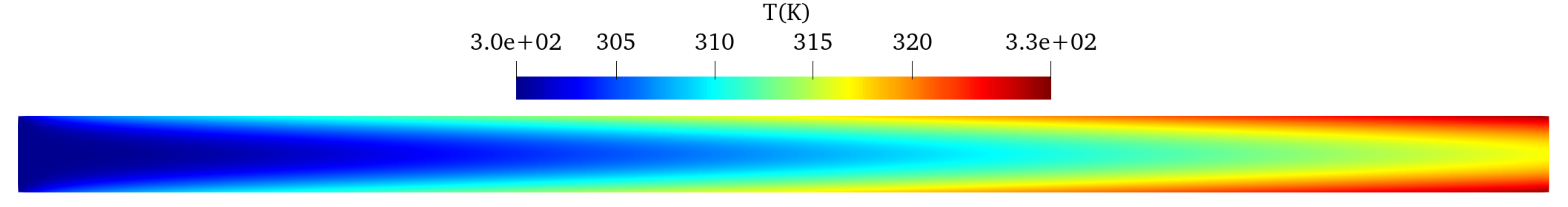}
        \caption{}\label{}
        \label{Temp_contours_channels_2}
    \end{subfigure}

    \begin{subfigure}[b]{0.95\textwidth}
        \centering
        \includegraphics[width=\textwidth]{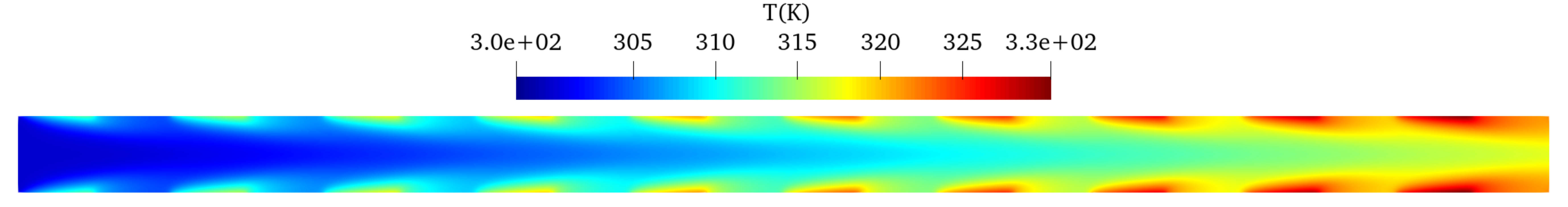}
        \caption{}\label{}
        \label{Temp_contours_channels_3}
    \end{subfigure}
       \caption{\centering Temperature field distribution along the channel, for the first (a), second (b) and third (c) simulated cases.}
       \label{Temp_contours_channels}
\end{figure}

The temperature field along the entire channel is shown in Fig. \ref{Temp_contours_channels} for the three simulated cases. Various notable differences can be observed. In Fig. \ref{Temp_contours_channels_1} the flow is fully developed at the channel inlet and the temperature distribution shows slight variations along the channel due to the heating process under the oscillating heat flux. In this case the warmer fluid regions near the walls are slightly bigger and the cold fluid core decreased at the channel outlet. However, this is the only case were the temperature of the core fluid is almost equal to the smaller input temperature value. The other two cases are for the developing flow under the constant and oscillating heat flux, Fig. \ref{Temp_contours_channels_2} and \ref{Temp_contours_channels_3}, respectively. In the case of the constant heat flux, the outlet mean temperature is the higher one due to the constant heating suffered along the channel. The temperature field follow a similar pattern for the the developing flow submitted to an oscillating heat flux, but the convective heat transfer into the center of the channel is less due to the heat flux intermittence.

The higher average Nusselt number was obtained for the developing flow under the constant heat flux ($\overline{Nu} = 8.92$), which is even higher than the theoretical value for the developed flow, $Nu = 8.24$, due to the effect of the development of the hydrodynamic and thermal boundary layers from the channel inlet. The average Nusselt numbers for the oscillating heat flux two case were lower, being equal to $\overline{Nu} = 6.67$ and $\overline{Nu} = 6.43$ for the developed and developing flow, respectively. In this case, the inlet cold flow developing region decreased the total heat transferred from the wall heaters. The present problem could mimic the heat transfer process in a refrigerating channel of electronic devices.


In addition, it is important to mention that the viscous dissipation was neglected in all the simulations of this section. This assumption is assumed based on the small values of Brinkman number, $B_r$ (Eq. \ref{Brinkman}). As the obtained highest value was $B_r = 10^{-5}$, the contribution of heat dissipation due viscous stress is insignificant for these cases, and can be safely neglected. 

\begin{equation}
    B_r = \frac{\zeta u_m^2}{k(T_w - T_m)}
    \label{Brinkman}
\end{equation}


In Appendices \ref{sec:appendixB}, \ref{sec:appendixC} and \ref{sec:appendixD} are shown simulations results obtained with the proposed dimensional LBM for other classical problems, considering an one dimensional convection-diffusion flow, an isothermal Pouseuille flow and a natural convection in an enclosure. For the two first cases (Appendices \ref{sec:appendixB} and \ref{sec:appendixC}), the obtained results are compared with those from analytical solutions. For the natural convection (Appendix \ref{sec:appendixD}), however, the numerical solutions are compared with benchmark results available from the literature \citep{De_vahl_davis}. In all the cases the dimensional LBM results were also compared with the conventional LBM. All tests show very small global errors, proving that the proposed LBM is physically coherent and accurate. Therefore, considering the results discussed in the present section and those provided in Appendices \ref{sec:appendixB}, \ref{sec:appendixC} and \ref{sec:appendixD}, it can be stated that the dimensional LBM is useful for simulating applied problems involving heat convection and diffusion. 

\FloatBarrier

\subsection{Static bubble}
\label{static_bubble}

In this section is studied a two-phase fluid system composed by one (liquid and vapor saturated water system) and two components (air-water system). A theoretical static problem related to the simulation of a bubble surrounded by a liquid in equilibrium is considered. This is a common benchmark test which will allow the evaluation of the conventional and dimensional LBM performance.

The problem consists in a circular (2D) bubble of radius $R$ surrounded by liquid, initialized at the center of a square domain. All the boundaries are considered as periodic and the order parameter $\phi$ (and the density $\rho$) profile between the phases is initialized by the Eq. \ref{diffuse_interface}, in order to avoid instabilities related with a sharp interface. With the evolution of the time, the system reaches the steady state, and the equilibrium density profile must match with the analytical solution, represented by Eq. \ref{density_prof} \citep{Zou_He_2013}. Also, the relation between the pressure variation at the interface ($\Delta P = P_{out} - P_{in}$) and the surface tension of the liquid $\sigma$ must follow the Laplace law, given by Eq. \ref{Laplace_law}. 

\begin{equation}
    \rho(x,y) = \frac{(\rho_l + \rho_g)}{2} - \frac{(\rho_l - \rho_g)}{2}\mbox{tanh}\left\{\frac{2\left[\sqrt{(x - x_c)^2+(y - y_c)^2}-R \right]}{W} \right\}
    \label{density_prof}
\end{equation}

\begin{equation}
    \Delta P = \frac{\sigma}{R}
    \label{Laplace_law}
\end{equation}

The verification of these two properties is used to evaluate and compare the performance of the LBM considering four different two-phase systems. The first one is a air-water system at $25^oC$ and $101.325 kPa$ (1 atm). The other three systems consist in a vapor bubble surrounded by liquid phase of saturated water in equilibrium, without phase-change, for three saturated temperatures: $100^oC$, $80^oC$ and $25^oC$, respectively. The involved thermodynamic and transport properties for each system are presented in Tab. \ref{properties_bubble}. The water-air system was simulated with the dimensional and conventional LBM, in order to compare these two solutions. The other three two-phase systems were just simulated with the dimensional LBM.

\begin{table}[h]
\centering
\begin{tabular}{lllll}
\hline
            & Air and water - 25$^o$C, 1 atm & Sat. water - 100$^o$C & Sat. water - 80$^o$C & Sat. water - 25$^o$C \\ \hline
$\rho_g (kg\ m^{-3})$        & 1.184                      & 0.598                   & 0.294                  & 0.023                  \\
$\rho_l(kg\ m^{-3})$        & 997.048                    & 958.349                 & 971.766                & 997.003                \\
$\rho_l/\rho_g$   & 842.1                     & 1602.6                 & 3305.3                & 43349.0               \\
$\nu_g(10^{-7}m^2\ s^{-1})$ & 155.770                     & 204.493                 & 392.919                & 4204.120               \\
$\nu_l(10^{-7}m^2\ s^{-1})$  & 8.927                      & 2.938                   & 3.643                  & 8.927                  \\
$\nu_g/\nu_l$     & 17.4                      & 69.6                   & 107.9                  & 470.9                  \\
$\zeta_g (10^{-5}Pa\ s)$ & 1.845                      & 1.223                   & 1.154                  & 0.970                  \\
$\zeta_l(10^{-5}Pa\ s)$  & 89.006                     & 28.158                  & 35.404                 & 89.004                 \\
$\zeta_l/\zeta_g$     & 48.2                      & 23.0                   & 30.7                  & 91.8                  \\
$\sigma(N\ m^{-1})$       & 0.072                     & 0.059                 & 0.063                & 0.072                \\ \hline
\end{tabular}
\caption{Thermodynamic and transport properties of the four fluid systems considered for the static bubble simulations, obtained from \citep{CoolProp}.}\label{properties_bubble}
\end{table}

The dimensions of the domain were taken as $1mm:1mm$, and for each system it was tested six different bubble radius: $0.250mm$, $0.225mm$, $0.20mm$, $0.175mm$, $0.150mm$ and $0.125mm$. The interface width was assumed as $W = 25.0\cdot10^{-6}m$, the mobility was taken as $M = 1.0 \cdot 10^{-5} m^2\ s^{-2}$ and the discrete time and space intervals were considered as $\Delta x = 5.0\cdot10^{-6}m$ and $\Delta t = 1.0\cdot10^{-7}s$ for all the simulations. For the air-water system it was applied the BGK collision operator for both LBM models, while for the other three systems, the MRT operator was implemented for the two-phase momentum equation, instead of Eq. \ref{eq_lattice_boltzmann_pressure}, in order to get more stability, and the BGK for the interface tracking LBE (Eq. \ref{eq_lattice_boltzmann_interface}).

\begin{figure}[h!]
    \centering
    \begin{subfigure}[b]{0.36\textwidth}
        \centering
        \includegraphics[width=\textwidth]{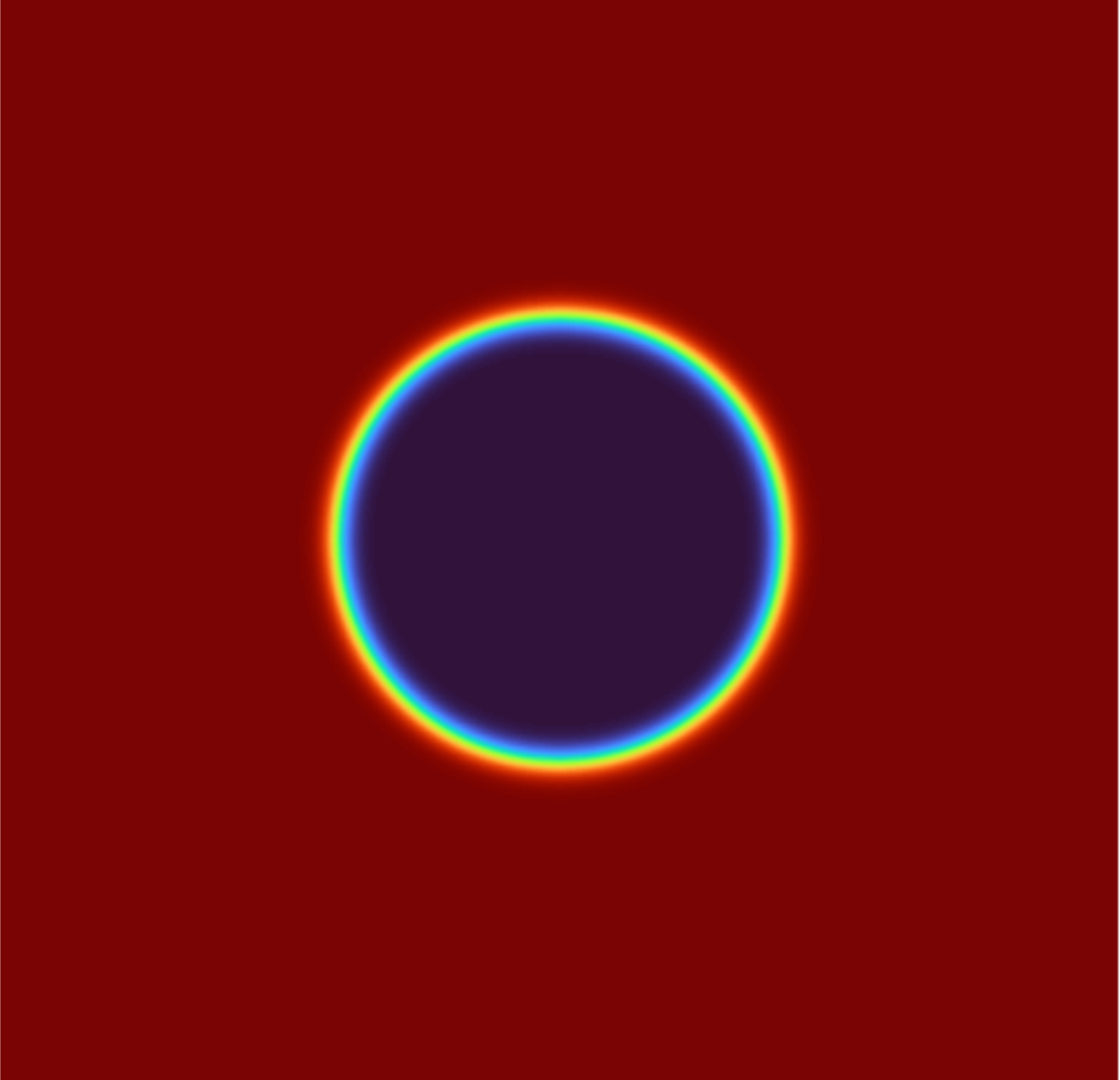}
        \caption{}\label{Repres_bubble}
    \end{subfigure}
    \hfill
    \begin{subfigure}[b]{0.45\textwidth}
        \centering
        \includegraphics[width=\textwidth]{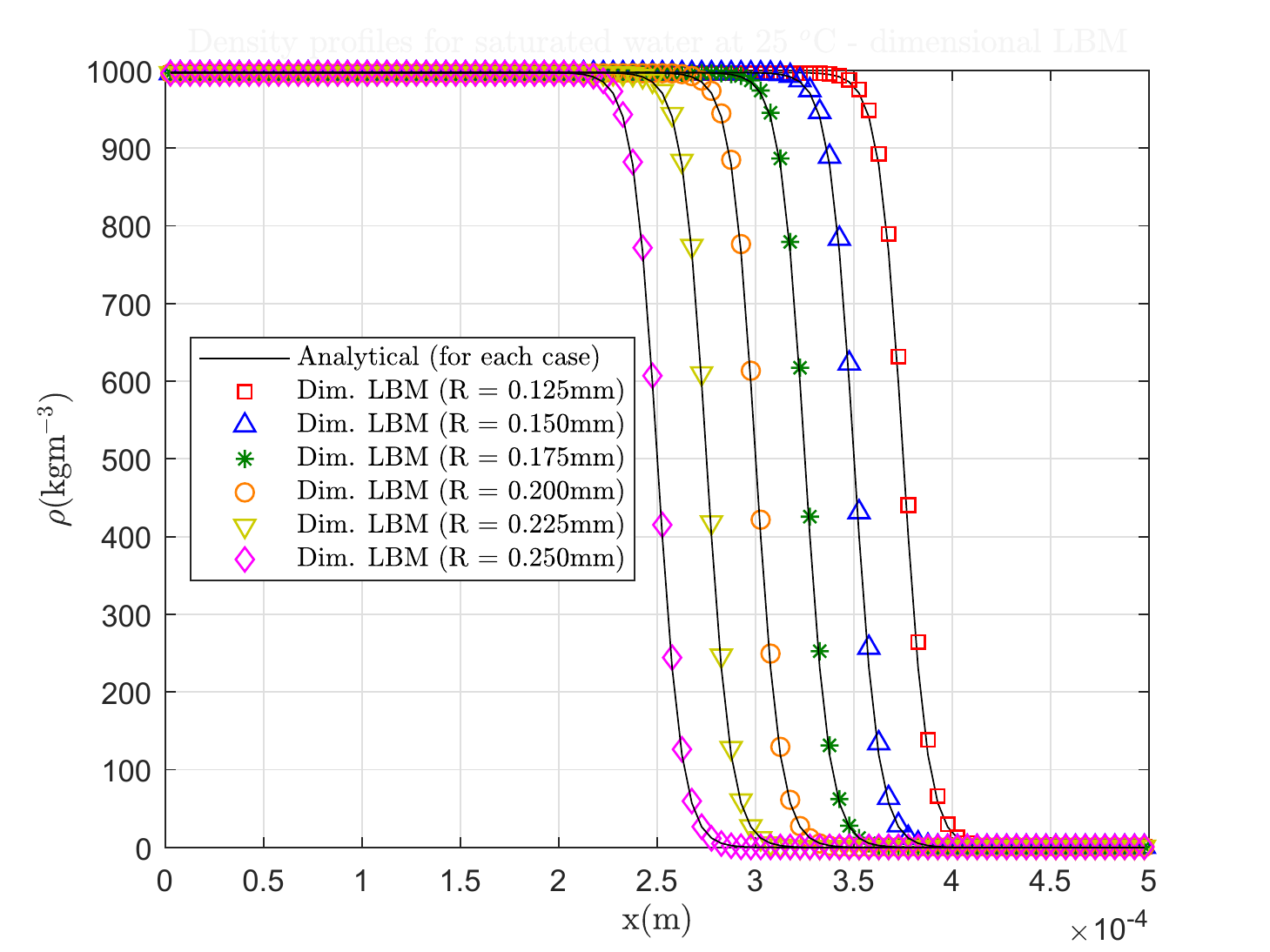}
        \caption{}\label{}
        \label{Density_profiles_ag_sat_25}
    \end{subfigure}
       \caption{\centering (a) Representation of the density profiles for saturated water at 25$^oC$, considering a bubble with radius $R=0.20mm$, being the blue region (inner part) occupied by the gas, and the red (outer part), by the liquid. (b) Density profiles at cross sections for each $R$ value simulated, for the saturated water at 25 $^o$C. (Obs: it was represented only a half of the cross section, because the domain is symmetric).}
       \label{Bubble_profile}
\end{figure}

A representation of the simulated domain density profile for the saturated water at $25^oC$, considering a bubble of radius $R=0.20mm$, is displayed Fig. \ref{Repres_bubble}, and the results of the density variation with $x$ for the six bubble radius tested with saturated water at 25$^o$C are presented in Fig. \ref{Density_profiles_ag_sat_25}. The global errors for the density profile in comparison with the analytical solution are shown in Table \ref{error_density_profile} for the four two-phase systems. It is possible to see that the errors are very low for all cases, showing the good accuracy of both LBM models. For the air-water system the dimensional and conventional LBM presented the same errors, showing that the dimensional LBM does not changed the stability and accuracy of LBM for this problem. 

It should be noted that the results presented in Fig. \ref{Bubble_profile} and Table \ref{error_density_profile} for the saturated water at 25$^o$C were obtained for very high density and viscosity ratios, namely: $\rho_l/\rho_g = 43349.0$, $\nu_g/\nu_l = 470.9$, and $\zeta_l/\zeta_g = 91.8$, (see Table \ref{properties_bubble}). These are very high ratios, not fully simulated in the open literature, indicating the precision and reliability of the proposed dimensional LBM.

\begin{table}[h!]
\centering
\begin{tabular}{lllll}
\hline
\multicolumn{1}{c}{R$^{-1}$ (m$^{-1}$)} & \multicolumn{1}{c}{\begin{tabular}[c]{@{}c@{}}Air-water - 25$^o$C, 1 atm\\ Dim. and Conv. LBM\\ (\%)\end{tabular}} & \multicolumn{1}{c}{\begin{tabular}[c]{@{}c@{}}Sat. water - 100$^o$C\\ (\%)\end{tabular}} & \multicolumn{1}{c}{\begin{tabular}[c]{@{}c@{}}Sat. water - 80$^o$C\\ (\%)\end{tabular}} & \multicolumn{1}{c}{\begin{tabular}[c]{@{}c@{}}Sat. water - 25$^o$C\\ (\%)\end{tabular}} \\ \hline
4000.0                        & 0.253                                                                                         & 0.253                                                                              & 0.253                                                                             & 0.253                                                                             \\
4444.44                       & 0.289                                                                                         & 0.286                                                                              & 0.286                                                                             & 0.286                                                                             \\
5000.0                        & 0.316                                                                                         & 0.316                                                                              & 0.316                                                                             & 0.316                                                                             \\
5714.29                       & 0.346                                                                                         & 0.346                                                                              & 0.347                                                                             & 0.347                                                                             \\
6666.67                       & 0.391                                                                                         & 0.392                                                                              & 0.392                                                                             & 0.392                                                                             \\
8000.0                        & 0.460                                                                                         & 0.462                                                                              & 0.462                                                                             & 0.461                                                                             \\ \hline
\end{tabular}
\caption{$E_2$ results between LBM and analytical solutions of the density profiles for each value of $R^{-1}$.}\label{error_density_profile}
\end{table}

\begin{figure}[h!]
    \centering
    \begin{subfigure}[b]{0.45\textwidth}
        \centering
        \includegraphics[width=\textwidth]{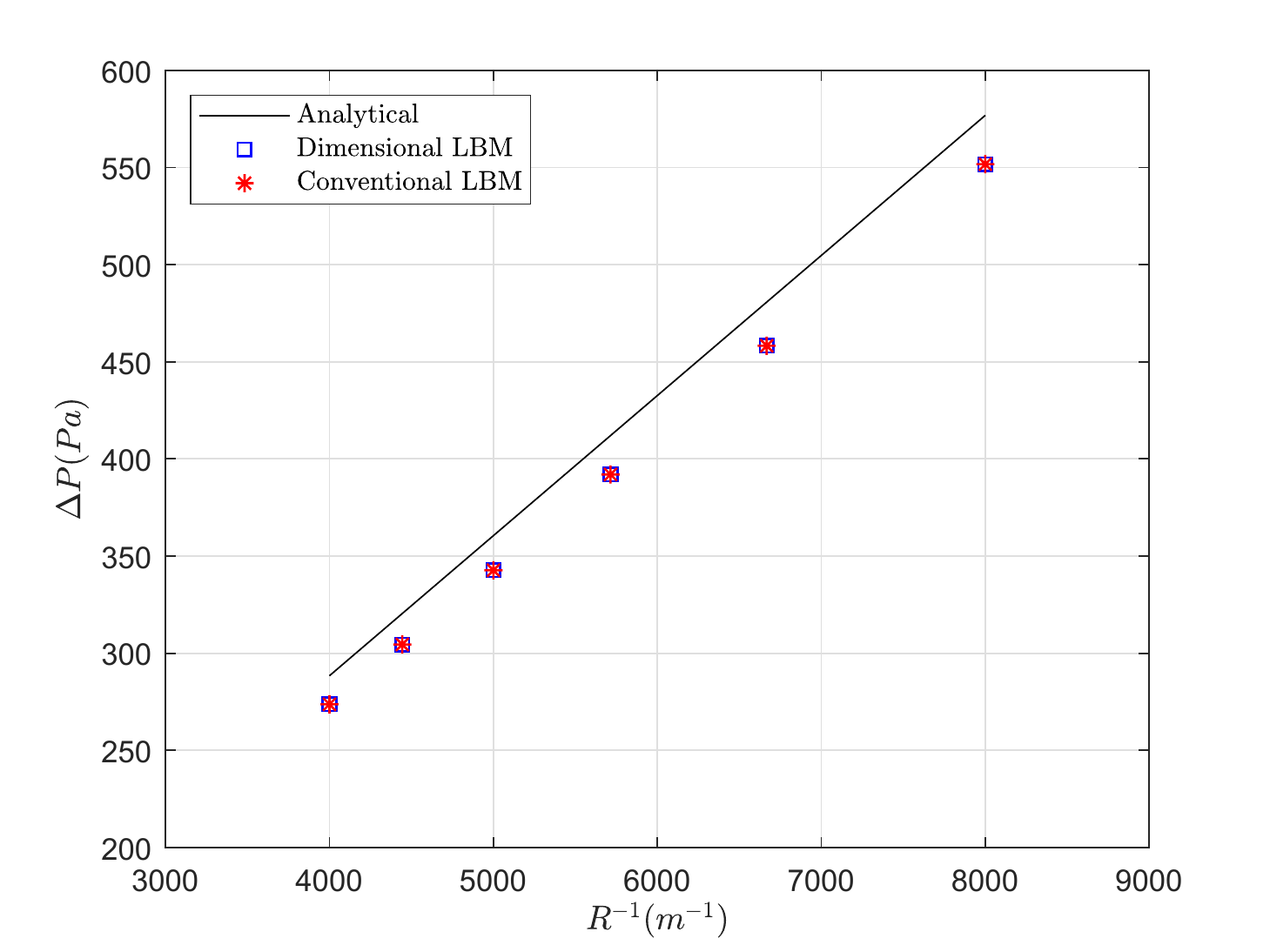}
        \caption{}\label{Law_YL_air_water}
    \end{subfigure}
    \hfill
    \begin{subfigure}[b]{0.45\textwidth}
        \centering
        \includegraphics[width=\textwidth]{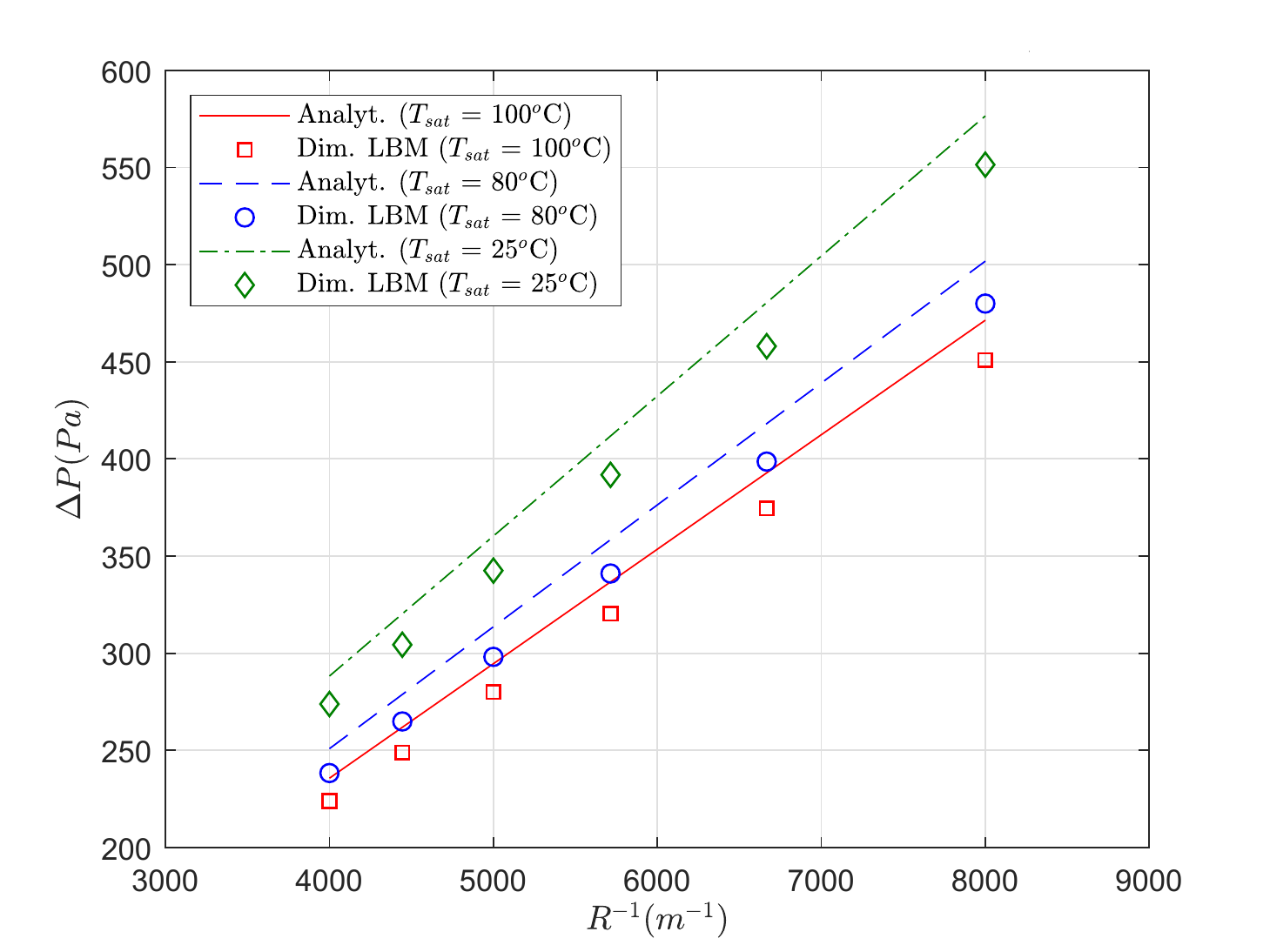}
        \caption{}\label{}
        \label{Law_YL_sat_water}
    \end{subfigure}
       \caption{\centering (a) Pressure difference variation with R$^{-1}$ obtained from the Laplace law and the conventional and dimensional LBM simulations for: (a) Air-water system and (b) Saturated water systems.}
       \label{Laplace_law_fig}
\end{figure}

Considering now the Laplace law verification, the pressure inside the bubble ($P_{in}$) was calculated as the average between all the gas nodes (where $\phi = 0.0$). Similarly, the outside pressure ($P_{out}$) was calculated by the average at the fluid part (nodes where $\phi = 1.0$). Having this in mind, the results from LBM simulations and those expected from the Laplace law are shown in Fig. \ref{Laplace_law_fig}, while the relative errors for each simulation are displayed in Table \ref{error_laplace_law}. Again, it is possible to note that the errors were the same for the dimensional and conventional LBM model in the case of air-water system, as expected. In addition, besides being greater than for the density profile, the errors observed for the $\Delta P$ are considerable low, showing the validity of all simulated results and highlighting the capability of the multiphase LBM presented in this work to simulate real fluids with high density and viscosity ratios. 

\begin{table}[h!]
\centering
\begin{tabular}{lllll}
\hline
\multicolumn{1}{c}{R$^{-1}$ (m$^{-1}$)} & \multicolumn{1}{c}{\begin{tabular}[c]{@{}c@{}}Air-water - 25$^o$C, 1 atm\\ Dim. and Conv. LBM\\ (\%)\end{tabular}} & \multicolumn{1}{c}{\begin{tabular}[c]{@{}c@{}}Sat. water - 100$^o$C\\ (\%)\end{tabular}} & \multicolumn{1}{c}{\begin{tabular}[c]{@{}c@{}}Sat. water - 80$^o$C\\ (\%)\end{tabular}} & \multicolumn{1}{c}{\begin{tabular}[c]{@{}c@{}}Sat. water - 25$^o$C\\ (\%)\end{tabular}} \\ \hline
4000.0  & 5.097 & 4.998 & 4.998 & 4.996 \\
4444.44 & 4.981 & 4.976 & 4.976 & 4.965 \\
5000.0  & 4.943 & 4.936 & 4.936 & 4.935 \\
5714.29 & 4.857 & 4.853 & 4.849 & 4.850 \\
6666.67 & 4.675 & 4.667 & 4.666 & 4.667 \\
8000.0  & 4.357 & 4.349 & 4.349 & 4.351 \\ \hline
\end{tabular}
\caption{$E_2$ results between the LBM models simulations and the Laplace law for the pressure difference in the static bubble problem.}\label{error_laplace_law}
\end{table}

A point to be addressed in relation to the simulation with the traditional LBM is connected with the used values of the dimensionless variables and numbers. Commonly the LBM is employed for a set of dimensionless numbers and properties ratios. In the majority of performed numerical test of the bubble in liquid in equilibrium stationary problem, the authors seeks to simulate a density ratio of $\rho_l/\rho_g = 1000$, representing the water-air system, but the employed viscosity ratios and surface tension do not correspond exactly to the physical values of the system.

In the present simulations of the air-water system at 25$^o$C, considering the non-dimensionalization process explained before, the following dimensionless parameters were used: $\tilde{\nu}_g = 1.682$, $\tilde{\nu}_l = 3.57\cdot 10^{-3}$, $\tilde{\sigma} = 5.7648$, $\tilde{\rho}_g = 1.184$, $\tilde{\rho}_l = 997.084$ and $\tilde{M} = 0.040$. As can be seen the dimensionless surface tension value in this problem, $\tilde{\sigma} = 5.7648$, is much more higher than the commonly used, about $0.0001 \le \tilde{\sigma} \le 0.2$ \citep{Liang_2018,Liang_2019,He_2019,Hagani_2021}. This value was used in order to maintain the physical correct value of $\sigma = 0.072N\ m^{-1}$. This indicates the necessity of using the LBM considering the real physical properties of the simulated systems. This will allow to study the method limitations and advantages from a more realistic point of view. This is the main aim of proposing and using the dimensional LBM. In this LBM model the real physical conditions are automatically considered. The difficulties are related to the simulation of the problems and the obtainment of converged solutions, as is the case with any numerical simulation of a complex problem. 

\FloatBarrier

\subsection{Layered Poiseuille flow}

In this last section, a dynamic multiphase test, namely the layered Poiseuille flow, is accomplished for evaluating the performance of the dimensional LBM. This problem consists in a flow between two parallel plates channel, with one phase occupying the lower half-part of the channel and the other phase, the upper half-part. Then, the fluid is submitted to a constant force field which accelerates the components or phases of the fluid in the channel length direction. When the viscous forces and the force field reach the equilibrium, the system attains the steady state and a constant velocity profile in $x$ direction can be observed.


The problem is studied considering the stationary solution and assuming that the dynamic viscosity varies with the channel height position $y$. Respecting the diffuse interface model followed in the LBM, the NSE can be simplified and described by Eq. \ref{NSE_for_layered_poiseuille}. The profile assumed for the dynamic viscosity is given by Eq. \ref{viscosity_prof}. To obtain a reference solution, the problem was solved considering a central second order FD method, allowing for the comparison between with the solution provided by the LBM. 

It is important to mention that, if it is considered a sharp interface between the phases, the problem has an analytical solution \citep{Ren_2016,Liang_2017}. However, in order to be more coherent in the comparisons between the LBM models and the reference solution, it was preferred the previously described FD scheme, which allows the diffuse interface consideration, represented by Eq. \ref{viscosity_prof}, as it is assumed in the multiphase LBM models. This choice was also made by other authors in the literature \citep{Fakhari_2017,Zhang_2022}. The spatial interval used for the FD scheme was $\Delta y_{FDM} = 4.0 \cdot 10^{-8}m$, in order to respect the convergence criteria established for the FD solutions.

\begin{equation}
    \frac{d}{dy}\left[ \zeta(y)\frac{du(y)}{dy}\right] + F_x = 0
    \label{NSE_for_layered_poiseuille}
\end{equation}

\begin{equation}
    \zeta(y) = \frac{(\zeta_l + \zeta_g)}{2} - \frac{(\zeta_l - \zeta_g)}{2}\mbox{tanh}\left(\frac{2y -H }{W} \right)
    \label{viscosity_prof}
\end{equation}

The same four two-phase systems of sec. \ref{static_bubble} are simulated, considering both LBM models for the air-water system, and just the conventional LBM for the other three cases consisting in saturated water at different temperatures (100$^o$C, 80$^o$C and 25$^o$C). The channel height is taken as $H = 0.50mm$ and as the steady state solution does not depend on the channel length, the total length of the channel was assumed to be the corresponding to $10 \Delta x$, in order to reduce the simulation time. For the problem with LBM, it was selected the $D2Q9$ velocity scheme, with the BKG collision operator for the interface tracker equation (Eq. \ref{eq_lattice_boltzmann_interface}) and MRT for the momentum one. The interface width and the mobility were again assumed to be $W = 25.0\cdot10^{-6}m$ and $M = 1.0 \cdot 10^{-5} m^2\ s^{-1}$. The spatial and time discretization interval selected were $\Delta x = 1.25\cdot10^{-6}m$ and $\Delta t = 1.25\cdot10^{-7}s$ for all the simulations. The driven force of the problem was taken as $F_x = u_c(\zeta_l + \zeta_g)/H^2$, being $u_c = 1.0 \cdot 10^{-4} m\ s^{-1}$ the velocity at the center of the channel.

Furthermore, instead of using the linear relation between $\nu(\mathbf{x})$ and $\phi(\mathbf{x})$, mentioned previously in sec. \ref{multiphase_LBM} for calculating the local $\tau$ values as a function of $\nu(\mathbf{x})$, now the kinematic viscosity is obtained as $\nu(\mathbf{x}) = \zeta(\mathbf{x})/\rho(\mathbf{x})$, where $\zeta(\mathbf{x})$=$\phi(\mathbf{x})(\zeta_l - \zeta_g) + \zeta_g$. This assumption was made in order to obtain better results for the relaxation parameters transition in the interface, as pointed by \cite{Zu_He_2013} and \cite{Fakhari_2017}.

\begin{figure}[h!]
    \centering
    \begin{subfigure}[b]{0.45\textwidth}
        \centering
        \includegraphics[width=\textwidth]{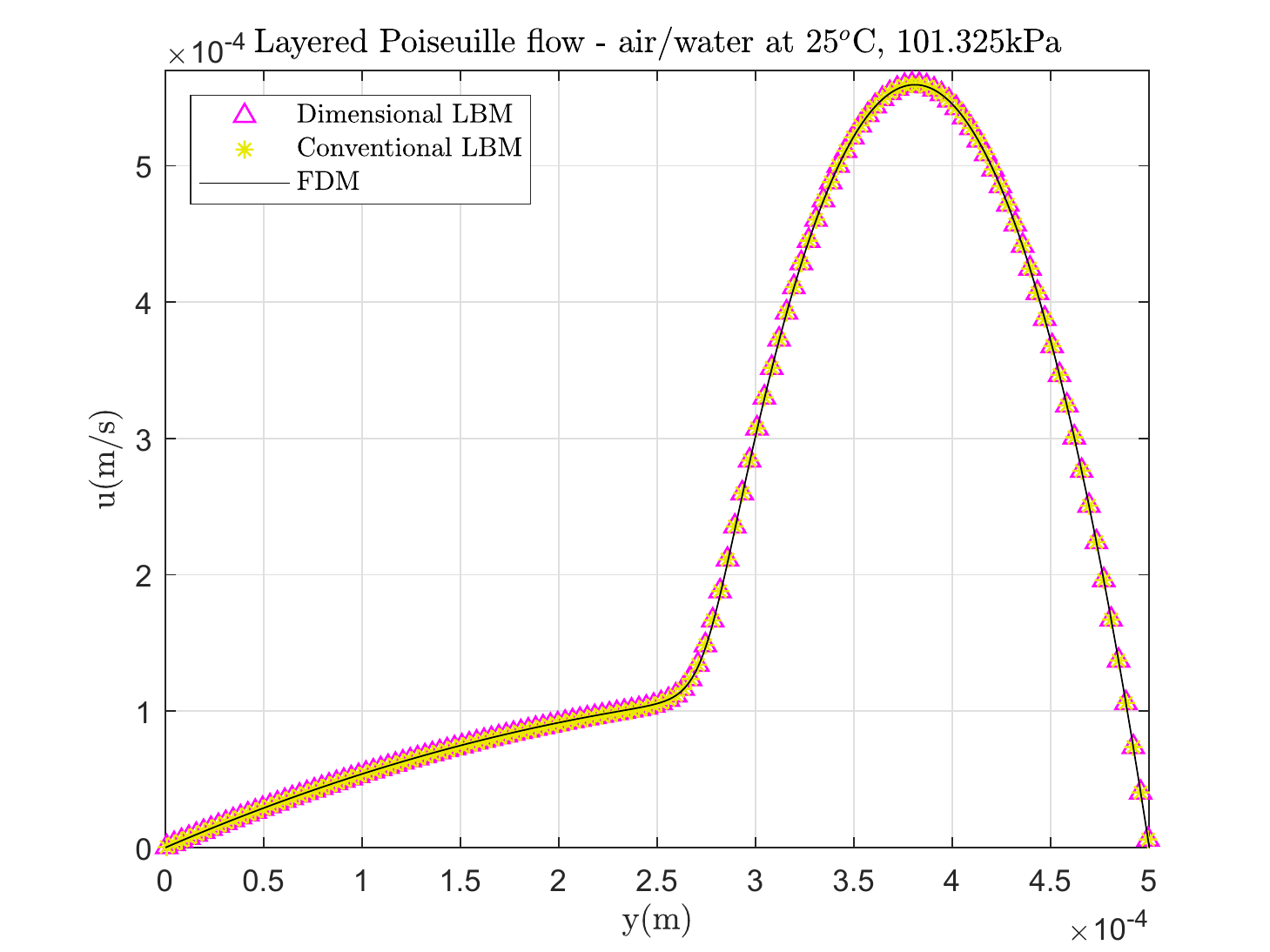}
        \caption{}\label{Layered_poiseuille_air_water}
    \end{subfigure}
    \hfill
    \begin{subfigure}[b]{0.45\textwidth}
        \centering
        \includegraphics[width=\textwidth]{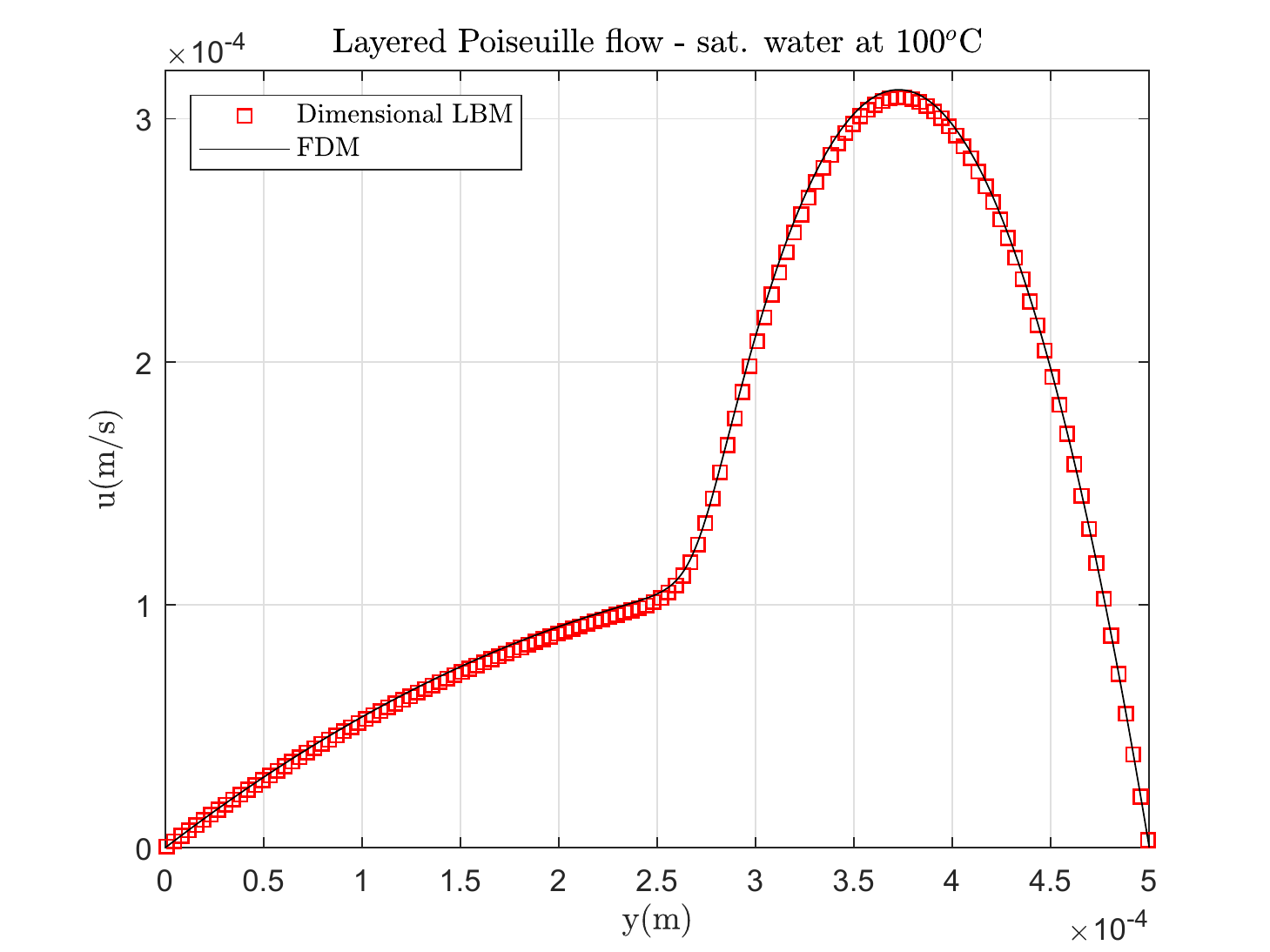}
        \caption{}\label{}
        \label{Layered_poiseuille_Tsat_100}
    \end{subfigure}

    \begin{subfigure}[b]{0.45\textwidth}
        \centering
        \includegraphics[width=\textwidth]{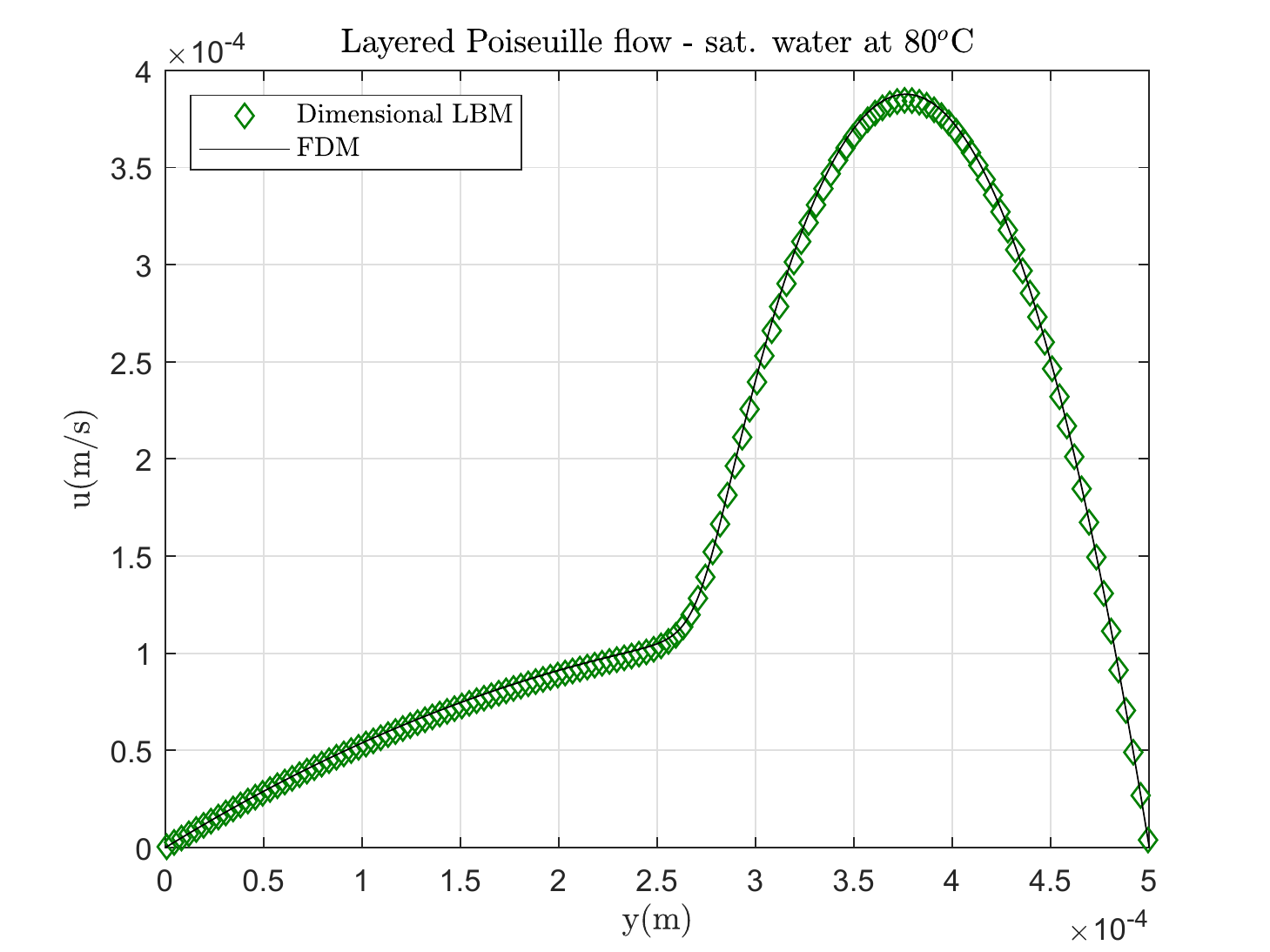}
        \caption{}\label{Layered_poiseuille_Tsat_80}
    \end{subfigure}
    \hfill
    \begin{subfigure}[b]{0.45\textwidth}
        \centering
        \includegraphics[width=\textwidth]{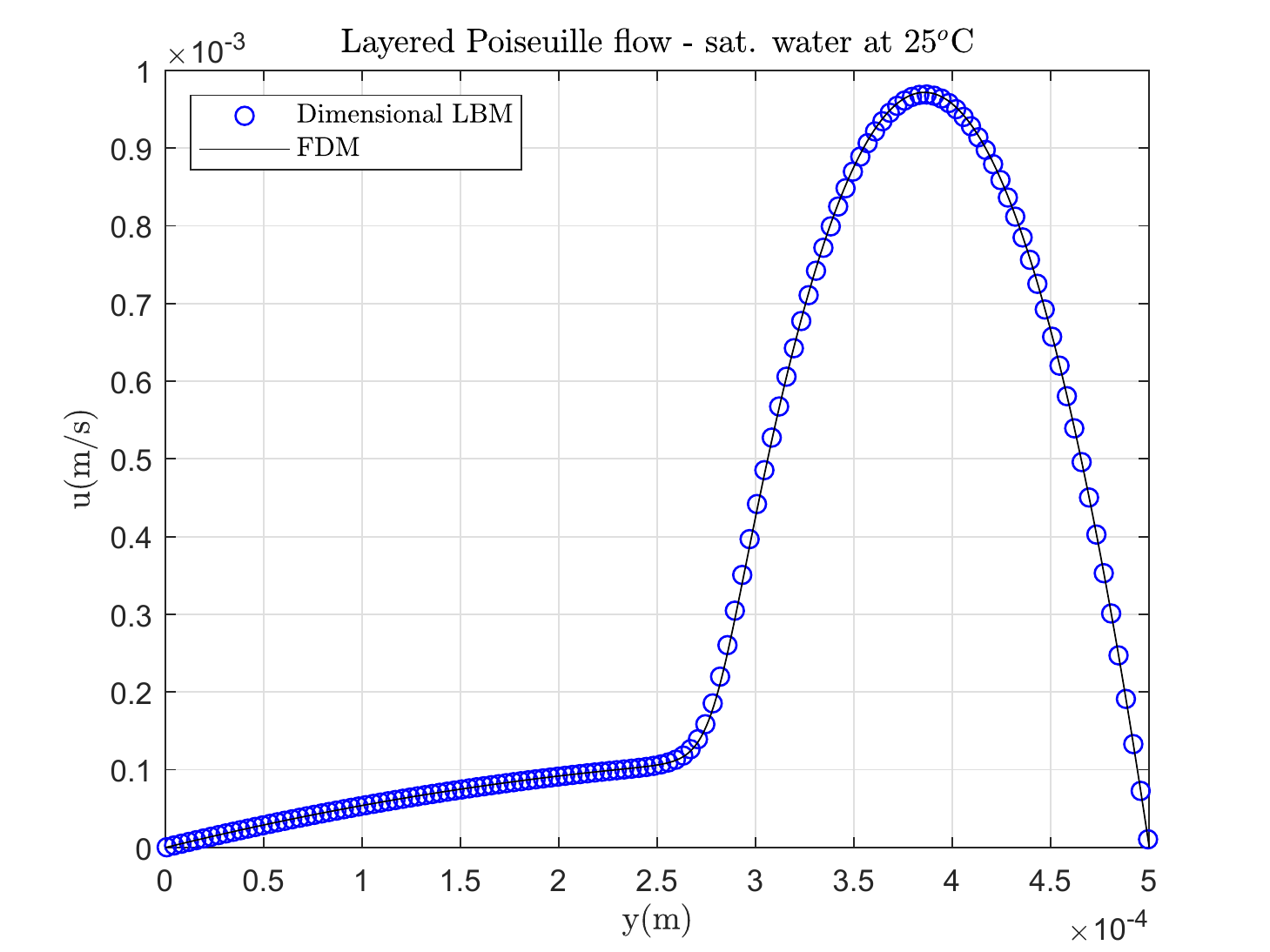}
        \caption{}\label{}
        \label{Layered_poiseuille_Tsat_25}
    \end{subfigure}
   \caption{\centering Velocity profiles for the layered Poiseuille flow given by both LBM and FDM for: (a) air-water system, (b) saturated water at 100$^o$C, (c) saturated water at 80$^o$C, and (d) saturated water at 25$^o$C.}
   \label{Layered_poiseuille_fig}
\end{figure}

The numerical solutions obtained with LBM and FD method are presented in Fig. \ref{Layered_poiseuille_fig} and the respective global errors between LBM results and FD solutions are provided in Tab. \ref{errors_layered_poiseuille}. Once again, both LBM models presented the same errors for the air-water system, indicating the physical coherence of the dimensional LBM. The dimensional LBM model also showed a very good accuracy for the other three cases, related with the saturated water system. It is interesting to observe that the highest global error was obtained for the saturated water at 100$^o$C, which is the system with the smaller kinematic viscosity ($\nu_l = 2.938 \cdot 10^{-7}m^2\ s^{-1}$). This behavior is related to the fact that low kinematic viscosities lead to low $\tau$ values, which may cause lost of accuracy due to the out grow of small instabilities that still do not affect the convergence. In this case of almost pure shear flow the viscosity magnitude exerts more influence over the simulation results than the density ratio. In fact, the global error for the saturated water at 25$^o$C is the smallest one for the water system, even if this case is characterized by the highest density ratio equal to $\rho_l/\rho_g = 43349.0$, which is very high.

\begin{table}[h!]
\centering
\begin{tabular}{lllll}
\hline
\multicolumn{1}{c}{} & \multicolumn{1}{c}{\begin{tabular}[c]{@{}c@{}}Air-water - 25$^o$C, 1 atm\\ both\end{tabular}} & \multicolumn{1}{c}{Sat. water - 100$^o$C} & \multicolumn{1}{c}{Sat. water - 80$^o$C} & \multicolumn{1}{c}{Sat. water - 25$^o$C} \\ \hline
$E_2(\%)$                   & 0.282                                                                                          & 1.628                                 & 1.311                                & 0.817                                \\ \hline
\end{tabular}
\caption{Global relative errors, $E_2$, between LBM simulations and FD solutions for the layered Poiseuille two-phase system cases.}\label{errors_layered_poiseuille}
\end{table}

\begin{figure}[h!]
    \centering
    \includegraphics[width=0.5\textwidth]{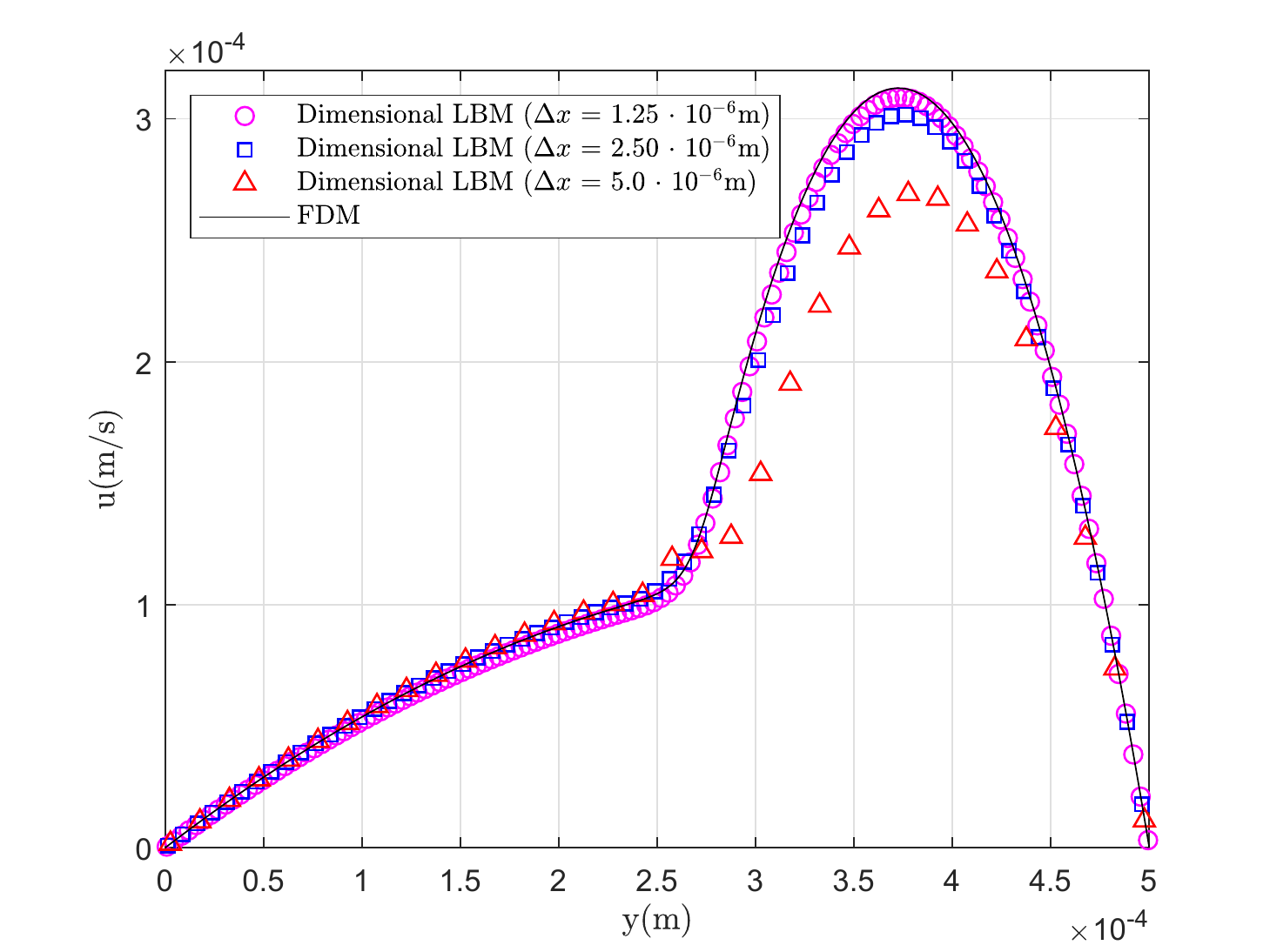}
    \caption{\centering (a) Numerical solutions of the layered Poiseuille flow given by the dimensional LBM for the saturated water system at 100$^oC$, considering three different spatial grids.}
    \label{Grid_ref_sat_water_100}
\end{figure}

Now, in order to illustrate the facilities of the dimensional procedure for performing  numerical simulations, it is shown a simple analysis of grid refinement considering the worst case of the four tested systems, saturated water at 25$^0$C, as an example. First, it was considered a grid with $\Delta x = 5.0 \cdot 10^{-6}m$ and $\Delta t = 1.0 \cdot  10^{-7}s$. This mesh size resulted in results with poor accuracy, providing a global error of $E_2 = 16.132\%$ (see Fig. \ref{Grid_ref_sat_water_100}). Then, the mesh size was refined by a half, trying to improve the solution. In order to respect the LBM stability criteria it was also necessary to change the discrete time interval, using the following new discrete intervals $\Delta x_{new} = 2.50 \cdot 10^{-6}m$ and $\Delta t_{new} = 0.25 \cdot  10^{-7}s$. The global error for this mesh was reduced to almost $E_2 = 3.804\%$. After this step, it was again performed other mesh size refinement, using $\Delta x_{new2} = 1.25 \cdot 10^{-6}m$ and $\Delta t_{new2} = 0.125 \cdot  10^{-7}s$. For this finer mesh it was obtained the smallest error, equal to $E_2 = 1.628\%$, as expected.




A higher refinement was not performed considering the necessary time for performing the simulations and the obtained small global errors for all systems, showed in Table \ref{errors_layered_poiseuille}. The finest mesh size was employed for all the simulations, as expected. From the above explanation it is clear that the proposed dimensional LBM completely avoids the use of the dimensionalization process, focusing only on the setting of the proper spatial and time discretization intervals. The obtained results indicate that the used multiphase LBM is also appropriate for treating dynamic two-phase problems, considering also real very high density and viscosity ratios. These results are rare to be found in the open literature.

\FloatBarrier

\section{Conclusions}
\label{sec:conclusions}


It was presented a new implementation methodology of the lattice Boltzmann method, which considers all the variables in physical units, instead of performing the conversion unit process to the lattice space and vice-versa, as usually. To validate and analyze the applicability of the proposed LBM, several common engineering problems were solved. They include: one-dimensional heat conduction with heat generation; two-dimensional forced convection in a channel under both an oscillating and a constant heat flux, for both developed and developing flows; two-dimensional multiphase stationary bubble in a liquid phase, and two-dimensional multiphase layered Pouseuille flow, both considering real fluids with very high density and viscosity ratios. The solutions for three additional problems including: the one-dimensional advection-diffusion equation, the isothermal channel flow and the natural convection, where briefly presented in Appendices \ref{sec:appendixB}, \ref{sec:appendixC} and \ref{sec:appendixD}. All the numerical results were compared with analytical solutions, when available, and with those given by FD scheme, otherwise. In all cases the simulated results presented a very good accuracy. The following main conclusions are provided.



I. The obtained results confirmed that the dimensional LBM can be safely used for simulating many common transport phenomena involving single-phase fluid flow and heat transfer processes, as well as hydrodynamic static and dynamic two-phase and two component flows. The proposed method produced accurate results, which were very similar to those obtained with the conventional LBM for the same problems. The method can be expanded to three dimensions without major difficulties and also for the use of other collision operators.

II. Giving the discussed results, it is seem of paramount importance to highlight that in this work we were also able to simulate static and dynamic two-phase problems considering real fluids with high density and viscosity ratios, with values over $\rho_l/\rho_g = 43300$ and $\nu_g/\nu_l = 470$, respectively, presenting good accuracy. At the best of authors knowledge, these are new results in the phase-field multiphase Lattice Boltzmann area, which show the power of the LBM model described in sec. \ref{multiphase_LBM} based on the use of Allen-Canh equation, and also the facilities given by the dimensional approach proposed in this paper.


III. The use of the proposed dimensional LBM enables the development of numerical simulations for applied transport phenomena problems using physical units directly, providing results of the same accuracy in relation to the conventional LBM. In fact, the non-dimensionalization process can make the LBM application for simulating applied problems more laborious, demanding additional steps for its implementation. Therefore, the proposed method could enhance the LBM use as simulation tool for an ample spectrum of problems where it is applied. 



In this paper, the dimensional LBM was successfully employed for solving various single-phase fluid and heat transfer problems, and two hydrodynamic two-phase tests using the phase-field LBM. However, in the open literature there are several other LBM models, mainly for simulating multiphase and multicomponent flows, which were not addressed here. Further studies will be developed in order to apply the proposed methodology to some of these models, essentially to the methodologies capable to simulate liquid-gas flows which phase-change, mainly considering two-phase flows in channels and microchannels, boiling in cavities and channels, among others.


\section*{Acknowledgments}

The authors acknowledge the support received from FAPESP (São Paulo Research Foundation, grants 2019/21022-9 and 2016/09509-1) and CNPq (National Council for Scientific and Technological Development, process 305941/2020-8).

\appendix
\renewcommand{\thesection}{\Alph{section}}
\numberwithin{equation}{section}
\numberwithin{figure}{section}
\numberwithin{table}{section}

\section{Inverse transformation matrices}
\label{sec:appendixA}

The inverse of the transformation matrix for the $D2Q9$ velocity scheme can be given by Eq. \ref{Minv_matrix}, in the dimensionless case, and by Eq. \ref{Minv_matrix_dim}, for the dimensional LBM.

\begin{equation}
    [\mathbf{M}]^{-1} = \left (\begin{matrix}
    \frac{1}{9} & \frac{-1}{9} & \frac{1}{9} &   0   &   0    &   0   & 0    & 0      &   0 \\
    \frac{1}{9} & \frac{-1}{36} & \frac{-1}{18} & \frac{1}{6} & \frac{-1}{6} &   0   &   0 & \frac{1}{4} &   0\\
    \frac{1}{9} & \frac{-1}{36} & \frac{-1}{18} &   0   &   0    & \frac{1}{6} & \frac{1}{6} & \frac{-1}{4} &   0 \\
    \frac{1}{9} & \frac{-1}{36} & \frac{-1}{18} & \frac{-1}{6} & \frac{1}{6} &   0   &   0 & \frac{1}{4} &   0 \\
    \frac{1}{9} & \frac{-1}{36} & \frac{-1}{18} &   0   &   0    & \frac{-1}{6} & \frac{1}{6} & \frac{-1}{4} &   0 \\
    \frac{1}{9} & \frac{1}{18} & \frac{1}{36} & \frac{1}{6} & \frac{1}{12} & \frac{1}{6} & \frac{1}{12} &   0   & \frac{1}{4} \\
    \frac{1}{9} & \frac{1}{18} & \frac{1}{36} & \frac{-1}{6} & \frac{-1}{12} & \frac{1}{6} & \frac{1}{12} &   0   & \frac{-1}{4} \\
    \frac{1}{9} & \frac{1}{18} & \frac{1}{36} & \frac{-1}{6} & \frac{-1}{12} & \frac{-1}{6} & \frac{-1}{12} &   0   & \frac{1}{4} \\
    \frac{1}{9} & \frac{1}{18} & \frac{1}{36} & \frac{1}{6} & \frac{1}{12} & \frac{-1}{6} & \frac{-1}{12} &   0   & \frac{-1}{4} 
    \end{matrix}\right)
    \label{Minv_matrix}
\end{equation}

\begin{equation}
    [\mathbf{M}]_{dim}^{-1} = \left (\begin{matrix}
    \frac{1}{9} & \frac{-1}{9c^2} & \frac{1}{9c^4} &   0   &   0    &   0   & 0    & 0    &   0 \\
    \frac{1}{9} & \frac{-1}{36c^2} & \frac{-1}{18c^4} & \frac{1}{6c} & \frac{-1}{6c^3} &   0   &   0 & \frac{1}{4c^2} &   0\\
    \frac{1}{9} & \frac{-1}{36c^2} & \frac{-1}{18c^4} &   0   &   0    & \frac{1}{6c} & \frac{1}{6c^3} & \frac{-1}{4c^2} &   0 \\
    \frac{1}{9} & \frac{-1}{36c^2} & \frac{-1}{18c^4} & \frac{-1}{6c} & \frac{1}{6c^3} &   0   &   0 & \frac{1}{4c^2} &   0 \\
    \frac{1}{9} & \frac{-1}{36c^2} & \frac{-1}{18c^4} &   0   &   0    & \frac{-1}{6c} & \frac{1}{6c^3} & \frac{-1}{4c^2} &   0 \\
    \frac{1}{9} & \frac{1}{18c^2} & \frac{1}{36c^4} & \frac{1}{6c} & \frac{1}{12c^3} & \frac{1}{6c} & \frac{1}{12c^3} &   0   & \frac{1}{4c^2} \\
    \frac{1}{9} & \frac{1}{18c^2} & \frac{1}{36c^4} & \frac{-1}{6c} & \frac{-1}{12c^3} & \frac{1}{6c} & \frac{1}{12c^3} &   0   & \frac{-1}{4c^2} \\
    \frac{1}{9} & \frac{1}{18c^2} & \frac{1}{36c^4} & \frac{-1}{6c} & \frac{-1}{12c^3} & \frac{-1}{6c} & \frac{-1}{12c^3} &   0   & \frac{1}{4c^2} \\
    \frac{1}{9} & \frac{1}{18c^2} & \frac{1}{36c^4} & \frac{1}{6c} & \frac{1}{12c^3} & \frac{-1}{6c} & \frac{-1}{12c^3} &   0   & \frac{-1}{4c^2} 
    \end{matrix}\right)
    \label{Minv_matrix_dim}
\end{equation}

\FloatBarrier

\section{1D advection-diffusion equation}
\label{sec:appendixB}

In this example is solved the advection-diffusion equation in a one-dimensional domain. It is considered air at mean temperature of $335.50 K$, with the following properties: $\rho = 1.052kg\ m^{-3}$, $c_p = 1008.174 J\ kg^{-1}K^{-1}$ and $k = 0.029 W\ m^{-1}K^{-1}$. Initially, the domain is at $T_{ini} = 298.0K$, and suddenly the right boundary is submitted to a temperature of $T_L = 373.0 K$, while the other extremity is kept at $T_0 = 298.0K$. The domain length is $L = 1.0m$ and the air is moving with a constant speed $u = 0.001m\ s^{-1}$.


The macroscopic equation which represents the physical problem is expressed by Eq. \ref{energy_eq_advection_diffusion_1D} and the corresponding analytical solution for the steady-state condition is given by Eq. \ref{analytical_advection_diffusion_1D}.

\begin{equation}
   u\frac{\partial T}{\partial x} = \alpha \frac{\partial^2 T}{\partial x^2}
    \label{energy_eq_advection_diffusion_1D}
\end{equation}

\begin{equation}
   T(x) = T_0 + (T_L - T_0)\left[\frac{\mbox{exp}\left(\frac{\rho c_p u x}{k} - 1 \right)}{\mbox{exp}\left(\frac{\rho c_p u L}{k} - 1 \right)} \right]
    \label{analytical_advection_diffusion_1D}
\end{equation}

For this particular problem, given its simplicity for treating boundary conditions in the one-dimensional case, it was used the wet-node scheme for the boundaries, instead of the link-wise. This last scheme was used in the rest of all simulations performed in the paper. Then, using the $D1Q3$ velocity scheme, the BCs were implemented according to Eq. \ref{BCs_1D_advection_diffusion}. The collision operator considered was the traditional BGK operator, represented in Eq. \ref{eq_lattice_boltzmann_BGK_T}.

\begin{equation}
   \begin{cases}
       g_1(0,t+\Delta t) = T_0 - g_0(0,t+\Delta t) - g_2(0,t+\Delta t),\mbox{ for }x=0.0;\\
       g_2(L,t+\Delta t) = T_L - g_0(L,t+\Delta t) - g_1(L,t+\Delta t),\mbox{ for }x=L;
   \end{cases}
    \label{BCs_1D_advection_diffusion}
\end{equation}

\begin{figure}[h!]
    \centering
        \centering
        \includegraphics[width=0.5\textwidth]{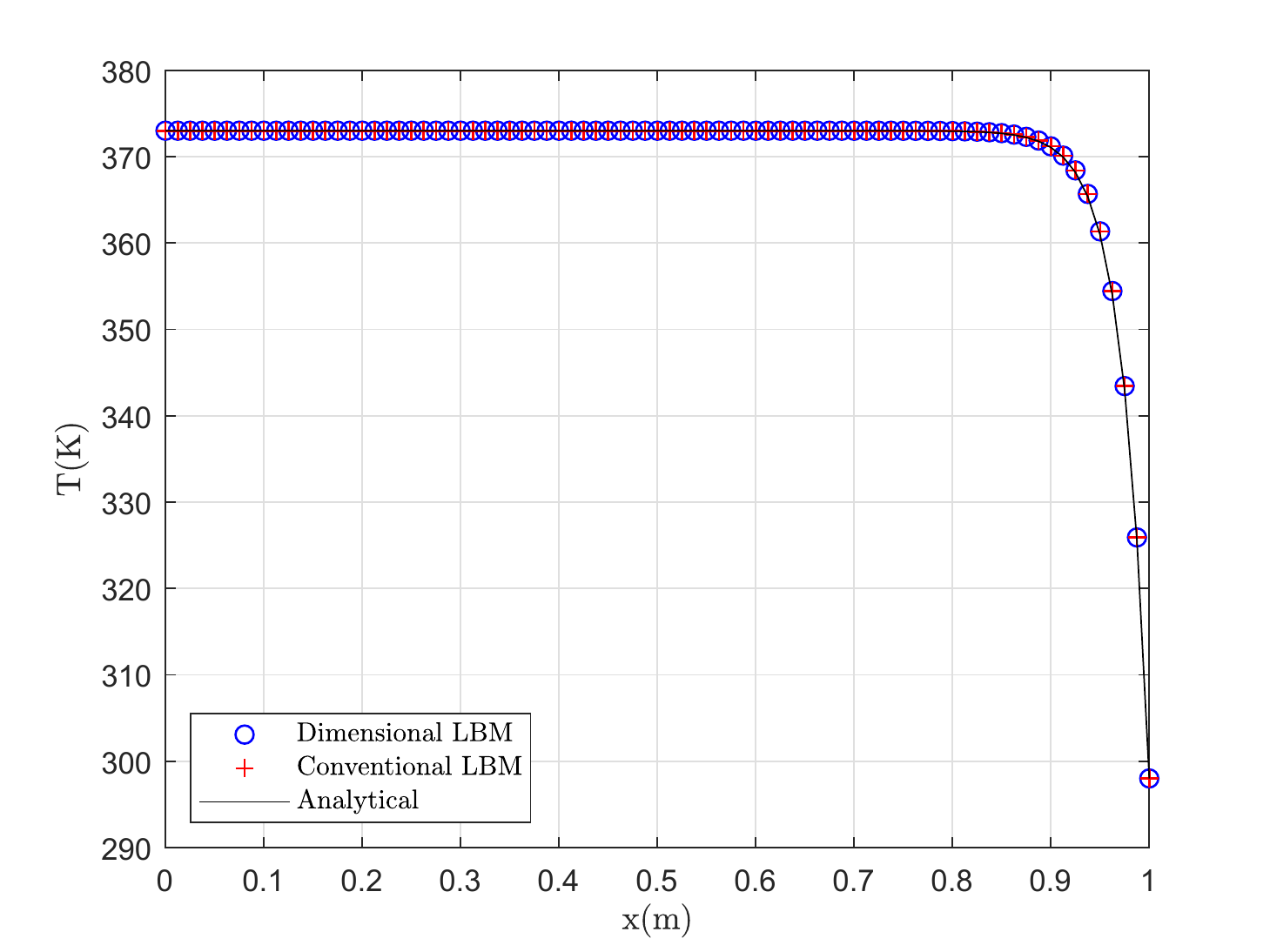}
       \caption{\centering Steady-state temperature distribution for the one-dimensional advection-diffusion problem.}
       \label{advection_diffusion}
\end{figure}

The LBM models are solved considering $\Delta x = 0.0125 m$ and $\Delta t = 0.10 s$. The numerical solutions with LBM and the analytical solution are all displayed in Fig. \ref{advection_diffusion}. The comparison of the solutions resulted in a global error of $E_2 = 0.030 \%$ for both LBM models in relation to the theoretical solution.


The provided results verify the correctness of the dimensional LBM.

\FloatBarrier

\section{Isothermal channel flow}
\label{sec:appendixC}

The LBM models were also applied for simulating an isothermal Poiseuille flow between two parallel plates. The distance between the plates was assumed $H = 0.50 mm$ and as the analytical solution doesn't depends of the channel length (given by Eq. \ref{velocity_prof}), the length was taken equal to 10 computational cells ($10 \Delta x$), considering the $D2Q9$ velocity scheme with $\Delta x = 5.0\cdot10^{-6}m$ and $\Delta t = 1.0\cdot10^{-7}s$. 

\begin{figure}[h!]
    \centering
        \centering
        \includegraphics[width=0.5\textwidth]{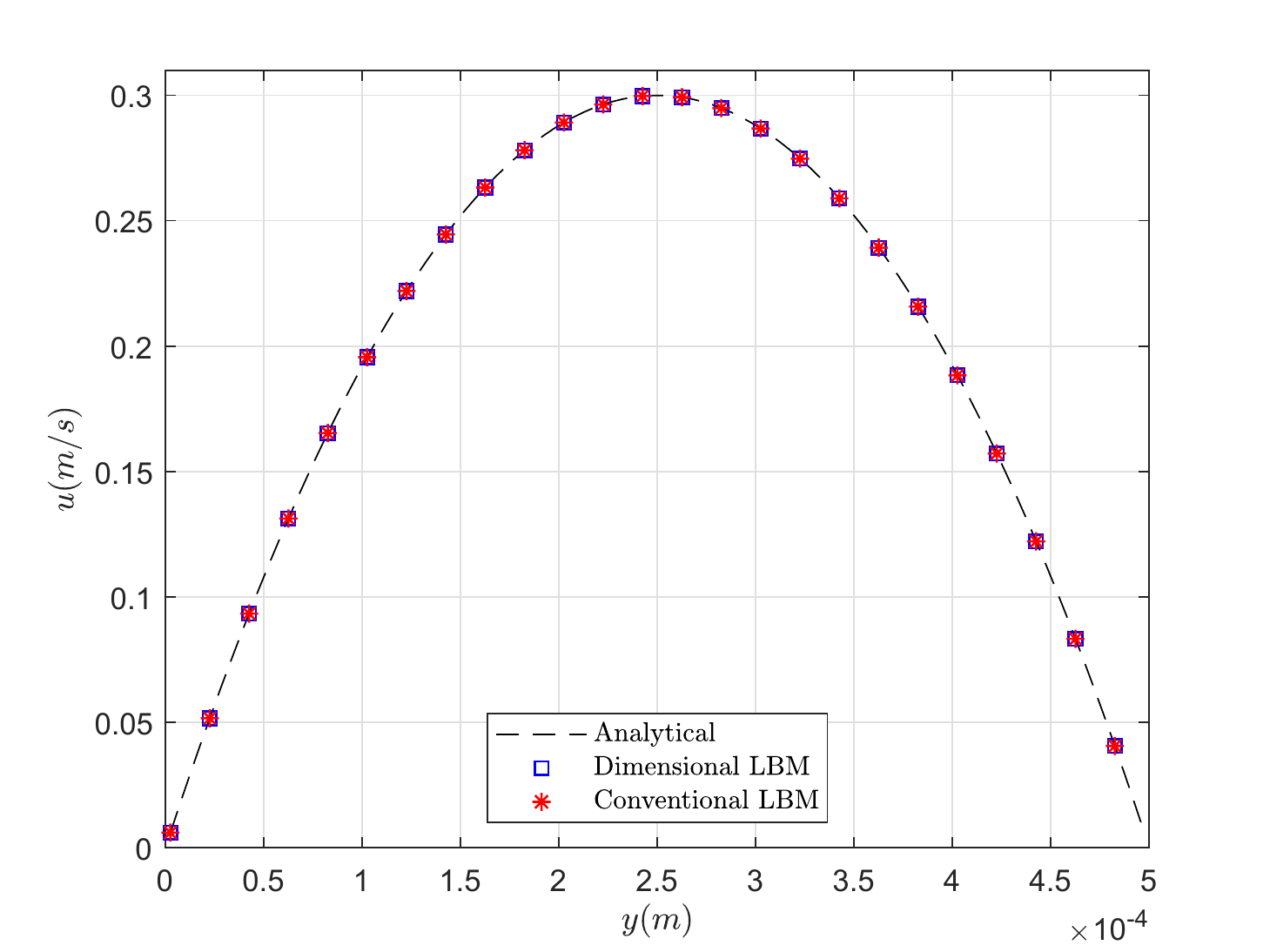}
       \caption{\centering Steady-state velocity profiles of the isothermal Poiseuille flow between two parallel plates obtained for the dimensional and conventional LBM simulations and the analytical solution.}
       \label{u_poiseuille}
\end{figure}

The mean velocity of the channel was assumed to be $u_m = 0.20 m\ s^{-1}$ and the driven force in $x$ direction, as $F_x = 12u_m\zeta/H^2$. The fluid is water at a mean temperature of $301K$, whose properties are given in Tab. \ref{water_properties_channel}. The results are shown in Fig. \ref{u_poiseuille}, and the global errors of both LBM models in comparison with analytical solution were found equal to $E_2^{dim} = E_2^{conv} = 0.011\%$, showing a very good accuracy.

\FloatBarrier

\section{Natural convection in a square enclosure}
\label{sec:appendixD}

The natural convection in a square enclosure with length $L = 0.0130m$ containing air initially at $T_{ini} = 293.85K$ is considered in this Appendix. The left wall of the domain is kept at a higher constant temperature equal to $T_h = 373.15 K$, while the right one remains at the initial temperature of $T_c = 293.85K$. The top and bottom walls are considered as insulated. The air properties are calculated at a reference temperature of $T_{ref} = 333.50K$: $\rho = 1.059kg\ m^{-3}$, $c_p = 1008.045 J\ kg^{-1}K^{-1}$, $k = 0.029 W\ m^{-1}K^{-1}$, $\alpha = 2.702 \cdot 10^{-5} m^2\ s$, $\nu = 1.90 \cdot 10^{-5} m^2\ s$ and $\beta_{exp} = 3.004 \cdot 10^{-3} K^{-1}$ (thermal expansion coefficient).
 
The temperature difference between the walls causes a mass flux because of the density difference between the hot and the cold fluids. In order to consider this effect without changing the density of the fluid in the simulations, it is assumed a buoyancy force given by Eq. \ref{buoyancy_force} \citep{Merzhab_2010,Mohamad_2010,Wang_2013}. This is the so called Boussinesq approximation. In this equation, $\overline{\rho}$ is the reference density, calculated at the reference temperature $T_{ref}$, and $\mathbf{g} = (0,-9.81)m\ s^{-2}$ is the gravitational acceleration. 

\begin{equation}
    \mathbf{F_b}(\mathbf{x},t) = -\overline{\rho} \beta_{exp} \left[T(\mathbf{x},t) - T_{ref} \right]\mathbf{g}
    \label{buoyancy_force}
\end{equation}


It was considered the $D2Q9$ velocity set, with the BGK collision operator for the momentum LBE and the MRT for the thermal LBE. For the stationary walls, it was used the bounce-back BC for the momentum distribution function (Eq. \ref{BB_vel}). Moreover, for the fixed temperature BCs (left and right walls) it was employed the anti-bounce-back rule (Eq. \ref{anti_BB_T}), and the top and bottom walls were modeled as thermally insulated, just applying the BB rule (Eq. \ref{BB_T}) with zero heat flux ($q'' = 0$). 

The problem can be characterized by the Rayleigh number (Eq. \ref{Rayleigh}), which will be considered as $Ra = 10^4$ for the first case and $Ra = 10^6$, for the second one. In this last test, to obtaing $Ra = 10^6$ without changing the mean temperature of the fluid, it was considered a new size of the square cavity, euqal to $L = 0.60 m$, and the wall temperatures were changed to $T_h = 373.85 K$ and $T_c = 293.15K$, keeping $T_{ref} = 333.50K$. Therefore, the air properties in both tests were kept constant and unchanged.

\begin{equation}
    Ra = \frac{|\mathbf{g}|\beta_{exp}L^3(T_h - T_c)}{\nu \alpha}
    \label{Rayleigh}
\end{equation}

\begin{figure}[h!]
    \centering
    \begin{subfigure}[b]{0.45\textwidth}
        \centering
        \includegraphics[width=\textwidth]{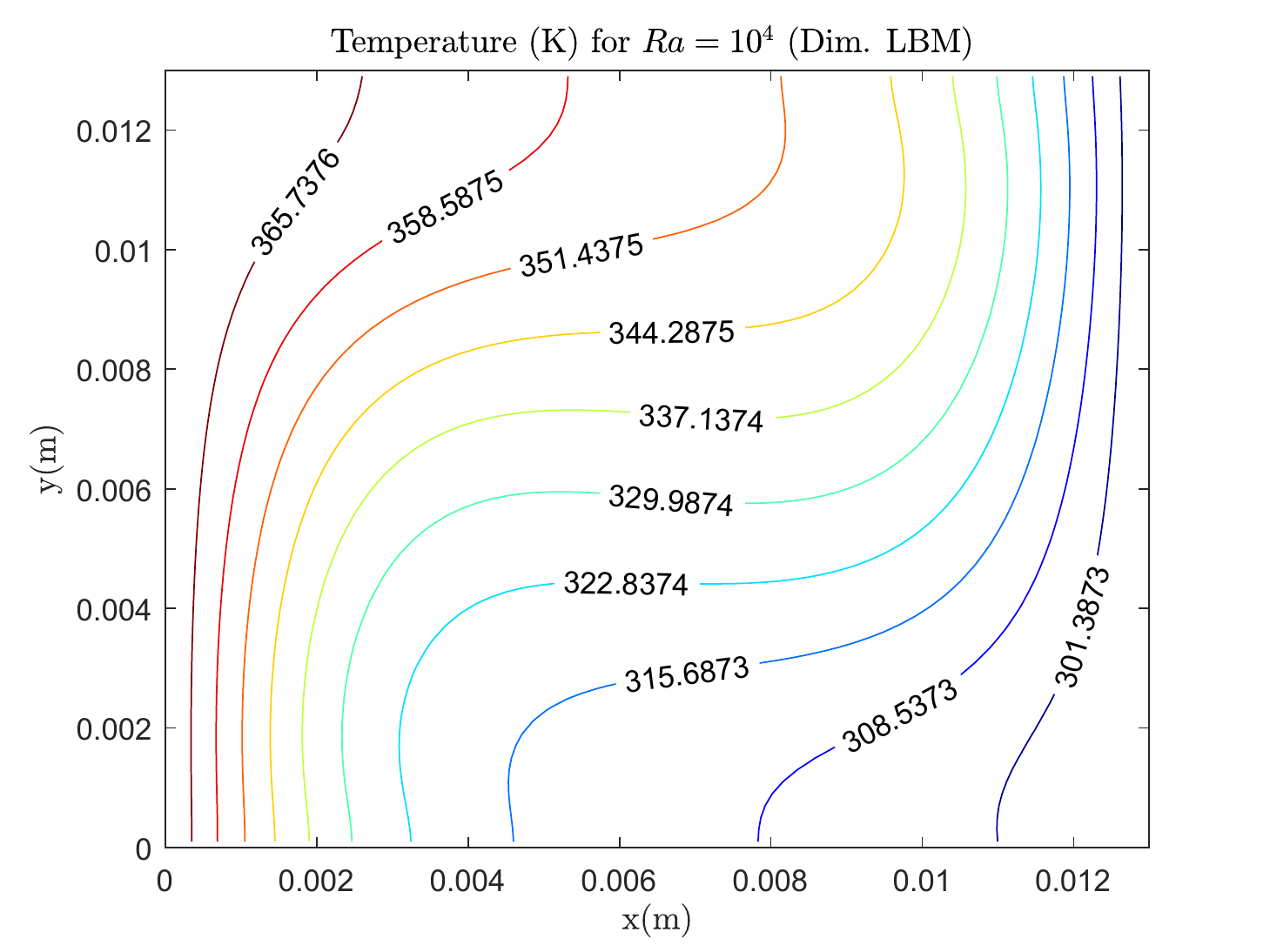}
        \caption{}
    \end{subfigure}
    \hfill
    \begin{subfigure}[b]{0.45\textwidth}
        \centering
        \includegraphics[width=\textwidth]{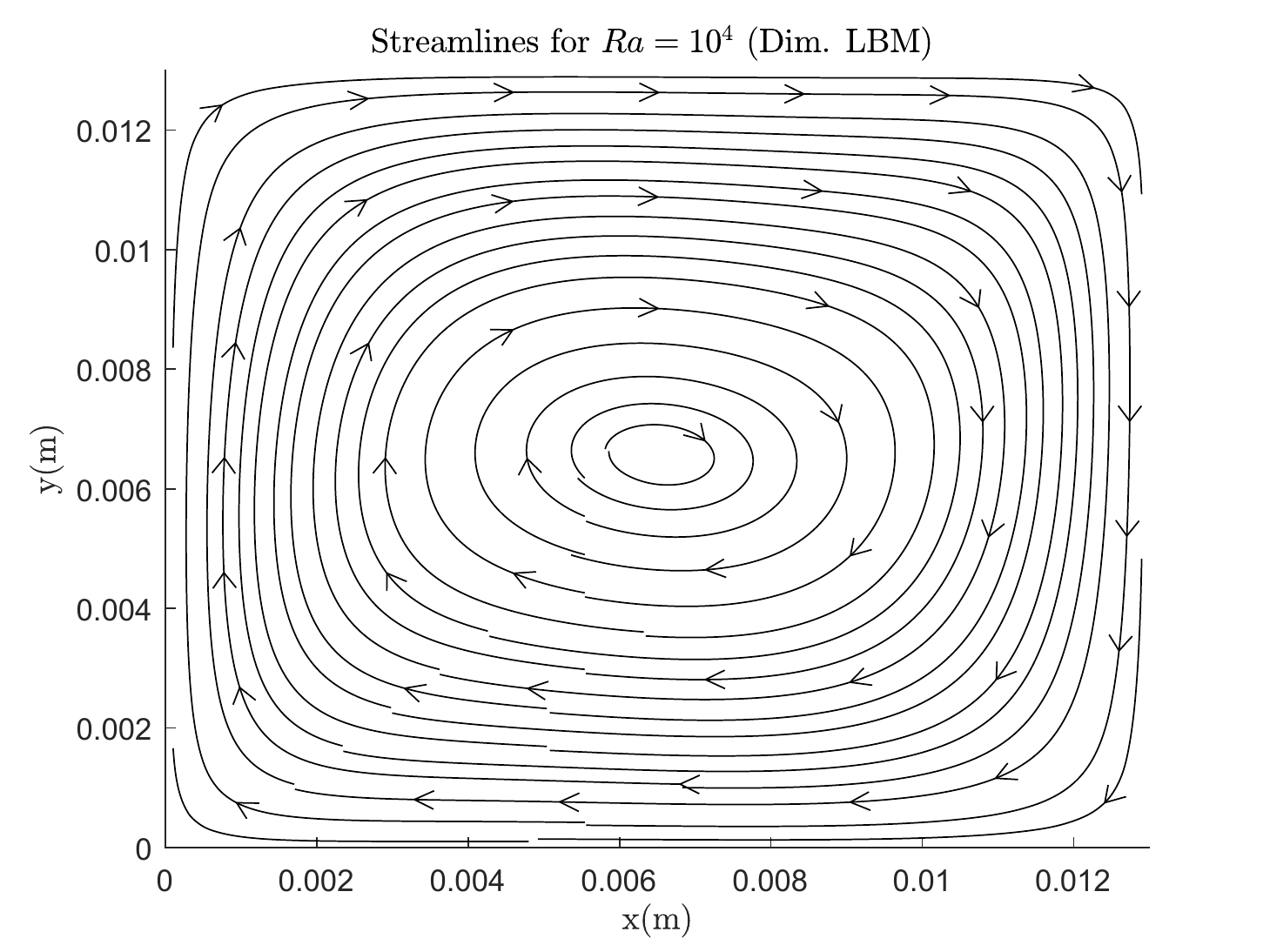}
        \caption{}
    \end{subfigure}
       \caption{\centering Simulated temperature contours (a) and streamlines (b) for $Ra = 10^4$ with the dimensional LBM.}
       \label{Conv_nat_Ra_10e4}
\end{figure}

\begin{figure}[h!]
    \centering
    \begin{subfigure}[b]{0.45\textwidth}
        \centering
        \includegraphics[width=\textwidth]{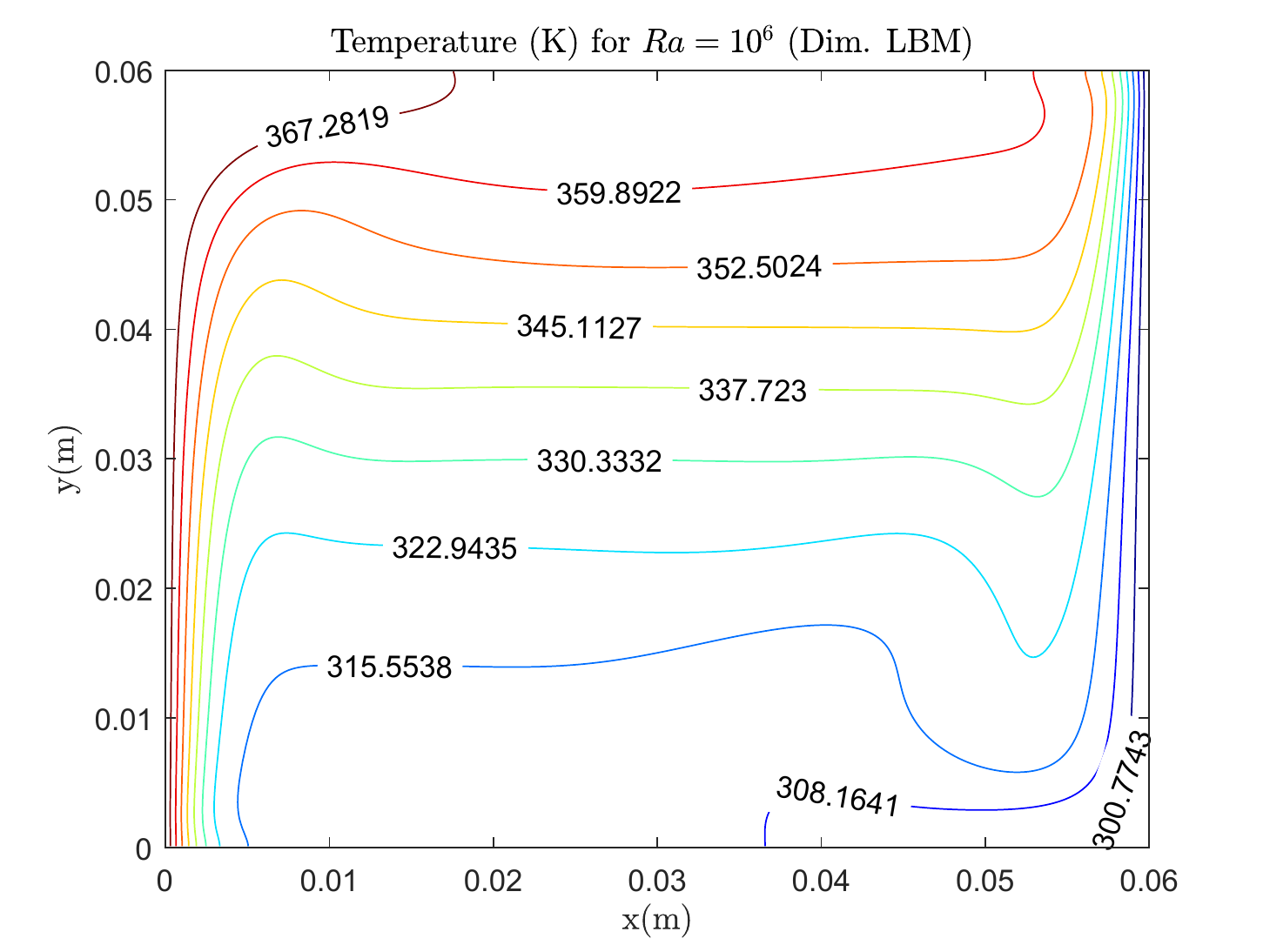}
        \caption{}
    \end{subfigure}
    \hfill
    \begin{subfigure}[b]{0.45\textwidth}
        \centering
        \includegraphics[width=\textwidth]{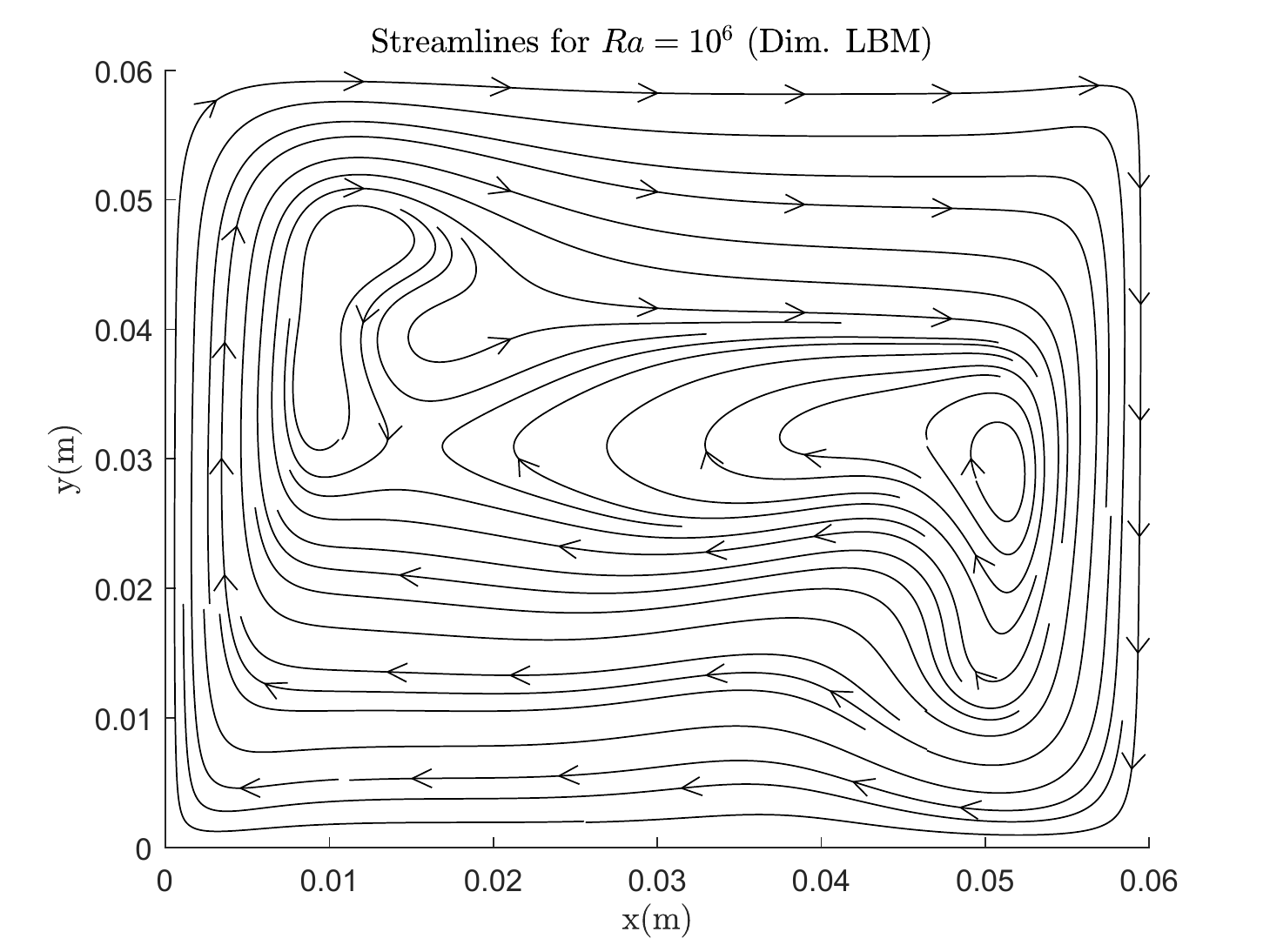}
        \caption{}
    \end{subfigure}
       \caption{\centering Simulated temperature contours (a) and streamlines (b) for $Ra = 10^6$ with the dimensional LBM.}
       \label{Conv_nat_Ra_10e6}
\end{figure}

For both cases it was considered discrete space and time intervals equal to $\Delta x = 2.0 \cdot 10^{-4}m$ and $\Delta t = 2.0 \cdot 10^{-4}s$, respectively. The steady-state results for the temperature contours and streamlines are presented in Figs. \ref{Conv_nat_Ra_10e4} and \ref{Conv_nat_Ra_10e6}. In order to evaluate the dimensional LBM, its solution is compared with the results from conventional LBM, and both numerical solutions are validated through a comparison with the benchmark solutions found in the literature \citep{De_vahl_davis}. All these solutions are presented in Tab. \ref{results_nat_conv}. Both LBM models shown a good agreement with the benchmark expected values, presenting very small global errors, $E_2$.

\newpage

\begin{table}[h!]
\centering
\begin{tabular}{ccccc}
\hline
          & \multicolumn{2}{c}{$Ra = 10^4$} & \multicolumn{2}{c}{$Ra = 10^6$} \\
          & $\overline{Nu}$   & Error(\%)  & $\overline{Nu}$   & Error(\%)  \\ \hline
Benchmark \citep{De_vahl_davis} & 2.243 & - & 8.800 & - \\
Dim. LBM & 2.242 & 0.045 & 8.805 & 0.057 \\
Conv. LBM & 2.242 & 0.045 & 8.794 & 0.068 \\ \hline 
\end{tabular}
\caption{Calculated average Nusselt numbers from the simulated results by both LBM models and the benchmark solution \citep{De_vahl_davis}.}\label{results_nat_conv}
\end{table}

\FloatBarrier

\clearpage

\bibliographystyle{unsrtnat}

\bibliography{references}

\end{document}